\documentclass[useAMS,onecolumn,doublespacing,referee,usenatbib]{mn2e}

\usepackage[dvips]{graphicx,epsfig,color}
\usepackage{amsmath,amssymb}
\usepackage{subfigure}

\makeatletter

\newcommand{\Rmnum}[1]{\expandafter\@slowromancap\romannumeral #1@}
\makeatother

\newcommand\budda{{\sc BUDDA}}

\newcommand\Ks{$K_s$}

\newcommand\etal{et al.}
\newcommand\galfit{{\scshape Galfit}}
\newcommand\gasphot{{\scshape Gasphot}}
\newcommand\gasptd{{\scshape Gasp2d}}
\newcommand\gimtd{{\scshape Gim2d}}

\newcommand\bie{{\rm BIE}}
\newcommand\sn{{S/N}}

\newcommand\sersic{S\'{e}rsic} 
\newcommand\galphat{{\sc Galphat}}
\newcommand\devauc{de Vaucouleur}

\definecolor{Purple}{cmyk}{0.45,0.86,0,0}

\title[New insights on galaxy structure from GALPHAT \Rmnum{1}]{New
  insight on galaxy structure from GALPHAT \Rmnum{1}. Motivation,
  methodology, and benchmarks for \sersic\ models}

\author[Ilsang Yoon, Martin D. Weinberg and Neal Katz]{Ilsang
  Yoon$^{1}$\thanks{E-mail:
    iyoon@astro.umass.edu (IY)}, Martin D. Weinberg$^{1}$ and Neal Katz$^{1}$ \\
  $^{1}$Department of Astronomy, University of Massachusetts, Amherst,
  MA 01003-9305, USA}

\begin{document}

\pagerange{\pageref{firstpage}--\pageref{lastpage}} \pubyear{2010}

\maketitle

\label{firstpage}

\begin{abstract}
  We introduce a new galaxy image decomposition tool, \galphat\
  (GALaxy PHotometric ATtributes), which is a front-end 
  application of the Bayesian Inference Engine (BIE), 
  a parallel Markov chain Monte Carlo package, to provide 
  full posterior probability distributions and reliable confidence 
  intervals for all model parameters.  
  The BIE relies on \galphat\ to compute the
  likelihood function.  \galphat\ generates scale-free cumulative
  image tables for the desired model family with precise error
  control.  Interpolation of this table yields accurate pixelated
  images with any centre, scale, and inclination angle.  \galphat\
  then rotates the image by position angle using a Fourier shift
  theorem, yielding a high speed and accurate likelihood computation.
  
  We benchmark this approach using an ensemble of simulated \sersic\
  model galaxies over a wide range of observational conditions:
  the signal-to-noise ratio \sn, the ratio of galaxy size to the PSF and the 
  image size, and errors in the assumed PSF; and a range of structural
  parameters: the half-light radius $r_e$ and the \sersic\ index $n$. 
  We characterise the strength of parameter covariance in \sersic\ 
  model, which increases with \sn\ and $n$, and the results strongly 
  motivate the need for the full posterior probability distribution in 
  galaxy morphology analyses and later inferences.

  The test results for simulated galaxies 
  successfully demonstrate that, with a careful choice of Markov chain 
  Monte Carlo algorithms and fast model image generation, \galphat\ is 
  a powerful analysis tool for reliably inferring morphological 
  parameters from a large ensemble of galaxies over a wide range of 
  different observational conditions.
\end{abstract}


\section{Introduction}
The formation and evolution of galaxies is an outstanding problem in
Astronomy and galaxy morphology remains a key observational attribute in
the quest to increase our understanding of galaxy evolution.  
The increasing sensitivity
and resolution of planned surveys will enable tests of evolution
theories from the epoch of formation to the present.  However,
selection effects and features peculiar to one's choice of models will
affect any interpretation.  Therefore, to exploit the promise of
survey data, we need to verify that our conclusions are reliable.  The
tools described in this paper are a step in this direction.

Early $\Lambda$CDM hierarchical galaxy formation theory and
simulations placed galaxies in the Hubble sequence by following a combination
of merger histories and gas accretion \citep{white1991,steinmetz2002}. Later,
``zoom-in'' resimulations of individual galaxies have produced more
realistic galaxy morphologies
\citep{abadi2003a,abadi2003b,sommer-larsen2003,
  governato2004,robertson2004,zavala2008}.  Although challenging,
combinations of semi-analytic models and direct simulations
\citep{benson2010,croft2009,parry2009,scannapieco2010} have quantified the
distribution of galaxy morphology with redshift.

These recent theoretical studies have been motivated by large-scale
spectroscopic and image surveys.  In the local Universe, millions of
galaxies have been detected in the Sloan Digital Sky Survey
\citep[SDSS,][]{york2000} and the Two Micron All Sky Survey
\citep[2MASS,][]{skrutskie1997,skrutskie2006}.  Recent analyses have
used a range of models from single-component \sersic\ profiles
\citep{sersic1963} to more sophisticated bulge and disc-bar models to
characterise the structural properties of local galaxy morphology
\citep{blanton2003, allen2006, gadotti2009}.  In the more distant
Universe, COSMOS \citep{scoville2007} provides an ample collection of
multi-wavelength galaxy images and spectroscopy to study the evolution
of galaxy morphological structure for $z\la1$ as a function of mass
and environment \citep{capak2007,cassata2007,kovac2010}.  Gas
accretion, disc instability, mergers, and supernova and black hole
feedback have been modelled to explain morphological evolution and it
is possible to assess their relative importance by quantitatively
comparing observed galaxy morphological structures to those predicted
from theory.  Furthermore, future large-scale multi-band imaging
surveys, such as the Large Synoptic Survey Telescope \citep{lsst2009},
combined with accurate photometric distances will provide a uniform
and consistent data set to study the evolution of the galaxy
population.

Algorithmic approaches to \emph{measure} galaxy morphology are recent
inventions and are usually based on mixture models of parametric
surface brightness distributions \citep{byun1995, simard1998, wadadekar1999,
peng2002, simard2002, macarthur2003, desouza2004, pignatelli2006, mendez2008}.
However, systematic biases owing to
an ignorance of uncertainties in the sky background and covariances
between model parameters complicate their interpretation. 
To circumvent these difficulties, we advocate embedding the galaxy morphology 
analysis into the broader context of inference and hypothesis testing.  
In this paper, we present such a Bayesian approach using the Markov chain Monte
Carlo (MCMC) technique, facilitated by embedding it within the
Bayesian Inference Engine \citep[BIE,][]{weinberg2010}.  To motivate
this approach, we first illustrate the inherent difficulties in galaxy
image decomposition in \S\S\ref{casestudy}--\ref{rescue}.

\subsection{Case studies}
\label{casestudy}
The following three examples explore the limitations of conventional
model fitting and the improvements gained using Bayesian inference
when inferring the photometric attributes of galaxies.

\subsubsection{Posterior distributions  versus best-fit parameters}
We pick two galaxies from our pool of \sersic\ profile simulated
galaxy images (see \S\ 1.1.2).  One has a high
signal-to-noise ratio, \sn$=100.01$, and the other has a low
signal-to-noise ratio, \sn$=10.46$.  The value of \sn\ is defined by
the ratio of the galaxy signal to the noise within the half-light radius (see
\S\ref{datasectionsn}).  Since we know the \sersic\ model parameter
values used to generate these galaxy images, we fix all parameters to their
true values, except for the \sersic\ index $n$, and calculate the $\chi^2$
likelihood for different values of $n$.

\begin{figure}
  \centering
  \subfigure[Likelihood]{\epsfig{figure=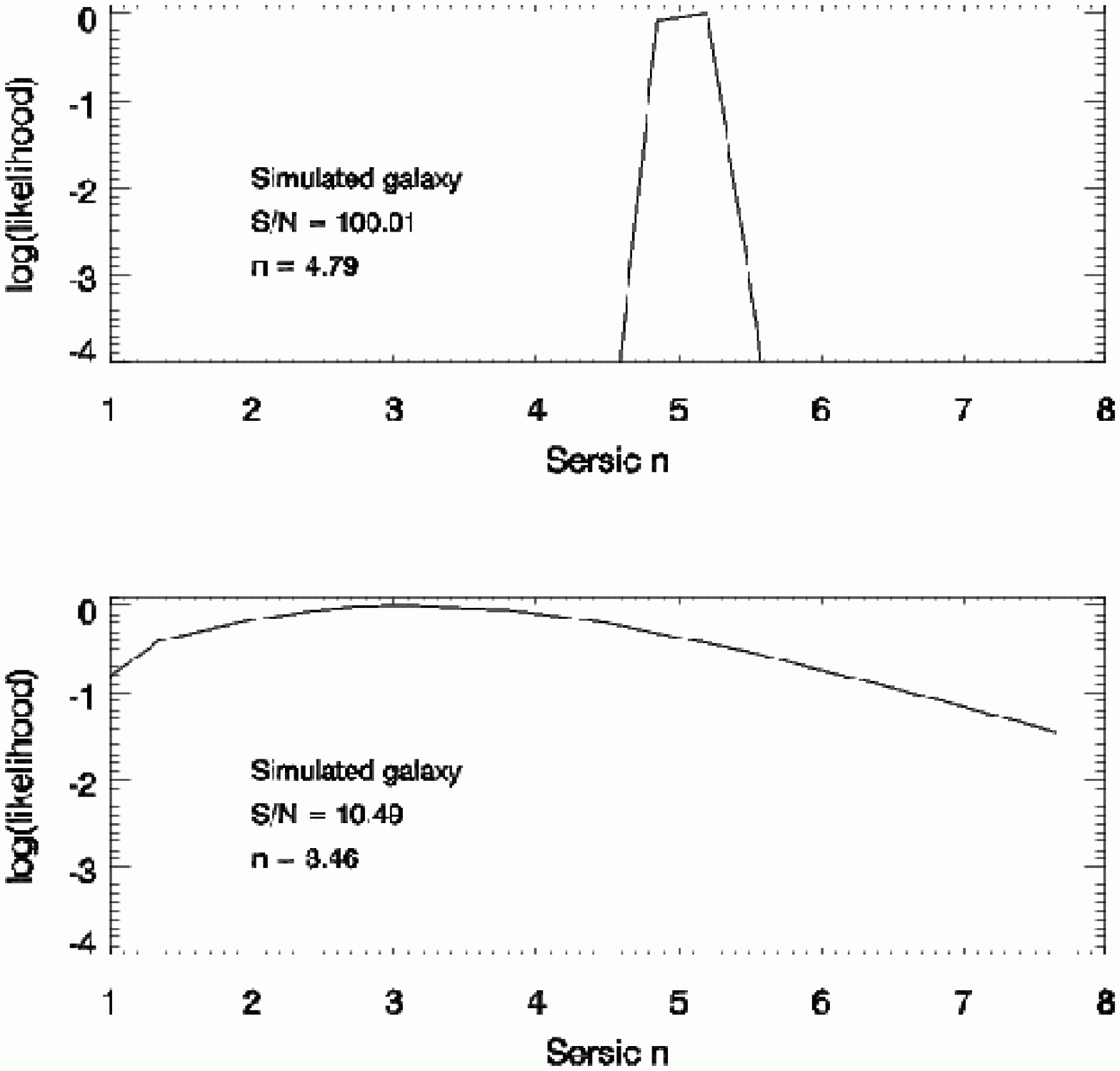,scale=0.47}}
  \subfigure[Posterior
  probability]{\epsfig{figure=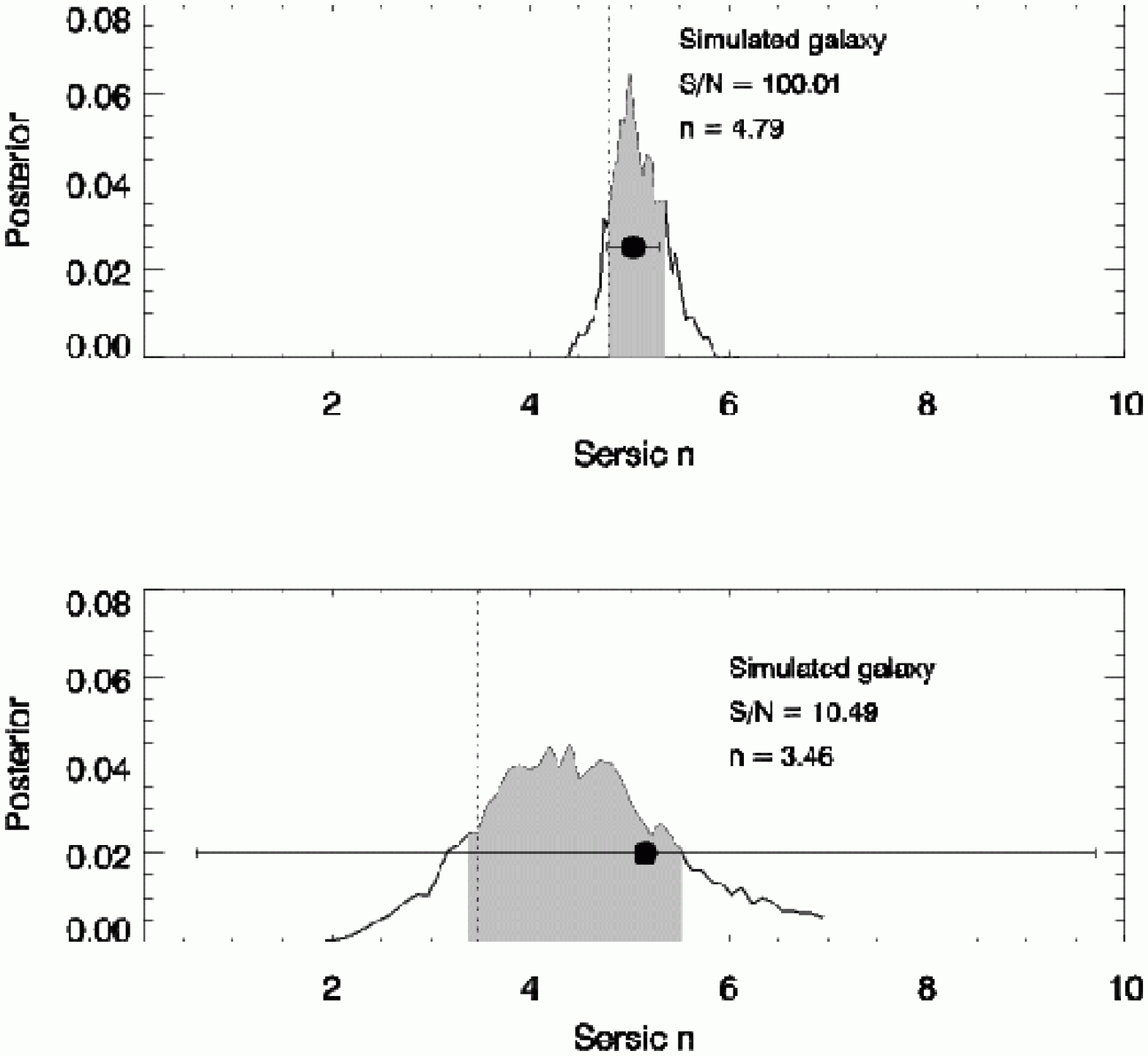,scale=0.47}}
  \caption{Panel (a): the likelihood as a function of \sersic\ index
    $n$ for two example galaxies with different \sn. Each likelihood
    value is normalised to the maximum.  The distribution for the
    high \sn\ galaxy (upper) is sharply-peaked and the distribution
    for low \sn\ galaxy (lower) is broadly-peaked.  Panel (b): the
    posterior probability density of $n$ for the two galaxies. The
    black dot with error bar is the best-fit parameter from \galfit\
    \protect{\citep{peng2002}}. The shaded region is the 68.3\%
    confidence interval and the vertical dotted line is the true value of
    $n$.  The conventional error estimate based on the second
    derivative of the likelihood is much too large for low \sn.}
\label{likelihood_posterior}
\end{figure}

Figure \ref{likelihood_posterior}a shows the likelihood as a function of
$n$ for each galaxy.  The upper panel plots the high
\sn\ galaxy and the lower panel the low \sn\ one.  For the high \sn\ galaxy
the likelihood has a very strong mode around $n=5$ with a change by 
factor of 4 in log, as $n$ varies from 4.5 to 5.5. However, for the 
low \sn\ galaxy the likelihood is very broad, smoothly changing by 0.4 
in log as $n$ varies from 3 to 5, and the likelihood profile is
not symmetric around the maximum.  In addition, the precise location
of the global maximum is not informative; a local analysis of the profile
using standard inverse-Hessian analysis would not give an accurate
estimate of the profile shape.  Finally, in general, a typical multi-dimensional
likelihood surface will have an even more complex landscape.

Figure \ref{likelihood_posterior}b shows the posterior probability of
each galaxy's $n$ for the given data, marginalised over the other
parameters as computed by \galphat.  The solid curve is the posterior
probability of $n$ and the shaded region corresponds to a 68.3\%
confidence interval\footnote{Here the confidence interval is estimated
  from the cumulative distribution function $F(\theta)$ of the
  marginalised parameter posterior probability density for the
  parameter $\theta$. The $100(1-\alpha)$ percent confidence interval
  $[\theta_1, \theta_2]$ has $F(\theta_1)=\frac{1}{2}\alpha$ and
  $F(\theta_2)=1-\frac{1}{2}\alpha$ where $0<\alpha<1$.}.  The true
value is indicated by the vertical dotted line, and the error bar
shows the result using \galfit\ \citep{peng2002}.  \galfit\ is a
widely-used galaxy image decomposition program, based on a maximum
likelihood (ML) approach, implemented as a $\chi^2$ minimisation
using the Levenberg-Marquardt algorithm \citep{press1998}.  The
posterior mode of $n$ is offset from the true value by 0.2 for the
high \sn\ galaxy and by 1.0 for the low \sn\ galaxy.  Such a bias
always occurs owing to random photon counting errors. Although the
bias is large for low \sn, each 68.3\% confidence interval of $n$
encloses the true value.  The best-fit value of $n$ from \galfit\ for
the high \sn\ galaxy is close to the posterior mode and the associated
error bar, corresponding to a 68.3\% confidence interval, encloses the
true value.  However, for the low \sn\ galaxy the best-fit parameter
from \galfit, using its simple minimisation algorithm, is more subject
to small-scale variations of the likelihood profile owing to sampling and
the best-fit parameter may also depend on the initialisation of the
downhill solver.  Furthermore, the inverse-Hessian estimate of the
variance in one dimension is simply the inverse of the second
derivative of the likelihood; geometrically, the faster the slope
varies with $n$, the smaller the variance.  This makes the error
reported by \galfit\ unrealistically large because there is no
significant variation in the tangent slope at the best-fit value of
$n=5.16$.  A reliable error estimate is crucial to quantifying trends
in the derived properties of galaxies.  The Bayesian MCMC approach
adopted here samples the full posterior distribution and yields
reliable error estimates of each model parameter given the data.
Hence, this provides a solid statistical base for an analysis of
galaxy morphology.

\subsubsection{Bias and prior assumptions}

\begin{figure}
\centering
\subfigure[Uniform prior
distribution]{\epsfig{figure=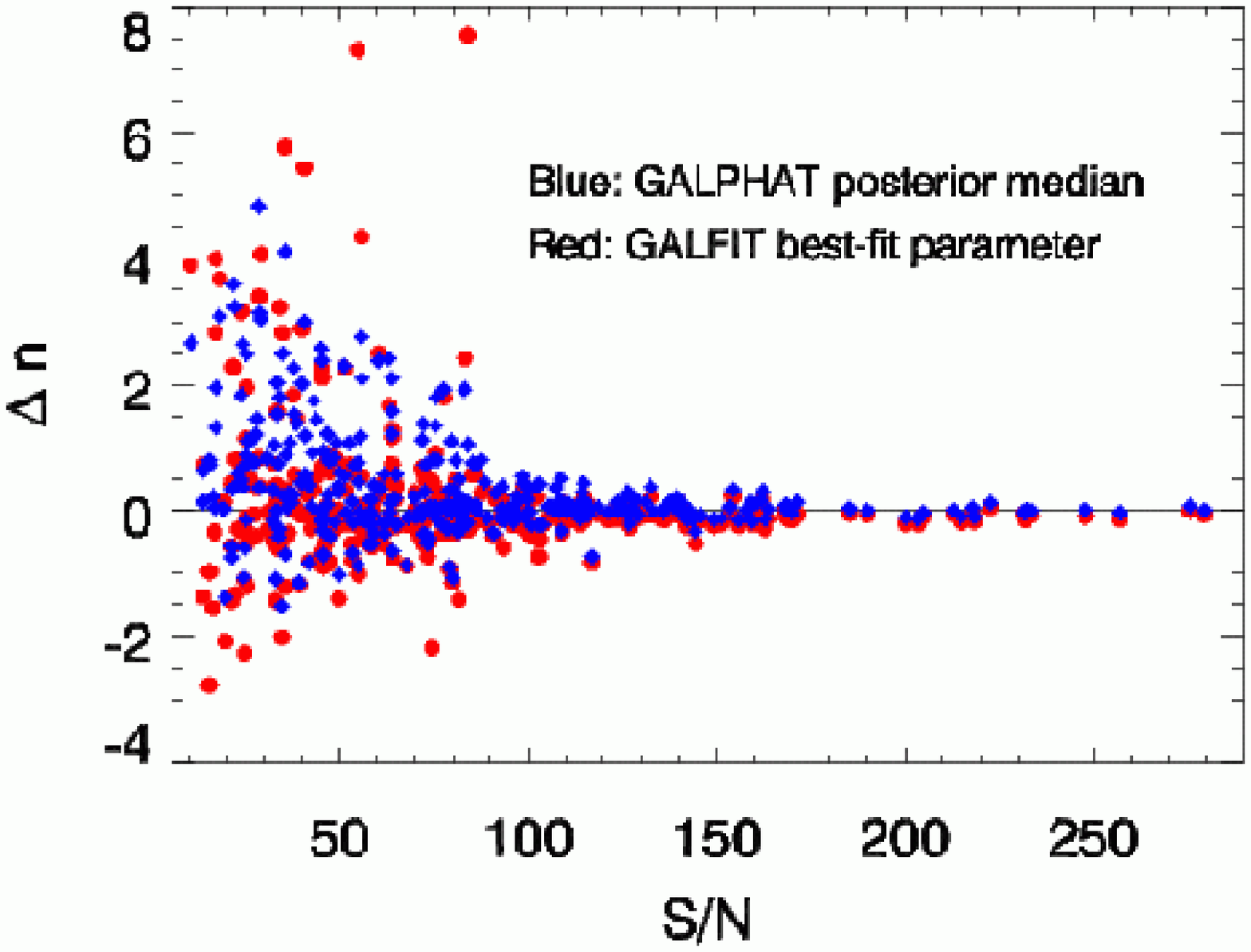,scale=0.48}}
\subfigure[Non-uniform prior
distribution]{\epsfig{figure=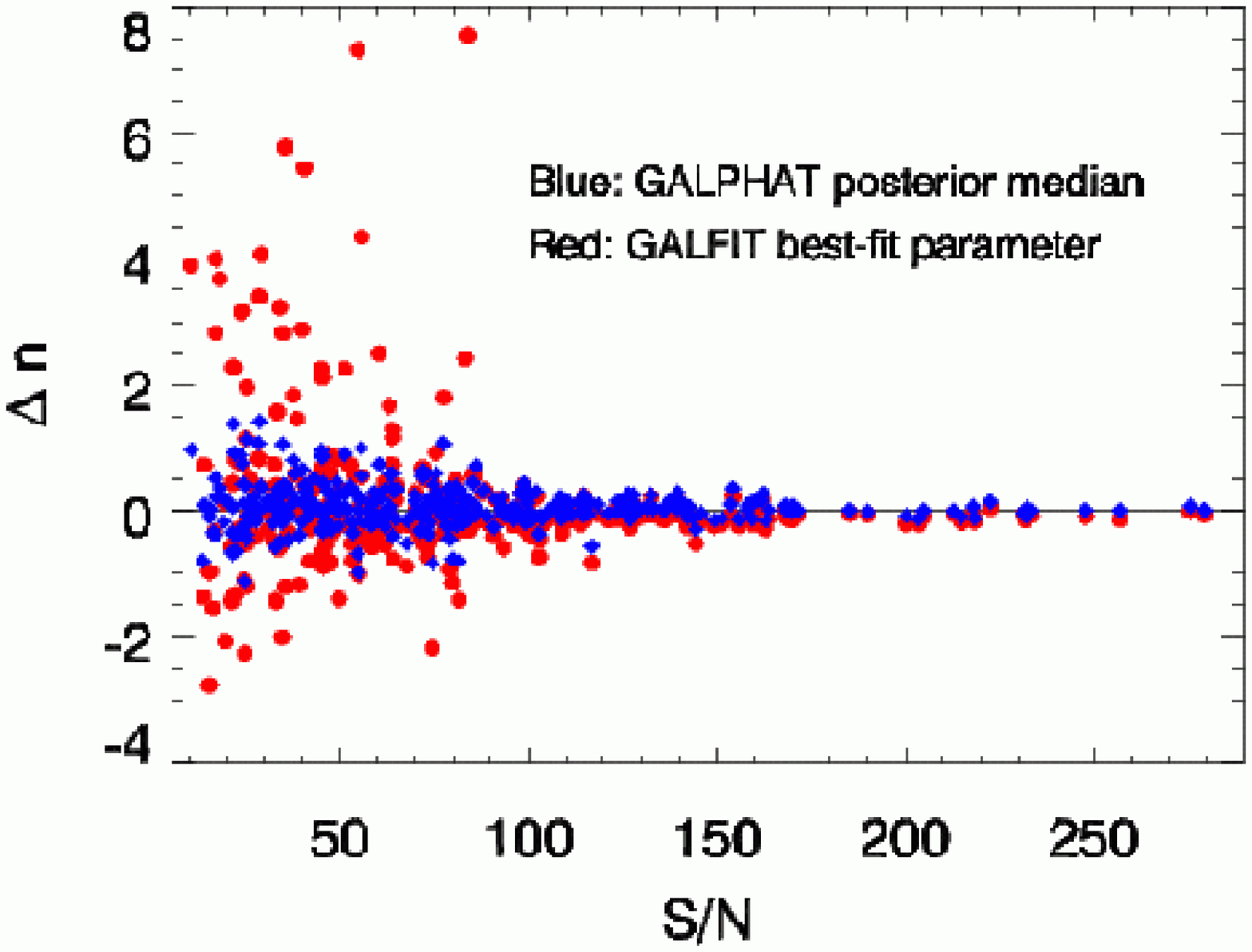,scale=0.48}}
\caption{The distribution of residual \sersic\ index $n$ for 325
  galaxies with $n \sim \mathcal{N}(4.0,0.33^2)$.  Panel (a):
  \galphat\ posterior median values (blue diamonds) and best-fit
  \galfit\ values (red circles).  \galphat\ uses uniform priors with
  the same parameter range used in \galfit\ for a fair comparison.
  Panel (b): same as (a) but with a non-uniform prior for $n$ in
  \galphat.  The prior distribution has a $\sigma$ roughly 13 times larger 
  than the input distribution. By introducing an
  informative prior, the parameter recovery is significantly improved
  with decreasing \sn. }
\label{ensemble_posterior}
\end{figure}

We now explore the distribution of the best-fit $n$ for 325 galaxies
with their \sersic\ model index $n$ sampled from a normal
distribution\footnote{In all that follows we will denote the normal or
  Gaussian distribution with mean $\mu$ and variance $\sigma^2$ as
  ${\cal N}(\mu, \sigma^2)$ and the uniform distribution between $a$
  and $b$ as ${\cal U}(a, b)$.  Other distributions will be introduced
  as needed.} $n \sim \mathcal{N}(4.0,0.33^2)$.  Figure
\ref{ensemble_posterior}a plots the residual value of the $n$ using
the posterior median of $n$ from \galphat\ (blue diamonds) with a
uniform prior for $n \sim \mathcal{U}(0.2, 10.0)$ and the best-fit
value from \galfit\ (red circles).  For the \galfit\ estimates, we used the
true values as the centroids of the distribution for the initial guess
and randomly perturbed the magnitude, galaxy radius $r_e$, axis ratio,
position angle, and sky background by $\pm0.5$, $\pm20\%$, $\pm10\%$,
$\pm15\%$ and $\pm1\%$, respectively, about these true values,
assuming a uniform distribution within these ranges.  In \galphat\ we assumed
uniform priors for these parameters within these same ranges. The initial
\galfit\ guess for the \sersic\ index is $n=2.5$ for all the galaxies.  As an
image's \sn\ decreases, the variance in the residual value of $n$
becomes larger and $n$ becomes preferentially overestimated owing to
the asymmetric shape of the likelihood profile (see
Fig. \ref{likelihood_posterior}). For low \sn, the \galfit\ values are
sensitive to the initial guess in contrast to \galphat, which samples
the parameter space using MCMC and hence is insensitive to the initial
guess.  The variance in the best-fit $n$'s from \galfit\ is slightly
larger than the variance in the \galphat\ posterior medians owing to
its insufficient accuracy in finding the correct global likelihood
minimum for low \sn\ data.

However, if we allow parameters to have an informative prior
distribution, the \galphat-derived posterior distribution improves
dramatically (Fig. \ref{ensemble_posterior}b).  We use a Weibull 
distribution\footnote{
  The Weibull distribution is
  \[
  P(x; \lambda, k) = \left(\frac{k}{\lambda}\right)
  \left(\frac{x}{\lambda}\right)^{k-1}
  \mbox{exp}\left[-(x/\lambda)^k\right].
  \]
}
for the prior probability of $n$ with $\lambda=7.0$ and $k=1.5$, 
whose deviation ($\sigma=4.3$) from the mean is still 13 times larger than 
the deviation of the true distribution but embodies our astronomical
experience from the literature. The variance in the posterior
median of the residual of $n$ for low \sn\ images decreases significantly.
With an appropriately chosen prior distribution, the Bayesian approach can
dramatically increase the quality of the inference over the entire
input catalogue.  Many people have aesthetic objections to the
Bayesian approach, because they view the selection of a prior as being
subjective and, therefore, arbitrary.  The statistics and
astrostatistics literature abound with philosophical discussions of
this issue.  It is our point of view that prior information will be
used by a researcher inevitably, so why not let the Bayes theorem tell us
how to use it quantitatively?  Conversely, wholesale adoption of
uniform priors, e.g. in the maximum likelihood method, is both arbitrary
and self-defeating, since we know that most parameters have both
physical constraints and previously measured distributions.  In other
words, ignoring one's own expert opinion is daft.

These issues are less significant for data with high \sn.  However,
astronomical surveys have widely-ranging \sn\ values, and the total
number of galaxies for flux-limited samples will always be dominated by
images with low or moderate \sn.  Therefore, improving our estimates
of structural parameters in the low \sn\ regime best uses the
available data and optimises scientific return.

\subsubsection{Scatter plots versus joint posterior distributions}

\begin{figure}
\centering
\subfigure[Scatter
plot]{\epsfig{figure=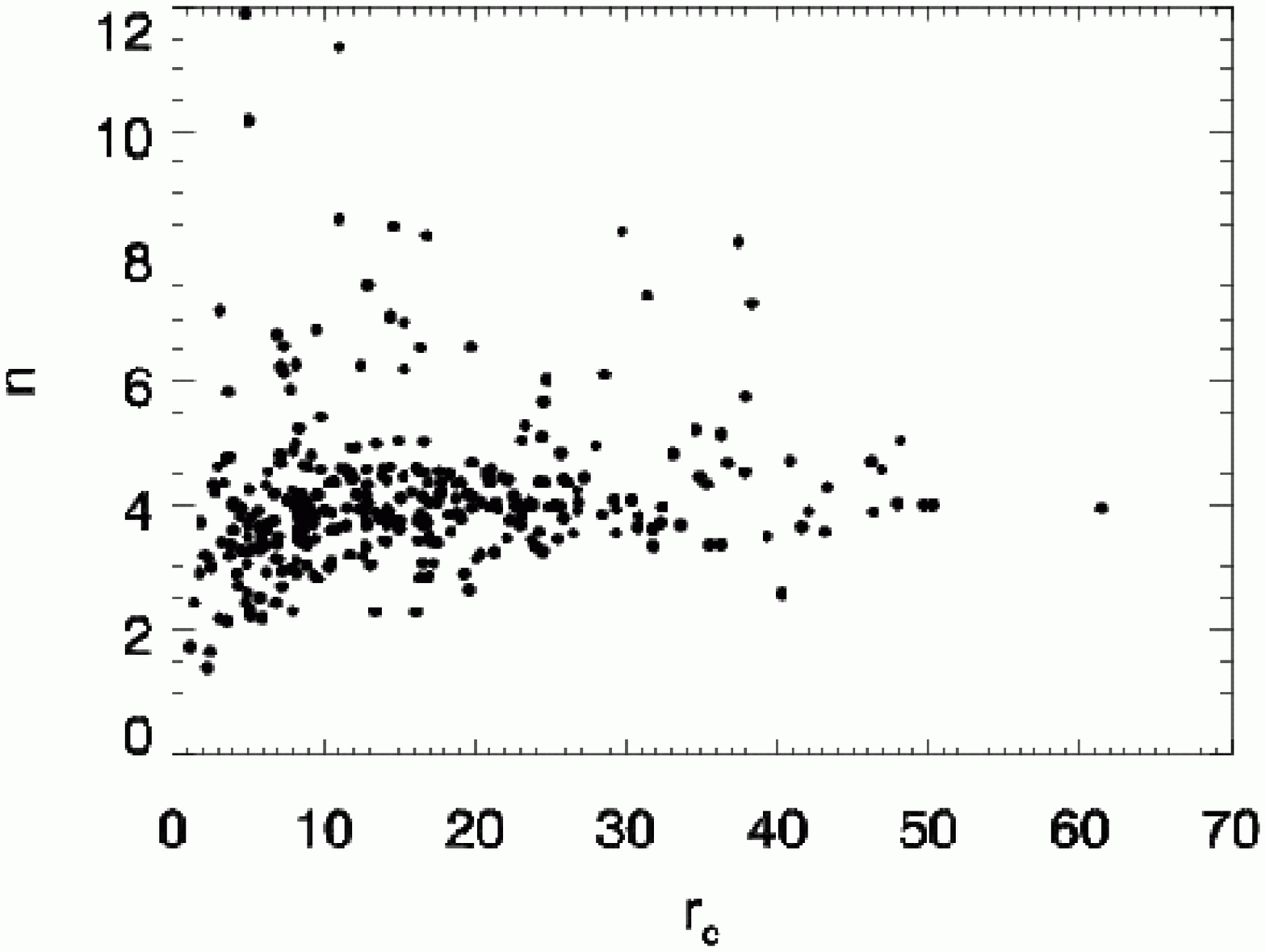,scale=0.48}}
\subfigure[Posterior
distribution]{\epsfig{figure=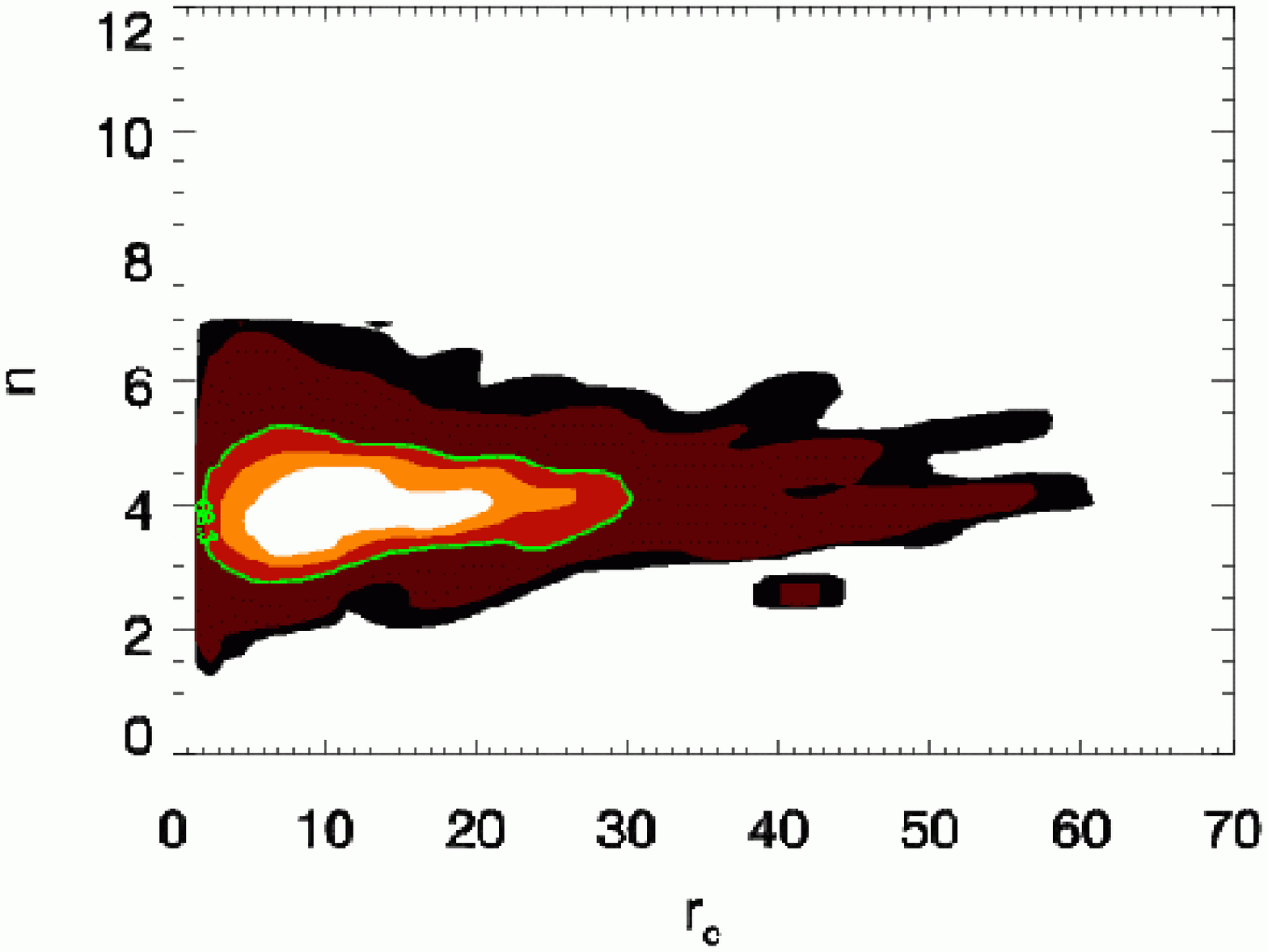,scale=0.48}}
\caption{The joint distribution of galaxy half-light radius $r_e$ and
  \sersic\ index $n$ for 325 galaxies.  The simulated galaxies were
  generated without any correlation between $r_e$ and $n$.  Panel (a) shows the
  conventionally used scatter plot using the best-fit model parameter
  from \galfit\ and Panel (b) shows the joint posterior of $r_e$ and
  $n$ from \galphat\ with 30, 50, 68.3 (green line), 95 and
  99\% confidence level (white to black).  }
\label{joint_posterior}
\end{figure}

Here we explore the joint distribution of galaxy half-light radius
$r_e$ and \sersic\ index $n$ for the same galaxy image sample from the last
sub-section.
Figure \ref{joint_posterior}a is a conventional scatter plot showing
the joint distribution of best-fit values of $r_e$ and $n$ from
\galfit.  Figure \ref{joint_posterior}b shows the joint posterior
distribution of $r_e$ and $n$ from \galphat.  The contour levels are
the 30, 50, 68.3 (green line), 95, and 99\% confidence values
(white to black). Since the galaxy sample
is generated without any correlations between $r_e$ and $n$, we should not
see any covariance between $r_e$ and $n$.  Comparing these two panels,
the scatter plot from \galfit\ shows a (spurious) systematic trend of $n$ with
$r_e$ while the joint posterior distribution from \galphat\ does not. Some 
\sersic-model parameters, e.g. $r_e$ and $n$, are reported to exhibit a
true covariance \citep{trujillo2001}, so such systematic trends in the
distribution of best-fit parameters as exhibited by \galfit\ 
will obscure or contaminate any intrinsic covariance.

The qualitative difference between the Maximum Likelihood
(ML)-inferred scatter plot and the Bayesian-inferred joint posterior
underlines our assertion that understanding parameter covariance, 
the use of thoughtful prior distributions, and a thorough error analysis are
essential to reliably testing hypotheses based on the morphologies of
a large number of galaxies.  Even so, the choice of a parametric
family induces a correlation between parameters and, therefore, a
single ``best-fit'' value does not adequately characterise the
knowledge acquired from the data.  Examples and conclusions such as
these motivate our using the entire posterior distribution in
parameter space for all of the galaxies that we wish to study.

\subsection{Bayesian MCMC to the rescue}
\label{rescue}

The main disadvantage of the Bayesian approach is its computational
expense.  Over the last 15 years, MCMC techniques have continued to
improve and we believe that these techniques are now suitable as
mainstream tools.  Here, we introduce a computationally-tractable
Bayesian MCMC approach that overcomes the difficulties outlined in the
previous sections.  Significant improvement over previous attempts to
understand galaxy evolution using morphology can be achieved using a
Bayesian approach for the following three reasons:
\begin{enumerate}
\item The ML method assigns the best-fit parameter value to the model
  that has the highest probability of generating the observed data.
  With sufficient data, this yields the correct result (assuming that
  some model in the family generates the data).  However, Nature gives
  us one realization of a particular galaxy and this leads to
  competing models that can generate the same data through different
  processes.  Rather than the ML question, we want to know the best
  model among the possible candidates, or \emph{what is the
    probability of the model for the observed data?}  Written in the
  language of conditional probability, this question gives us Bayes
  theorem and requires an estimate of the probability of the model
  before acquiring the data (the \emph{prior} probability).  The
  Bayesian formulation of the inference problem provides a natural
  foundation for Astronomy, where every single event is unique and
  observers can not test theories by changing the initial conditions
  of the Universe.  If data has high \sn\ and strongly supports a
  particular model, the inference should not be subject to the bias
  introduced by the prior distribution.  Conversely, if data has low
  \sn, the inference may be influenced by the prior assumptions, as
  intuitively expected.

\item We have seen that inferences based on best-fit analyses can be
  contaminated by intrinsic covariance in the chosen model family.  In
  addition, the topology of the likelihood function in a
  high-dimensional parameter space with a large number of free
  parameters is very complex in general.  Therefore, one needs to both
  find the true global extremum and assess the significance of this
  extremum with respect to nearby and possibly unanticipated extrema
  in the probability space. Bayesian MCMC provides the full posterior
  and the confidence levels of inferences for each model parameter. Hence one
  can investigate correlations or perform hypothesis tests with
  quantifiable confidence.

\item The adoption of specific functional families, e.g.
  \sersic\ profiles, may not provide an adequate explanation for the
  observed data or have sufficient power to classify differences between
  data sets.  In the Bayesian paradigm, one may consistently assess the
  explanatory power of different models even if they are not nested.
  For example, one may rigorously pose the question ``how strongly is
  the assumption of a single-component \sersic\ model supported by
  the galaxy image data compared to a two-component bulge and disc
  model?'' This provides a natural way of probabilistically
  classifying galaxies.
\end{enumerate}

To summarise, our Bayesian approach uses galaxy morphology as an intermediate
step in an overall inference problem for theories of galaxy evolution.
In contrast, by focussing on the best-fit parameters as the
data-reduction goal, and comparing the implied correlations to those
predicted by theories of galaxy formation, one runs the risk of
interpreting false correlations and one decreases the information
content of one's data by using only the best-fit parameter as a
summary value.  Motivated by the promise of dramatic improvement, we
have developed a novel image decomposition software package, \galphat\ (GALaxy
PHotometric ATtributes), based on a Bayesian Markov chain Monte Carlo
approach.  The general application to astronomical 
image analysis is not new: 
Bayesian MCMC has been used for X-ray surface brightness estimation of
galaxy clusters \citep{andreon2008}, for object detection \citep{carvalho2009, 
guglielmetti2009, hobson2003, savage2007}, for dynamical modelling of
a galaxy \citep{puglielli2010}, and for gravitational weak lensing
\citep{kitching2008, miller2007} and strong lensing analyses
\citep{vegetti2009}.  \galphat\ is designed for performing
morphological analysis of galaxy images similar to several widely-used
galaxy image decomposition packages such as: \budda\
\citep{desouza2004}, \galfit\ \citep{peng2002},
\gasptd\ \citep{mendez2008}, \gasphot\
\citep{pignatelli2006} and \gimtd\ \citep{simard2002,simard1998}, but
has a firm probabilistic foundation, providing full posterior
distributions of the parameters, which are
suitable for a variety of inference problems
such as model comparison, hypothesis testing, and correlation analyses.

One of our original motivations for developing \galphat\ is the large
scale analysis of galaxy morphological structures in 2MASS
\citep{skrutskie1997,skrutskie2006}.  The interpretation of
2MASS-galaxy properties has suffered from using unreliable best-fit
parameters obtained using conventional fitting algorithms.  An
ensemble of posterior distributions for a complete sample of local
galaxies becomes a rich database for hypothesis testing in two ways.
First, we may characterise the distributions of galaxy morphological
structures, e.g. the luminosity, the size and the shape, as a function
of environment, with rigorous statistical confidence levels.  Secondly
and more generally, we may compare galaxy formation theories based on
the morphological evidence.

This paper introduces \galphat\ and emphasises methods, features and
performance issues.  We demonstrate the feasibility of a large-scale
statistical inference based on galaxy morphology.  A more detailed
exploration of the influence of the prior distribution, explicit
examples of model comparisons between single \sersic\ and
two-component bulge and disc models, and inferences using an ensemble of
posterior distributions will be reported in a followup paper
\citep{yoon2010b}.  The paper is organised as follows. In
\S\ref{bgsection}, we describe the basic formalism of Bayesian
inference for galaxy image data and introduce the Bayesian Inference
Engine (\bie), which is used to sample the posterior distributions of
model parameters.  We describe the structure of \galphat\ in
\S\ref{algorithmsection} and our ensemble of simulated galaxy images
for calibrating \galphat\ in \S\ref{datasection}.  Comprehensive test
results are presented in \S\ref{resultsection}.  We summarise our
findings and conclusions in \S\ref{conclsection}.

\section{Bayesian Markov chain Monte Carlo}
\label{bgsection}

\subsection{Theoretical background}

Bayes theorem states that the probability of a model characterised by its
parameter vector $\theta$, given some data set $D$, is proportional to
the likelihood of the data for the given model multiplied by the prior
probability of the model
\begin{equation}
  P(\theta | D) = \frac{L(D|\theta) \pi(\theta)}{\int
    L(D|\theta)\pi(\theta)d\theta}
\label{eq:Bayes}
\end{equation}
where $P(\theta | D)$ is the posterior distribution, $L(D|\theta)$ is
the likelihood function, i.e. the probability of the data given
$\theta$, and $\pi(\theta)$ is the prior distribution of the parameter
vector $\theta$.  If $\pi(\theta)$ is a uniform distribution in a
compact subset of $\mathbb{R}^n$, we recover the ML method.

The goal is to characterise the posterior distribution by sampling
$P(\theta|D)$.  For simple cases with a few model parameters, one can
analytically calculate it or explore the probability space by
evaluating the posterior probability over a grid in parameter
space. However, for our \sersic\ model with 8 parameters this approach
is not feasible.  Fortunately, owing to the rapid improvement in
computational methodology, MCMC is feasible in high-dimensional
parameter spaces.  MCMC generates states by a first-order Markovian
process and the distribution of states asymptotically converges to the
target distribution $P(\theta|D)$ after a large number of iterations.

For the model selection problem, one may apply Bayes theorem to give
the probability of the theory $M$ based on the data $D$ given a prior
probability of the theory.  In our case, $M$ is the assumption of a
particular model family with a parameter vector $\theta$, e.g. the
\emph{theory} that galaxies are \sersic\ models.  The likelihood of a
theory is the probability of the data given the theory.
Algebraically, the probability of the data given the theory is the
marginalisation of the likelihood function over the prior probability
of the model parameter distribution: $P(D|M) = \int
L(D|M,\theta)\pi(\theta|M)d\theta$.  This quantity is a measure of how
well the evidence supports the theory.  In other words, the more
probable the evidence given the theory, the more the evidence supports
the theory.  Of course, one needs to know what the theory predicts to
know how well the evidence supports it and this is the job of the MCMC
simulation.  Now, let $P(M)$ be the prior probability of the theory.
Then, analogous to the Bayes theorem from equation (\ref{eq:Bayes}),
we may write the probability of the theory given the data as
\begin{equation}
  P(M|D) = \frac{P(D|M) P(M)} {\int P(D|M)P(M)\,dM}.
\label{eq:BayesM}
\end{equation}
Equation (\ref{eq:BayesM}) immediately gives an estimate of the posterior
odds of two different theories $M_1$ and $M_2$ parametrised by different
parameter vectors $\theta_1$ and $\theta_2$:
\begin{equation}
  \frac{P(M_1|D)}{P(M_2|D)} = \frac{P(M_1)}{P(M_2)} K_{12}
  \quad\mbox{where}\quad
  K_{12} \equiv \frac{P(D|M_1)}{P(D|M_2)}.
\end{equation}
The quantity $K_{12}$ is called the Bayes factor. The numerator and
denominator of $K_{12}$, $P(D|M_i)$, is called the marginal likelihood
for model $i$. If one does not favour one of the two theories a priori,
the term $\frac{P(M_1)}{P(M_2)}=1$ since $P(M_1) = P(M_2)$.
Therefore, the Bayes factor describes the increase of odds in
favour of one theory over another in the light of the data.  The Bayesian
model comparison does not depend on the particular parameters used by
each model. Instead, it considers the probability of the model
considering all possible parameter values.

However, it should be clear from equation (\ref{eq:BayesM}) that the
Bayes factor depends on the choice of the prior distribution.  If the
likelihood dominates, the effect of the prior is negligible, but if
the prior contributes to the posterior significantly as well as the
likelihood, an incorrect prior leads to a biased inference
\citep[e.g.][]{kass1993}.  The choice of prior must be considered
carefully. The effects of prior choice on \galphat\ will be addressed
in detail in our follow-up paper \citep{yoon2010b}.

\subsection{The Bayesian Inference Engine (\bie)}

The ability to realise the promise of this approach depends on an accurate
computation of the posterior distribution.  Although the Markov chain will
approach its steady-state distribution almost certainly, the number of
steps required, the \emph{mixing time}, is not known.  In addition,
the exploration of parameter space suffers from the \emph{curse of
  dimensionality}\footnote{The \emph{curse of dimensionality} is the
  exponential growth of hypervolume as a function of dimensionality
  \citep{Bellman1961}.}.  We need assurance that the Markov chain is
in a steady state beyond the local region in parameter space.  For
example, consider a posterior distribution with discrete, separated
modes; many chains will not be able to move between these modes,
resulting in an infinite mixing time and an incomplete posterior
distribution.

Various MCMC algorithms have been proposed to improve the convergence
of MCMC, and each of these have their own advantages and
disadvantages.  Beginning in 2000, a multi-disciplinary investigator
team from the Departments of Astronomy and Computer Science at the
University of Massachusetts designed and implemented the Bayesian
Inference Engine, a MCMC parallel software platform for performing
statistical inference over very large data sets.  The \bie\ uses a
scalable multiprocessor software architecture designed to operate on
modest cost, generally available hardware. MCMC algorithms and
Bayesian computation in general are ideally suited to multiprocessor
computation.  The \bie\ uses standard MPI and POSIX threads and,
therefore, will run in a broad spectrum of parallel or scalar
environments and can be easily ported to high-performance hardware for
production analysis.  Fundamentally, the \bie\ is a library, but the
package provides a command-line interpreter (CLI) with access to
nearly all of the import object classes.  This CLI was originally
intended for interactive or script-based prototyping with subsequent
stand-alone hard-coding.  However, most users simply use the CLI with
scripts.  The \bie's object-oriented design allows a researcher to
apply a wide variety of MCMC algorithms to the same target application
by changing several lines in a program or script.  The \bie\ currently
includes: the standard Metropolis-Hastings algorithm
\citep{hastings1970,metropolis1953}, the simulated tempering algorithm
\citep{neal1996}, the parallel tempering algorithm \citep{geyer1991},
the parallel hierarchical sampler \citep[PHS hereafter]{rigat2008},
the differential evolution algorithm \citep{braak2006}, and an
independent multiple chain algorithm.  For convergence testing, the
\bie\ implements the algorithm proposed by Giakoumatos \etal\
\citep{giakoumatos1999} for single-chain simulations and the Gelman-Rubin
\citep{gelman1992} convergence diagnostic for multiple-chain
simulations.  

The object-oriented design makes the \bie\ extensible; new MCMC
algorithms, new convergence algorithms, and new models or likelihood
functions can be implemented and added to the \bie\ at any time.  For
example, a typical user will typically develop new models for a
specific scientific problem.  At a later time, the user may use a
newly available MCMC algorithm without changing this model code.
Since MCMC computations are computationally expensive, the \bie\
provides a full serialisation and persistence system.  This system
saves the entire state of all the objects in the simulation.  For
example, the \bie\ automatically saves \emph{checkpoint} images.
Therefore, \galphat\ can restart from the very last MCMC step to
sample the posterior further for obtaining more MCMC samples when
needed, significantly saving computational resources.  Moreover, the
results of previously performed simulations can be restored on the fly
and compared or reused in new ways.  See \citet{weinberg2010} for
additional details.

\section{GALPHAT: algorithms and features}
\label{algorithmsection}

\galphat\ is implemented as a user-contributed likelihood function to
the \bie.  It reads two-dimensional galaxy image data from a FITS file,
generates a model image, and then computes the likelihood.  Because the
posterior simulation requires a large number of likelihood evaluations,
optimisation of the model image generation is essential to make the
analysis of an entire image catalogue feasible.
In this section, we describe the implementation
details and features of \galphat.

\subsection{Overview of our \galphat\  implementation}

The \bie\ controls the MCMC algorithm, the convergence testing, and
logging the sampled posterior distribution.  As needed, the \bie\
requests a likelihood evaluation from \galphat.  The flow chart for
\galphat\ is as follows:
\begin{enumerate}
\item \galphat\ reads the input FITS files (the galaxy image and the
  PSF) and the tabulated model images (see below) for later
  interpolation. 
\item As part of the MCMC simulation, the \bie\ calls the likelihood
  function with a parameter vector. Using these parameters, \galphat\
  interpolates and scales the image table using the scale radius and
  the minor/major axis ratio, and generates a model image in
  principle-axis coordinates.
\item \galphat\ convolves the model image with input PSF image in Fourier
  space using the FFTW package\footnote{\texttt{http://www.fftw.org}} and
  adds the sky background.
\item Finally, the model image is rotated by the position angle, 
  using a Fourier shift algorithm, in pixel coordinates.
\item \galphat\ returns the likelihood evaluation to the \bie.
\item Steps (ii)--(v) are repeated as necessary.
\end{enumerate}
We will describe the details for each important step below.
   
\subsection{Model generation}

Any symmetric galaxy model has six model-independent free parameters:
the centroid coordinates $(x,y)$, the axis ratio $q=b/a$, the position angle,
the scale length, and the total flux or magnitude.  For tests in this paper,
we use a \sersic\ model \citep{ciotti1991,graham2005,sersic1968}.
The \sersic\ model is a one-parameter model family described by the
index $n$.  As the index increases, the profile increases in
concentration: an exponential disc profile has $n=1$ and a \devauc\
profile has $n=4$.  The model has the following analytic form
\begin{equation}
\label{eq_sersic}
\Sigma (r) = \Sigma_e 
\mbox{exp}\left[-\kappa\left\{\left(\frac{r}{r_e}\right)^{1/n} - 1\right\}\right]
\end{equation}
where the effective radius, $r_e$, defines a scale length.  By
construction $r_e$ is equivalent to the half-light
radius, $r_{50}$.  The quantities $\kappa$ and $n$ are related through
the relation
\begin{equation}
\label{kappa}
\Gamma(2n) = 2\gamma(2n,\kappa)
\end{equation}
where $\Gamma$ is the complete gamma function and $\gamma$ is the incomplete
gamma function.  Approximate analytic expressions for $\kappa$ can reduce
the computation time. For $n>0.36$ we adopt the following asymptotic
expansion for $\kappa$, which is good to better than one part in $10^4$
\citep{ciotti1999,macarthur2003}:
\begin{equation}
  \kappa \approx
  2n-\frac{1}{3} + \frac{4}{405n} + \frac{46}{25515n^2} +
  \frac{131}{1148175n^3} - \frac{2194697}{30690717750n^4} + O(n^{-5}).
\end{equation}
For $n\le0.36$, we use the following polynomial fit
\citep{macarthur2003}, accurate to better than two parts in $10^3$:
\begin{equation}
  \kappa \approx 0.01945-0.8902n+10.95n^2-19.67n^2+13.43n^3.
\end{equation}

\begin{figure}
\centering
\subfigure[Surface brightness profile]{\epsfig{figure=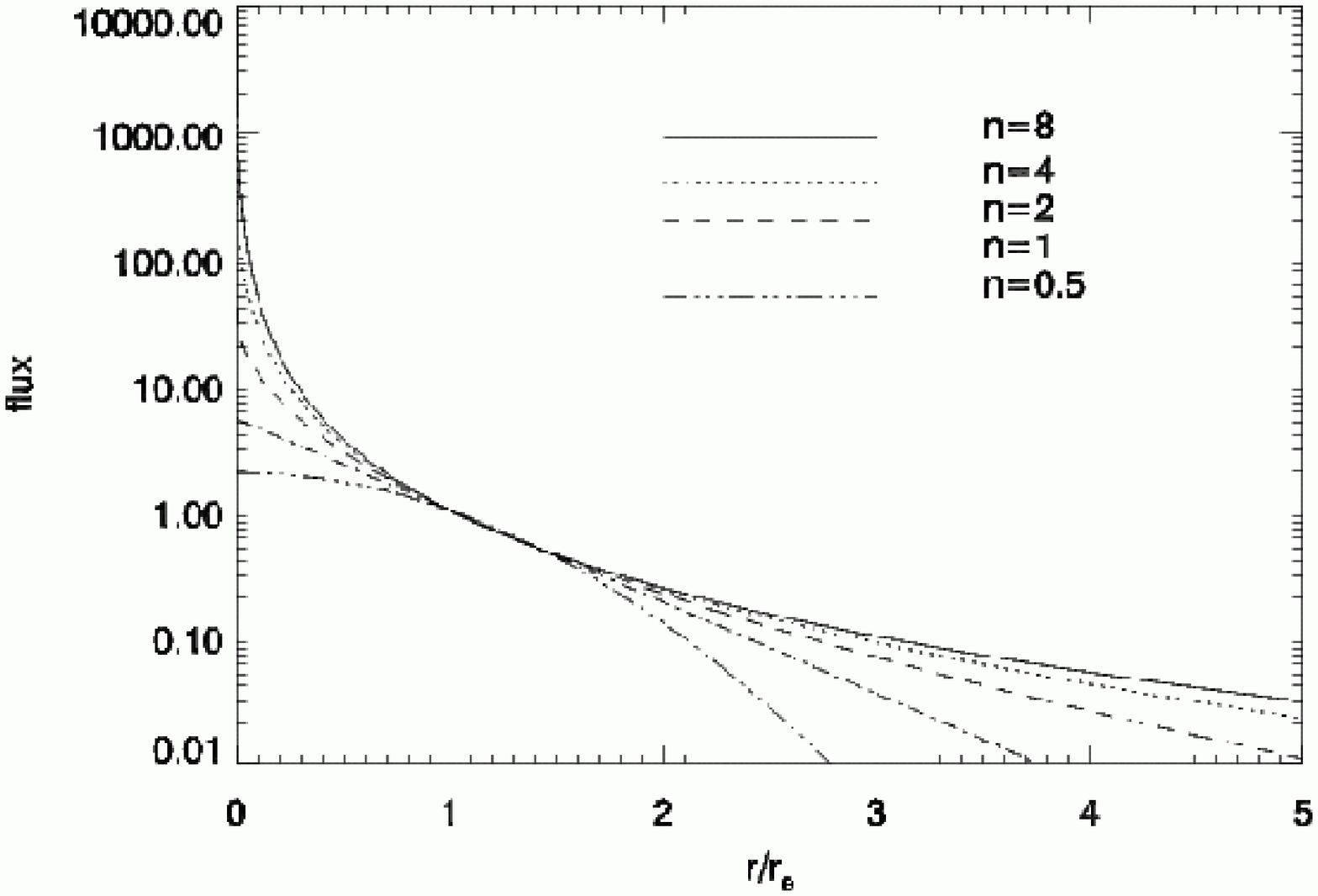,scale=0.47}}
\subfigure[Cumulative flux]{\epsfig{figure=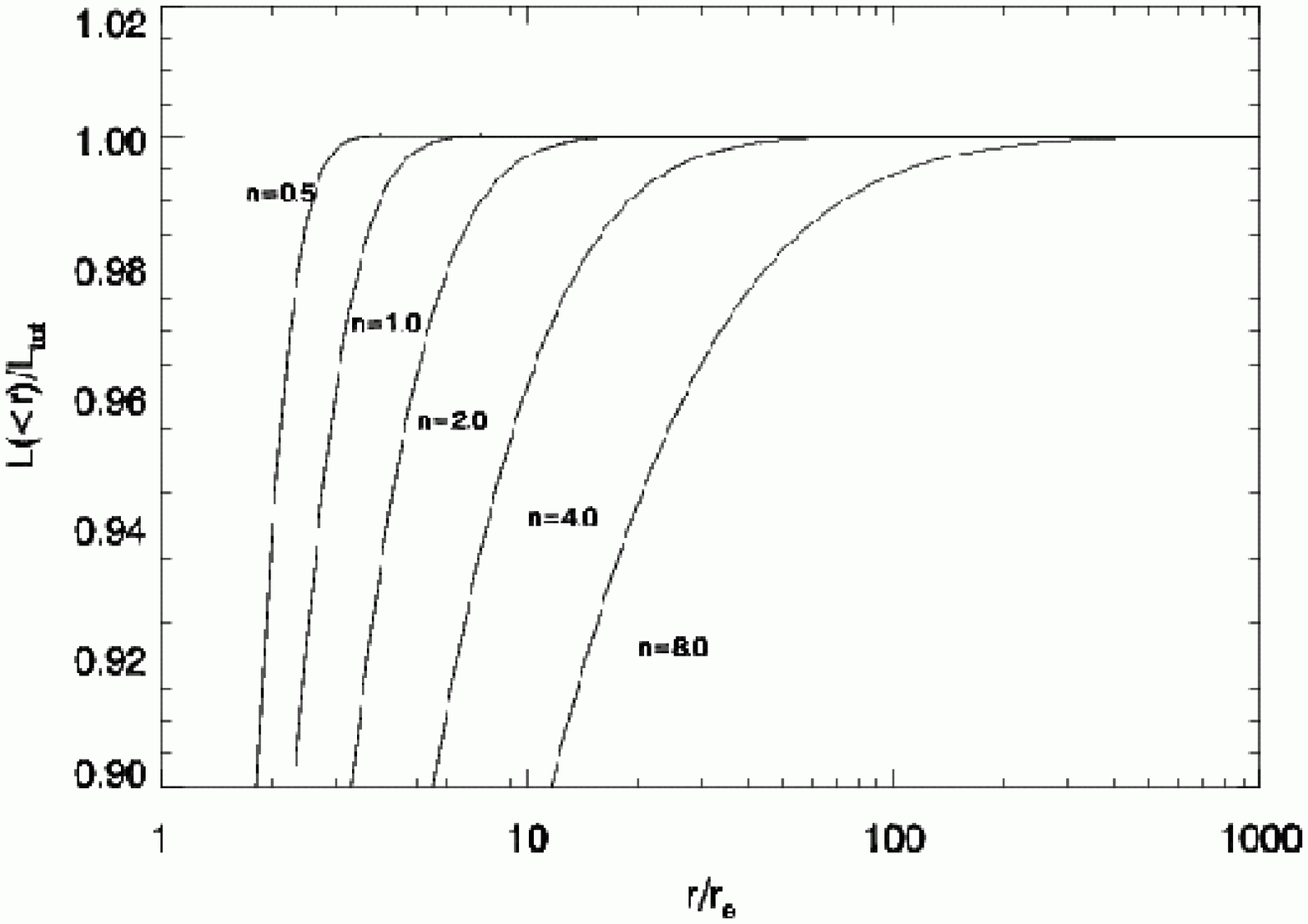,scale=0.49}}
\caption{Panel (a): \sersic\ surface-brightness profiles for n=0.5, 1,
  2, 4 and 8 (eq. \ref{eq_sersic}). The profiles are normalised to
  have equal flux density at $r=r_e$.  Panel (b): the fraction of
  light within $r$ for \sersic\ profiles with n=0.5, 1, 2, 4 and 8.
  For $n=8$, a few percent of the flux has $r>100r_e$! }
\label{sersic}
\end{figure}

Figure \ref{sersic}a shows \sersic\ profiles with different $n$,
normalised to have equal fluxes at $r_e$.  As $n$ increases, the
central profile steepens and the wings thicken.  The luminosity within
a radius $r$ is
\begin{equation}
\label{enclosedL}
L(< r) = \Sigma_e 2\pi r_e^2 q n
\frac{e^{\kappa}}{\kappa^{2n}}\gamma(2n,x)
\end{equation} 
where $x=\kappa(r/r_e)^{1/n}$ \citep{graham2005}.  Replacing
$\gamma(2n,x)$ with $\Gamma(2n)$ gives the total luminosity
$L_{tot}$ \citep{ciotti1991,ciotti1999,graham2005}.

The fraction of light within $r$ for different values of $n$ is shown
in Figure \ref{sersic}b.  As summarised in \citet{graham2005}, for an
exponential disc (i.e. $n=1$) profile, 99.1\% and 99.8\% of the flux
is within the inner $4r_e$ and $5r_e$, respectively.  For a
\devauc\ profile (i.e. $n=4$), 84.7\% and 88.4\% of the flux is within
the inner $4r_e$ and $5r_e$, respectively.  The sky background and
index $n$ become strongly covariant for small images and large-$n$
profiles, and this biases the estimate for $n$.

\subsubsection{Image tables}

For each image pixel, one typically assigns a flux value by directly
integrating the surface brightness profile $I(x,y)$ over the area of the pixel.
The value of pixel with an index of $(j,k)$ is
\begin{equation}
I_{jk} = \int^{x_{j+1}}_{x_j} dx \int ^{y_{k+1}}_{y_k} dy\,I(x,y).
\label{eq:Ijk}
\end{equation}
If $I(x,y)$ is not integrable in closed form, the numerical
integration becomes a computational bottleneck.  \galphat\ avoids
this by interpolating from a pre-prepared high-resolution table of
cumulative images.

Suppose that we have an object with brightness $\Sigma(x,y)$ over the
domain $[x_o,x_f]\otimes[y_o,y_f]$.  The two-dimensional cumulative
brightness distribution is
\begin{equation}
C(x,y) = \int^{x}_{x_o}dx\int^{y}_{y_o}dy\,\Sigma(x,y).
\end{equation}
The value of $\Sigma$ integrated over some arbitrary pixel is
then
\begin{equation}
I_{jk} = C(x_k,y_k)+C(x_j,y_j)-C(x_j,y_k)-C(x_k,y_j).
\label{fluxij}
\end{equation}
A high-accuracy tabulation of $C(x,y)$ allows one to use equation
(\ref{fluxij}) and to evaluate $C(x,y)$ by interpolation at minimal
computational cost.  For a \sersic\ model, we need a one-dimensional
grid in $n$ of images $C(x, y)$ with $r_e=1$ and these can be linearly
scaled for arbitrary $r_e$ as needed: $\hat{C}(x,y) =
C(xr_e,yr_e)$. Similarly, the pixel scale can be non-isotropically
scaled to obtain an arbitrary axis ratio $q=b/a$: $\hat{C}(x,y) =
C(xr_e,yr_e q)$.  In the end, we must interpolate over our model grid.
For \sersic\ models indexed by $n_i$, we use linear interpolation to
obtain an approximation of $C$ for $n_i\le n\le n_{i+1}$:
\begin{equation}
  \hat{C}(x,y; n) \approx \frac{1}{n_{i+1} - n_i}
  \left[
    (n_{i+1}-n)\hat{C}(xr_e,yr_eq; n_i) +
    (n-n_i)\hat{C}(xr_e,yr_eq; n_{i+1})
  \right].
\end{equation}
Since we may store the full set of tables in RAM, we choose to
increase the resolution of $n$ and the spatial resolution in the
table to obtain the desired accuracy, rather than increasing the order
of the interpolation.  For further efficiency, \galphat\ prepares two
tables, one uses a finer resolution, and the parameters governing the
generation of the tables can be adjusted by the user.  For the tests
presented here, we use one table for generating the inner part of
$C(x, y)$ ($r_e<10$) and another for the outer part ($10<r_e<100$),
having a resolution of 2000 and 1500 pixels, respectively, for each
given $n$, which is linearly distributed from 0.5 to 12.0, using 60
intervals.  Therefore, the region with $r_e<1$ is resolved using 100
pixels whose flux values are numerically integrated to high
accuracy. If we were to decrease the numerical error tolerance,
i.e. make it more accurate, when generating the table and were to use
more pixels, the model would become more accurate but would not
increase the computational cost.  It would only require more cache
memory for loading the tables.  The overall relative accuracy of the
image table is one part in $10^{6}$ for our \sersic\ models with
$n\in[0.5, 12.0]$.

\subsubsection{Rotation of the model galaxy}

One must rotate the realised profile with axis ratio $q$ to obtain the
desired position angle $\theta$.  A standard rotation using
interpolation would be too slow and would be insufficiently accurate
for our purposes.  Instead, \galphat\ rotates the model image in
Fourier space.  Consider a rotation by an angle $\theta$.  Any
rotation matrix may be decomposed into three shear operations
\citep{larkin1997}:
\begin{equation}
\mathsf{R}=
\left(
\begin{array}{cc}
\cos\theta & -\sin\theta\\
\sin\theta & \cos\theta
\end{array}
\right) = \mathsf{M_x M_y M_x} = 
\left(
\begin{array}{cc}
1 & -\tan\frac{\theta}{2}\\
0 & 1
\end{array}
\right) 
\left(
\begin{array}{cc}
1 & 0\\
\sin\theta & 1
\end{array}
\right) 
\left(
\begin{array}{cc}
1 & -\tan\frac{\theta}{2}\\
0 & 1
\end{array}
\right).
\label{eq:shift}
\end{equation}
The matrices $\mathsf{M_x}$ and $\mathsf{M_y}$ are shear operators
in the $x$ and $y$ directions, respectively.  Consider the function
$f(x,y)$ sheared in the $x$ direction by $a$: $f(x,y)\rightarrow
f(x+ay, y)$.
Using the Fourier shift theorem, 
\begin{equation}
{\mathbf U} \{f(x+ay,y)\} = {\mathrm{exp}}(-2\pi iuay) {\mathbf U} \{f(x,y)\}
\end{equation}
where $\mathbf{U}$ is the Fourier transform operator in $x$ and $u$ is
the transform variable.  Similarly, $f(x,y)$ sheared in the $y$
direction by $b$, $f(x,y)\rightarrow f(x,y+bx)$, has the Fourier
transform:
\begin{equation}
{\mathbf V} \{f(x,y+bx)\} = {\mathrm{exp}}(-2\pi ivbx) {\mathbf V} \{f(x,y)\}
\end{equation}
where $\mathbf{V}$ is the Fourier transform operator in $y$ and $v$ is
the transform variable.  Putting these together and performing the
inverse Fourier transform, the image sheared in the $x$ direction is:
\begin{equation}
I_x = {\mathbf U}^{-1} \{ {\mathrm{exp}}(-2\pi iuay) {\mathbf U}\{I(x,y)\}\}
\end{equation}
Next, the $x$-sheared image is sheared in the $y$ direction by another
Fourier transform and shift:
\begin{equation}
I_{yx}(x,y) = {\mathbf V}^{-1} \{ {\mathrm{exp}}(-2\pi ivbx) {\mathbf V}\{I_x(x,y)\}\}
\end{equation}
Lastly, the twice-sheared image is sheared again in the $x$ direction to
accomplish the rotation.  Hence, using equation (\ref{eq:shift}) the
rotated image may be written
\begin{equation}
I_\theta(x,y) = {\mathbf U}^{-1} \{ {\mathrm{exp}}(-2\pi iuay) {\mathbf U}\{I_{yx}(x,y)\}\}
\end{equation}
where $a=\tan(\theta/2)$ and $b=-\sin(\theta)$.  Computationally, the
rotation requires three 1D forward FFTs and three 1D inverse FFTs
performed on the 2D image and three 2D complex multiplications by the
phase factors ${\mathrm{exp}}(-2\pi iuay)$ and ${\mathrm{exp}}(-2\pi
ivbx)$.  We use the standard trigonometric recursion relations 
for evaluating $c_n\equiv\cos(2\pi n/N)$ and $s_n\equiv\sin(2\pi
n/N)$:
\begin{eqnarray}
c_0 &=& 1\\
s_0 &=& 0\\
c_{n+1} &=& c_n - (\alpha c_n + \beta s_n)\\
s_{n+1} &=& s_n + (\beta c_n - \alpha s_n)
\end{eqnarray}  
where $\alpha = 2\sin^2(\pi/N)$ and $\beta = \sin(2\pi/N)$.  
To shift the centre of the image to an arbitrary $x_0$ and $y_0$, one can
use $x-x_0$ and $y-y_0$ instead of $x$ and $y$ in the shear operations
above.  

Since the galaxy model image is smooth and the flux values go almost
to zero at the edges, aliasing should not cause any significant problems
but we pad the images with zeros for added safety.  In
practice, we increase the image size by $\sqrt{2}$ in each dimension
to provide a sufficient margin for image trimming after rotation.  We
convolve the interpolated, unrotated image with the PSF before
rotating the image in Fourier space.  We also reduce the dynamic range of
the surface brightness by a logarithmic mapping for large values of
$n$.  Then we rotate this PSF convolved image using the 3-shear
algorithm described above and apply the inverse logarithmic mapping if
necessary.  \galphat\ uses the FFTW package version 3.1.2 \citep{frigo2005}
for all its FFTs.

Since convolving with the PSF before rotation smooths out
high-frequency features in the profile, this image generation method
can produce very accurate images without any significant aliasing introduced
by the FFT rotation.  Furthermore, one could subdivide the pixels of the
image to perform the rotation computation and aggregate the pixels afterwards
to increase the accuracy of the rotation.
The error decreases exponentially with the number of
subdivisions.  For the test case described here, a subdivision by a
factor of two decreases the error by a factor of ten.  Of course, the
computation time increases as the square of the subdivision factor.

\begin{figure}
\centering
\subfigure[Exact image]{\epsfig{figure=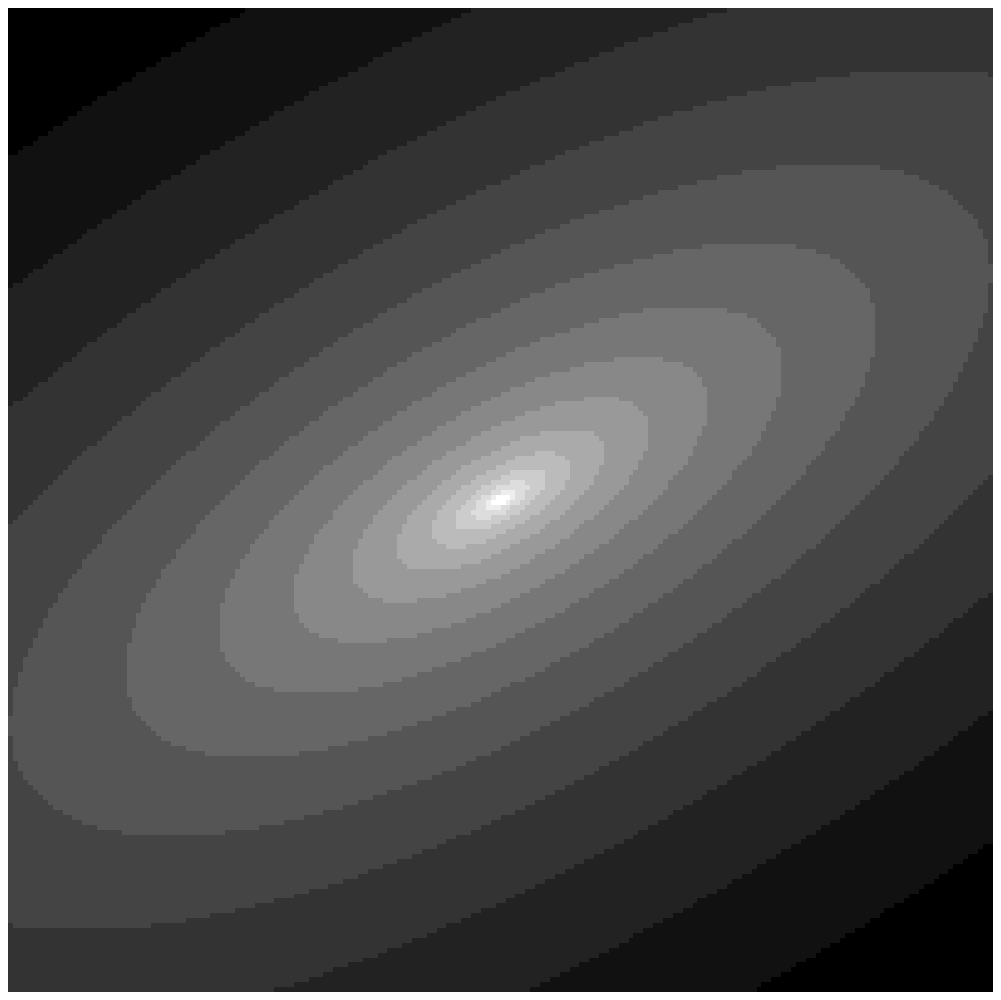,scale=0.45}}
\subfigure[Interpolated]{\epsfig{figure=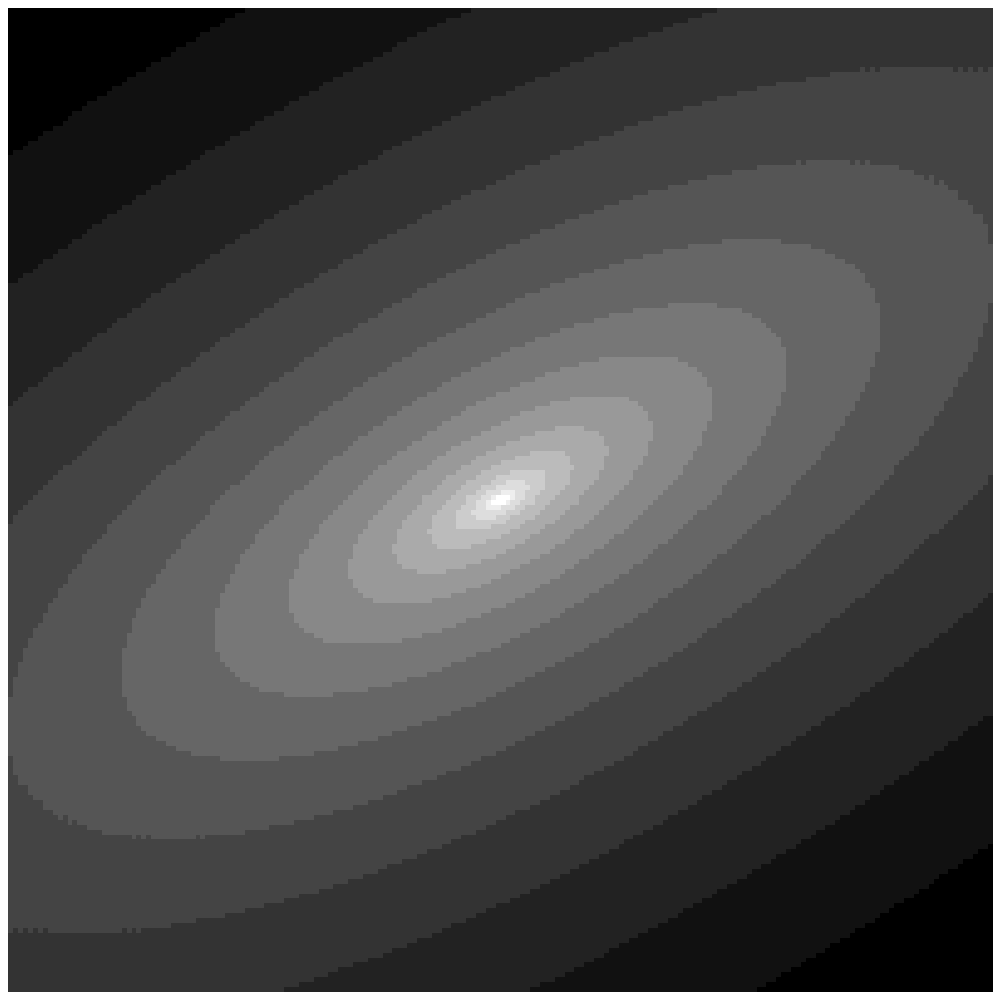,scale=0.45}}
\caption{A comparison between the integrated and the interpolated and
  rotated image of a normalised \sersic\ profile with $n=4$.  The axis
  ratio is 0.4 and the position angle is 30 degrees counterclockwise
  from the $x$ axis.  Panel (a) is an image produced by direct
  numerical integration of the profile and panel (b) is an image
  interpolated from table and rotated using the Fourier shift
  theorem as in \galphat.}
\label{tabimg}
\end{figure}

\begin{figure}
\centering
\subfigure[axis ratio $q=0.1$]{\epsfig{figure=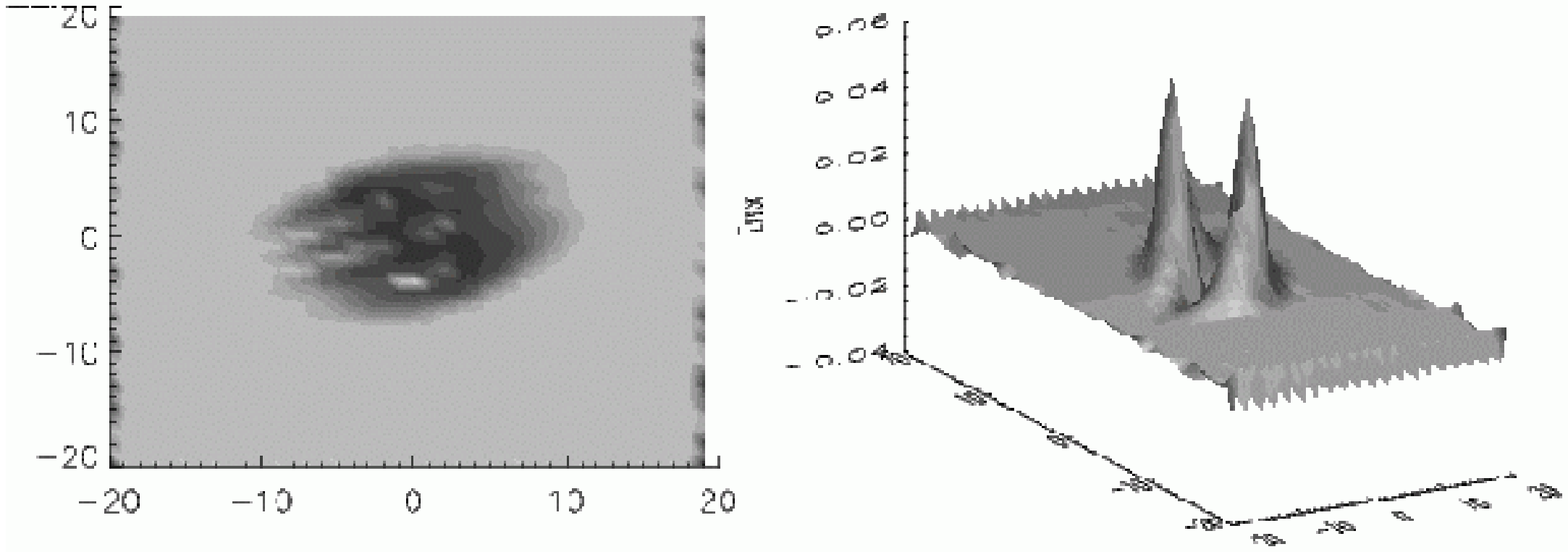,scale=0.7}}
\subfigure[axis ratio $q=0.4$]{\epsfig{figure=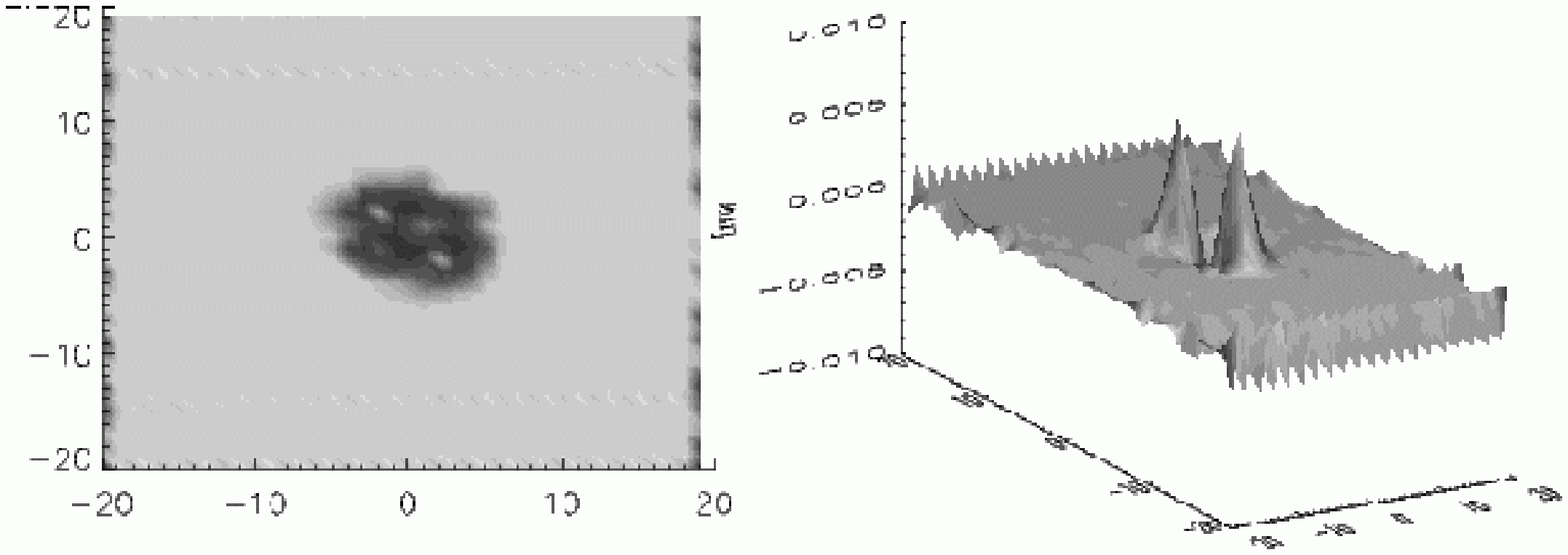,scale=0.7}}
\subfigure[axis ratio $q=0.7$]{\epsfig{figure=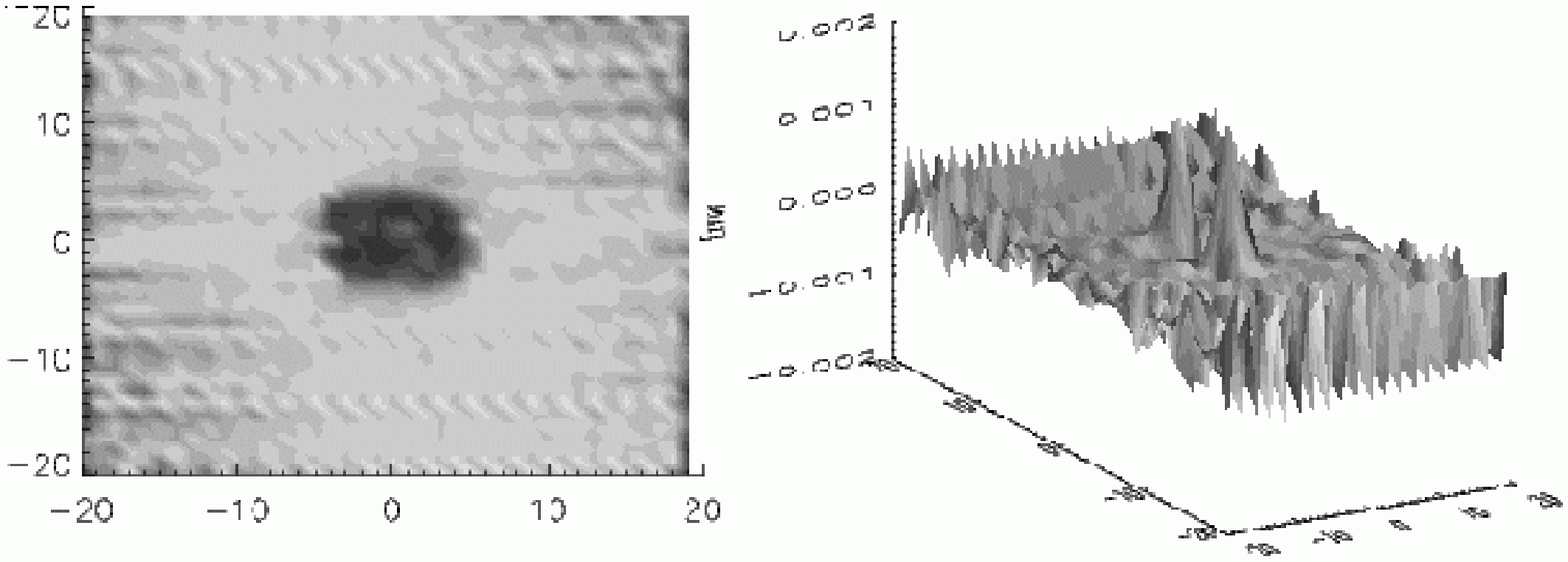,scale=0.7}}
\caption{Differences between an image produced through direct
  integration and that produced using the method in \galphat\ for a
  galaxy with $n=4$, $r_e=10$ pixels, and a position angle of
  $30^o$. The central region, $[-2r_e,2r_e]$, is shown to
  highlight the differences. 
  Panel (a),(b) and (c) are for $q=0.1$, $q=0.4$ and $q=0.7$ respectively.
  In each panel, the left column shows the face-on view of the relative 
  residual image and the right column shows a surface plot with the
  magnitude of the relative residual indicated on the $z$-axis.}
\label{errimg}
\end{figure}

As an example, we illustrate a $200\times200$ pixel $n=4$ \sersic\
model with $r_e=10$ pixels. We generate Figure \ref{tabimg}a by direct
numerical integration with a $\theta=30^o$ rotation and we generate Figure
\ref{tabimg}b using the image generation method in \galphat.
The two images appear the same.
Differences between the two methods only become obvious when one looks at
a relative residual image.
Figure \ref{errimg}a, \ref{errimg}b and \ref{errimg}c shows a comparison of 
the same galaxy (i.e. $n=4$ and $r_e =10$) generated using the methods of 
\galphat\ to the direct integrated image for three different axis ratios:
$q=0.1, 0.4, \mbox{and}\ 0.7$, respectively. We zoom in on the central region:
$[-2r_e,2r_e]$.
The left column in Figure \ref{errimg}a-\ref{errimg}c shows the face-on view of 
the relative residual image and the right column is the view
of the relative residual surface, with the magnitude of the residual
plotted in the $z$ direction.
The maximum relative difference decreases with increasing $q$ and
remains much less than 1\% except for the extreme case of a concentrated
galaxy with $n=4$ and $q=0.1$ (Fig. \ref{errimg}a), 
which is a very unrealistically small axis ratio for an
observed galaxy with an $n$ this large and still only
has a 5\% maximum error at the centre. The errors in the outer region are
negligible and the total flux is still conserved to better than one part in 
$10^{6}$ in all three cases.  Galaxies with smaller \sersic\ indices have
even smaller errors even when the axis ratio is small; such small axis
ratios are more realistic for observed galaxies when the \sersic\ index is
small.  The errors would be further reduced if we generated the images
using pixel subsampling.

\subsubsection{Computation time for model generation}

\begin{table}
\centering
\caption{\galphat\  model generation time}
\begin{tabular*}{0.99\textwidth}{@{}lllll@{}} \hline\hline
Image/model & CPU & CPU time (total) & CPU time (interpolation) & CPU time (FFT rotation) \\ \hline
190 by 190     & Quad-core AMD Opteron & 0.092 sec & 0.079 sec & 0.013 sec \\
\sersic & 2613 MHz              &           &           &           \\ \hline
\end{tabular*}
\label{mdlgentime}
\end{table}

The wall clock time for a posterior simulation depends on the model,
the image, and the MCMC algorithm.  A typical run using the PHS
algorithm requires of $\mathcal{O}(10^5)$ evaluations (see
\S\ref{runtimesection}).  Here we provide a CPU time estimate for the
generation of a single \sersic\ model (Table \ref{mdlgentime}) for an example
image with a size of $190\times190$ pixels.  The generation of a single
\sersic\ model requires 0.092 sec (on a single processor).  This is
more than an order of magnitude faster than would be required using
direct numerical integration over the pixels, and we still preserve
high numerical accuracy.  Most of the CPU time is spent interpolating
the image tables to obtain $I_{jk}$; the FFT rotation is a minor
contribution.  Since this image is larger than the mean galaxy size in
our 2MASS target sample, we conclude that \galphat\ is both
sufficiently fast and sufficiently accurate to be suitable for a
large-scale analysis over a large catalogue.  (see
\S\ref{runtimesection} for a discussion of the total MCMC run time).

\subsection{The likelihood function and the prior distribution}

CCD detectors count photons and this is well-described as a Poisson
process.  If the model predicts the flux $m_i$ for $i$th pixel then the
probability that we measure the flux $d_i$ for that pixel follows
a Poisson distribution, $P(d_i|m_i) = \exp(-m_i) m_i^{d_i}/d_i!$.
Assuming that each pixel is
independent, our likelihood function $L(D|\theta)$ is
\begin{equation}
L(D|\theta) = \prod^{N_{pix}}_{i=1} P(d_i|m_i)
\end{equation} 
where $N_{pix}$ is the total number of pixels and $\theta$ is the
parameter set of the \sersic\ model. To increase the numerical accuracy, we 
accumulate the logarithmic value of the probability.

As previously described, the prior distribution of model parameters
affects the inference.  A preliminary study with non-uniform prior
distributions confirms that the posterior maximum and confidence
values can be significantly changed for low \sn\ data that dominate
the galaxy population in a flux-limited survey.  However, for
simplicity, we have used uniform prior distributions for all the
parameters for the tests in this paper.  Again, the use of an informative
prior without careful consideration will almost certainly lead to a
biased result.  We present a detailed study of prior distributions for
inference and model selection for galaxy profiles in our followup
paper \citep{yoon2010b}.

\subsection{Sampling the posterior probability}
\label{mcmcsample}

It is difficult to ensure that the Markov chain correctly samples from
the entire posterior distribution in a high-dimensional parameter
space.  To combat these difficulties, the \bie\ uses a variety of
algorithms, each with features effective for different problems that
impede the efficiency of sampling and convergence.  One should
characterise the MCMC simulation on representative data using
different algorithms before starting any production runs.  We find
that the PHS algorithm performs best for our problem; it more
efficiently samples parameter space and more quickly reaches a steady
state compared with the other MCMC algorithms.  Here, we will briefly
introduce the PHS algorithm \citep{rigat2008}.

The PHS algorithm constructs an $n$-rung temperature ladder that
\emph{powers up} or \emph{heats} the target posterior probability
$\pi_0$:
\begin{equation}
  \pi_i = c_i\,\pi_0^{1/T_i}  = c_i\,e^{(\log\pi_0)/T_i}\quad
  \mbox{for}\quad n=0,\ldots,n
\end{equation}
where $1=T_0<T_1<\cdots<T_n$ and $c_i$ is a normalisation constant.
The number of chains can be chosen by the user as well as the maximum
temperature considering the dimensionality of the model.
Each Monte Carlo step has two
parts: 1) each chain is updated using a standard Metropolis-Hastings
step; and 2) chains at different temperatures may be swapped 
by one of two algorithms: i) each chain state is updated at fixed
temperature or swapped with a chain state at an adjacent temperature
following a fixed swapping probability (standard parallel chains); or
ii) an exchange is proposed between the cold (fiducial) chain and one
of the warmer chains and the remaining chains are updated at fixed
temperature.  At the end of the run, the fiducial chain with $T_0$
samples the posterior distribution.

\subsection{Chain convergence}

Monte Carlo simulations of the posterior distribution may suffer from
two classes of difficulties: 1) the Markov chain is mixing poorly in a
particular mode in the posterior distribution, and this leads to a
large number of dependent states; and 2) the posterior distribution
may have two or more discrete modes with similar posterior probability
and the Markov chain can not move between them.  The first difficultly
is easily diagnosed by observing very low or very high acceptance
rates and is addressed by tuning the Metropolis-Hastings transition
probability.  The second difficulty is addressed by a variety of
hybrid MCMC algorithms, implemented in the \bie, designed to move
between modes.  The parallel chain algorithm, and tempering in
general, decreases the contrast of the \emph{hills and valleys} in the
posterior distribution, which allows occasional large excursions
between modes.  

To test for convergence of the cold chain in the PHS algorithm,
i.e. the chain with $T_0$, we use a subsampled convergence diagnostic
\citep{giakoumatos1999} for this single chain.  The chain is cleaved
and used to compare in-chain and inter-chain variances, similar to the
Gelman-Rubin \citep{gelman1992} test. The ratio of these two variances should
approach unity for a converged steady-state chain. We have tested
\galphat\ over a wide variety of synthetic image data typical of
observed galaxies, using the empirically determined transition
probability, and have confirmed that our MCMC algorithm samples the
posterior with a reasonable acceptance rate of $\geq 25\%$ for good
mixing and a swapping rate of $\geq 25\%$ for efficient mode
exploration.

\section{Simulated galaxy images}
\label{datasection}

To characterise the performance of \galphat, we generated an ensemble
of 3000 isolated, synthetic \sersic\ galaxy images representative of
survey observations.  We vary both the \sn\ from 5 to 100 and $r_e$ to
probe the extremes of barely resolved galaxies and that of galaxies
that extend beyond the image frame. The PSF is a Gaussian with a 2.96
FWHM in pixels, which we convolve with the model images.  We use a
Poisson noise model and a gain factor of 8.0 [$\mbox{e}^{-}$/ADU].  Both
of these choices are motivated by 2MASS images.  We describe the
details of our choices below.

\subsection{Varying \sn}
\label{datasectionsn}

We define the signal-to-noise ratio as the ratio of the flux from the
galaxy within the half-light radius to the noise from the sky
background and the galaxy within the same area:
\begin{equation}
\mbox{\sn} = \frac{\langle\rho\rangle}{\sqrt{\langle\rho\rangle +
\langle\rho_{sky}\rangle}}
\label{eqsn}
\end{equation}
where $\langle\rho\rangle$ is the total electron count of the galaxy
profile within the area $\pi r_e^2 q$ and $\langle\rho_{sky}\rangle$ is
the background within the same area.

For each of $\sn\in\{5, 10, 20, 50, 100\}$, we generate 100 \sersic\
model galaxies with a randomly chosen combination of uniformly
distributed $r_e\in[6,20]$ $n\in[0.7,7.0]$, axis ratio $q\in[0.1,
1.0]$, and position angle $\mbox{PA}\in[0^\circ,90^\circ]$. We fix the
sky background to 300 [ADU].  Once we choose $r_e$ and $q$, we
determine the galaxy's magnitude for the given \sn\ value and
magnitude zero point using equation (\ref{eqsn}).

\subsection{Varying  $r_e$}
\label{datasectionre}

The inference of $r_e$ is biased if the galaxy is small compared to the PSF.
To test this, we generate 100 \sersic\ model galaxies for all
combinations of $\sn\in\{5, 10, 20, 50, 100\}$ and $r_e\in\{ 0.5, 1.0,
2.0, 4.0, 8.0\}$ (in units of $\frac{1}{2}$ FWHM of the PSF) using the same
distributions of $n$, $q$, and PA as in
\S\ref{datasectionsn}. We compute the Poisson counting errors and the sky
background as in \S\ref{datasectionsn}.

\section{GALPHAT performance testing}
\label{resultsection}

Parametric surface brightness models invariably result in parameter
covariance, and this covariance can be exacerbated by instrumental and
selection errors. The Bayesian MCMC approach explicitly incorporates
parameter covariance, noise sources and other selection effects
including data \sn, PSF convolution and the sizes of the galaxy and
the image, to yield a reliable inference, as illustrated in
\S\ref{casestudy} and \S\ref{rescue}.

Using the simulated galaxies from \S\ref{datasection}, we investigate
the effect of observational attributes such as the \sn, the galaxy's
$r_e$ compared to the PSF FWHM, the image size compared to galaxy's
$r_e$, and errors in the assumed PSF FWHM.
Although we generated 40,000 converged MCMC states for each image, we
do not use the full 40,000 MCMC states to construct the posterior
since we want a more conservative estimate of the burn-in period than
that diagnosed by the convergence test.  As previously described, we
use the PHS algorithm \citep{rigat2008} for all of our tests and
determine convergence using the subsampled convergence test
\citep{giakoumatos1999}. We tune the width of the Metropolis-Hastings
transition distribution to yield, roughly, a 25\% acceptance rate and
25\% chain swapping rate for each chain.  The prior distribution of
the galaxy centroid is normal with a mean at the image centre and a
$\sigma=$3 pixels.  The sky background prior is also normal with a
mean of the input sky background and a $\sigma=$0.5 ADU.  The prior
distributions for the other model parameters have a uniform
probability within a finite range.

\subsection{Examples of single fits}
Before we present our results for an ensemble of galaxies, we present the
results of \galphat\ fits to four galaxies.
From the simulated galaxy sample, we picked four example galaxies, two
that represent elliptical galaxies ($n\approx4$) and two that represent 
exponential disc ($n\approx1$) galaxies.  One galaxy of each type has a 
low \sn\ of 5 and one has a high \sn\ of 100.

\begin{figure}
\centering
\epsfig{figure=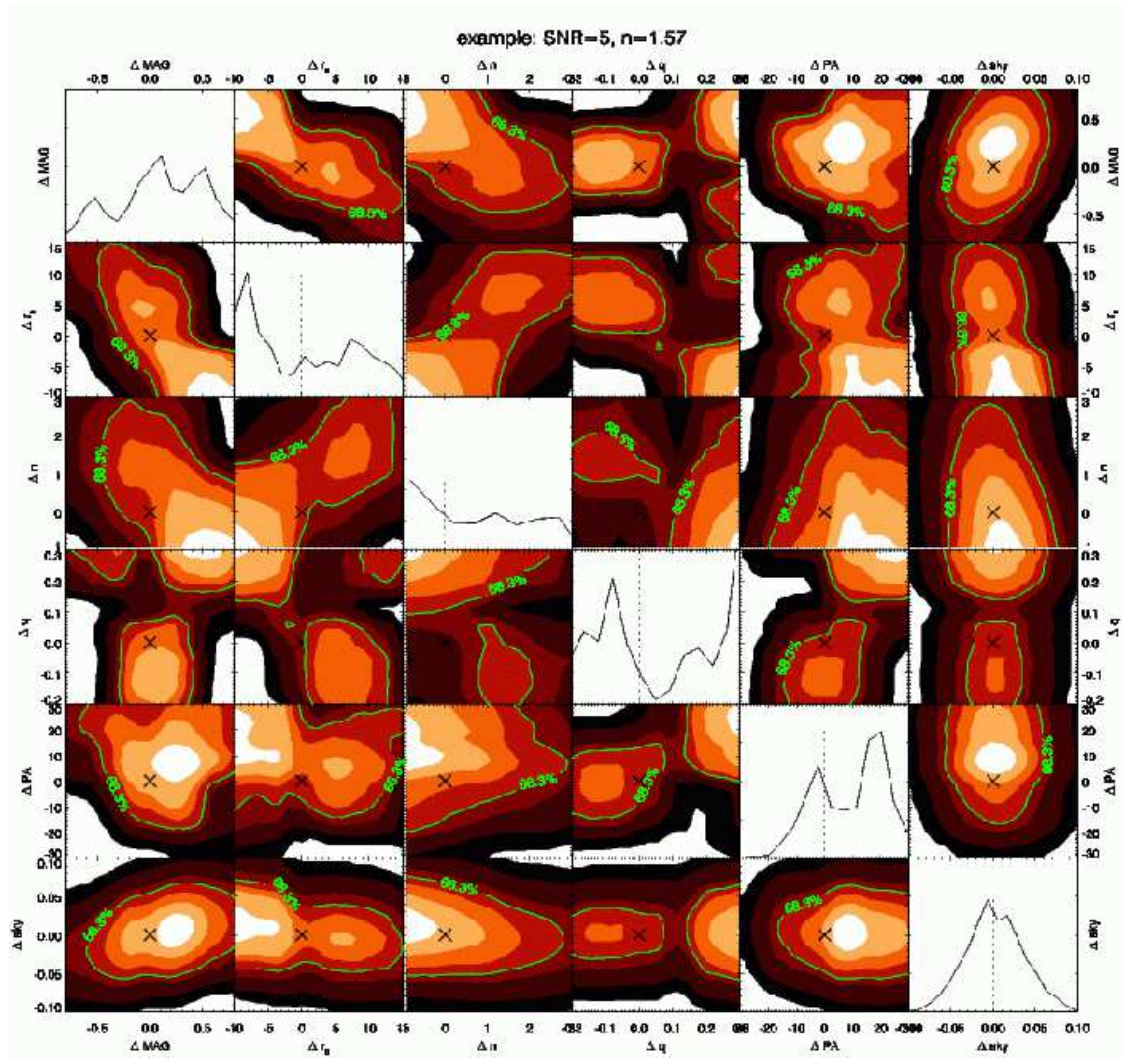,scale=0.6}
\vspace{5pt}
\caption{Posterior marginals for a galaxy with $\sn=5$ and $n=1.57$
  (i.e. an exponential-disc galaxy).  The full marginal distribution
  for each model parameter residual is shown on the diagonal with the
  vertical dotted line indicating the zero point.  Joint marginal
  pairs of parameter residuals are shown on the off-diagonals.  The
  seven colour contours represent the 10, 30, 50, 68.3, 80, 95 and
  99\% confidence levels and the green solid line is the 68.3\%
  confidence level.  The locations of zero points are indicated by
  $\times$ symbols.  Posteriors are not unimodal and are spread over large
  range in parameter space.  Although very weak covariances exist
  among magnitude, half-light radius ($r_e$), \sersic\ index ($n$) and
  sky background, parameter uncertainties are largely dominated by
  Poisson random noise.  }
\label{corr_sn5_n1_eg}
\end{figure}

\begin{figure}
\centering
\epsfig{figure=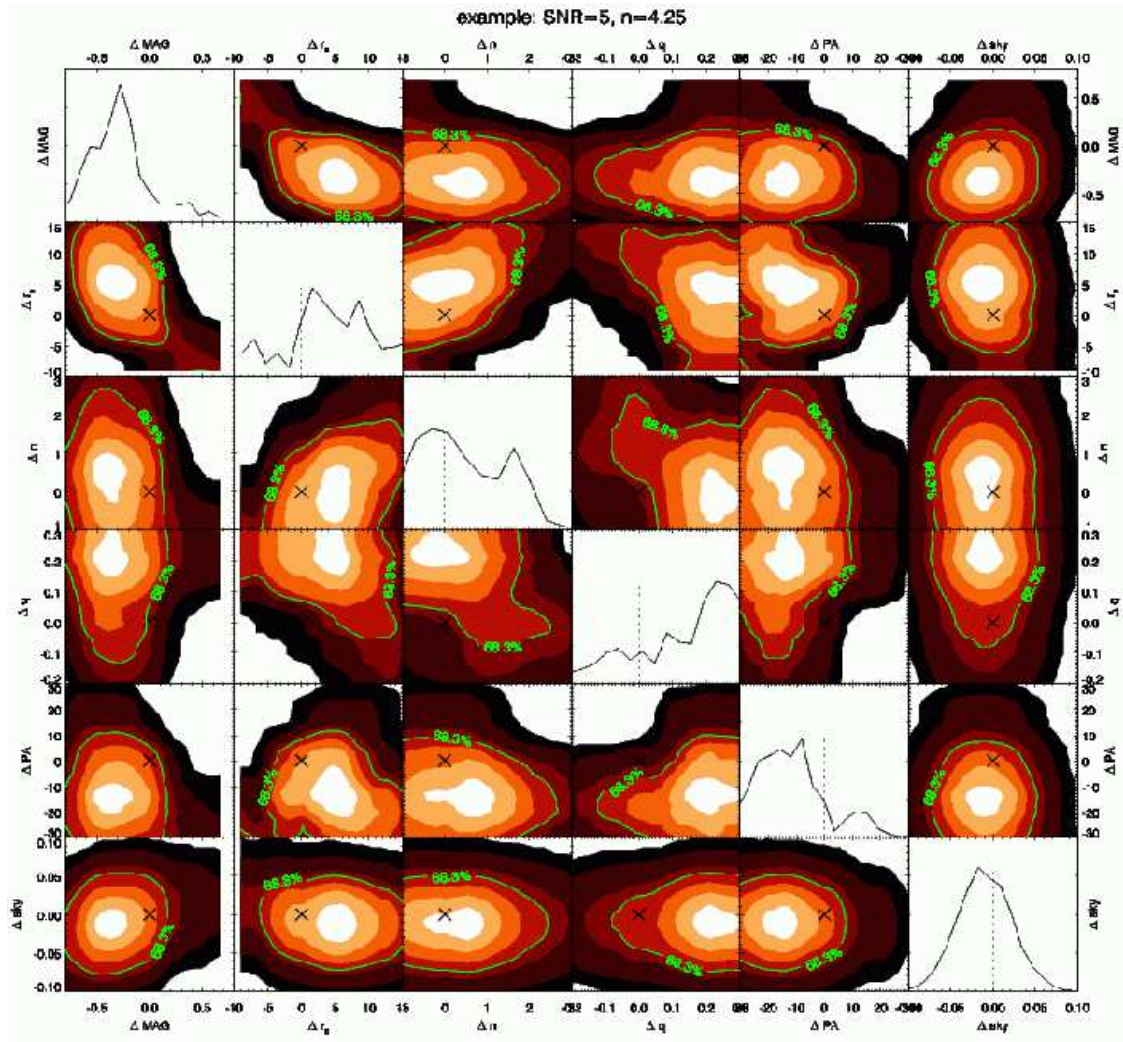,scale=0.6}
\vspace{5pt}
\caption{Posterior marginals for a galaxy with $\sn=5$ and $n=4.25$
  (i.e. an elliptical galaxy).  See the caption for
  Fig. \ref{corr_sn5_n1_eg}.  As for the exponential disc galaxy,
  parameter covariance is not significant and the posteriors spread
  out owing to noise.  }
\label{corr_sn5_n4_eg}
\end{figure}

\begin{figure}
\centering
\epsfig{figure=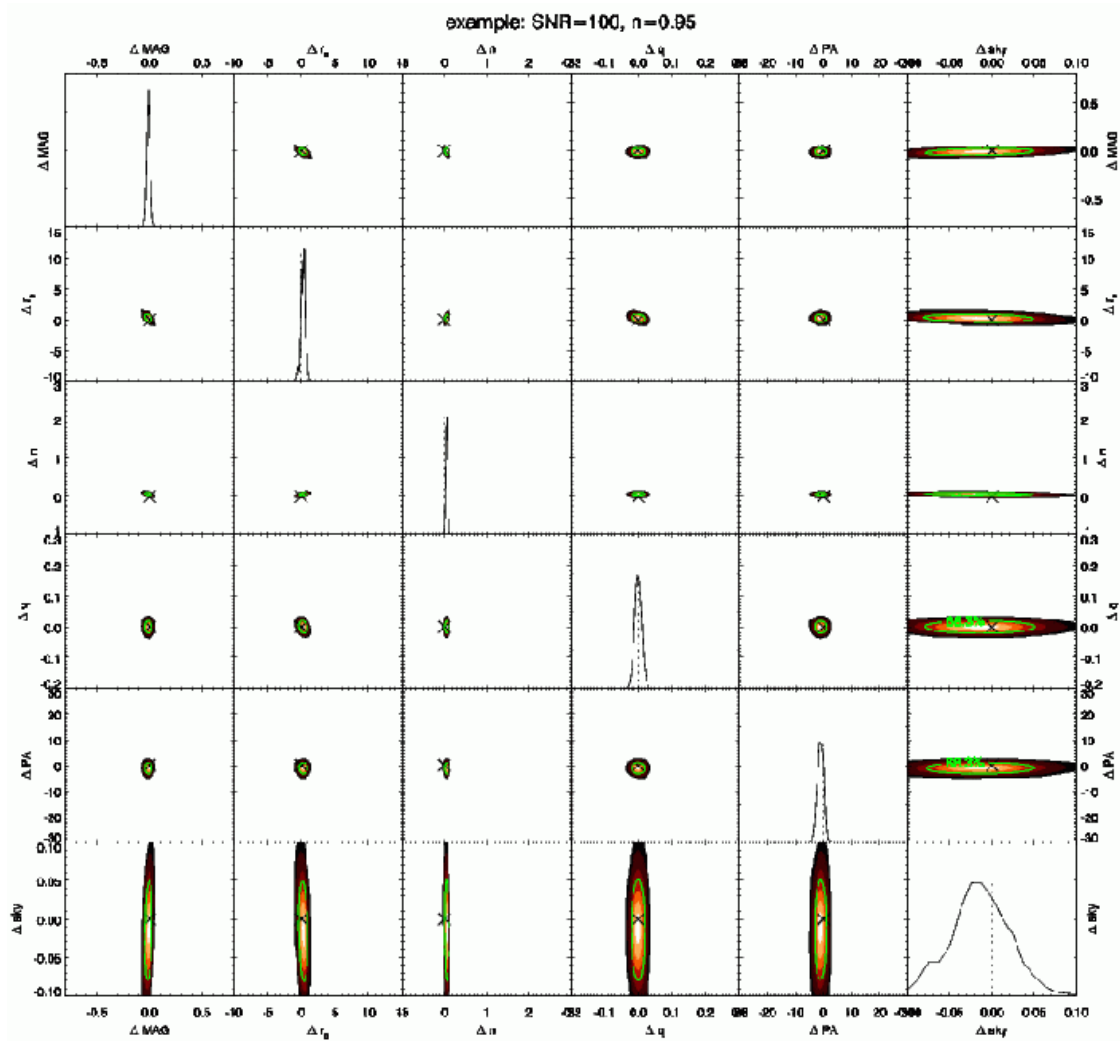,scale=0.6}
\vspace{5pt}
\caption{ Posterior marginals for a galaxy with $\sn=100$ and $n=0.95$
  (i.e. an exponential-disc galaxy).  See the caption for
  Fig. \ref{corr_sn5_n1_eg}.  Unlike the low \sn\ case, strong
  parameter covariances exist among magnitude, $r_e$, $n$, and the sky
  background, and the parameter posteriors are confined in a narrow
  region close to the true value in parameter space.  }
\label{corr_sn100_n1_eg}
\end{figure}

\begin{figure}
\centering
\epsfig{figure=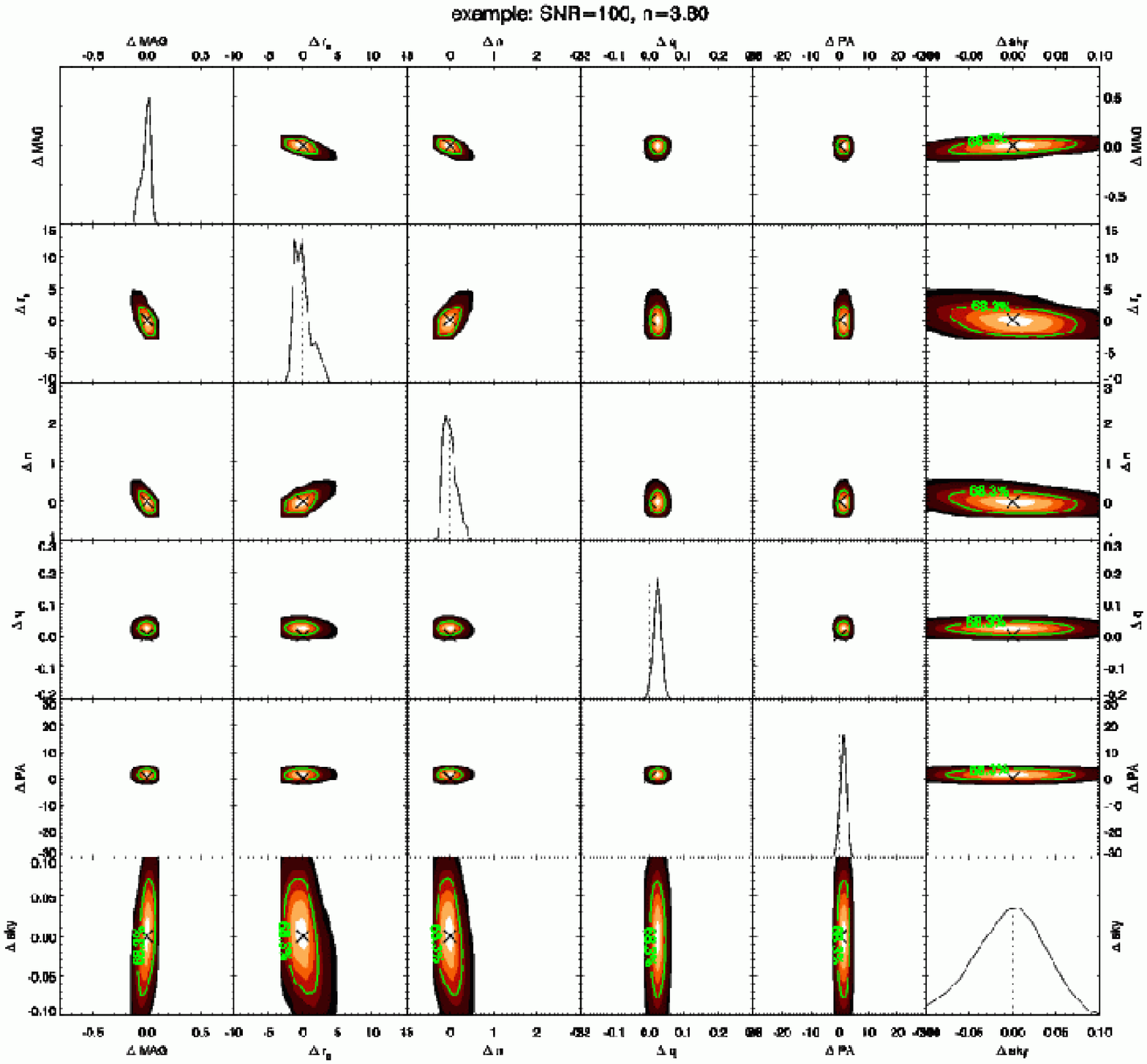,scale=0.6}
\vspace{5pt}
\caption{ Posterior marginals for a galaxy with $\sn=100$ and $n=3.80$
  (i.e. an elliptical galaxy).  See caption for
  Fig. \ref{corr_sn5_n1_eg}.  While parameter posteriors are compact
  and parameter uncertainty is dominated by parameter covariance, as
  for the high \sn\ exponential disc galaxy, there is a stronger parameter
  covariance among magnitude, $r_e$, $n$, and the sky background than
  for the exponential disc.  This leads to inflated errors for
  individual parameters determined from the marginalised distribution.
}
\label{corr_sn100_n4_eg}
\end{figure}

In Figures \ref{corr_sn5_n1_eg}--\ref{corr_sn100_n4_eg} we show the
marginalised posteriors of the galaxy magnitude (MAG), the
half-light radius ($r_e$), the \sersic\ index ($n$), the axis ratio ($q$),
the position angle (PA), and the sky background (sky). 
Figures \ref{corr_sn5_n1_eg} and \ref{corr_sn5_n4_eg} show the results
at low \sn\ for an exponential disc-like galaxy and an elliptical-like
galaxy, respectively, and
Figures \ref{corr_sn100_n1_eg} and \ref{corr_sn100_n4_eg} 
show the results at high \sn\ for an exponential disc-like galaxy and an 
elliptical-like galaxy, respectively.
In each figure, the full marginal distribution 
for each model parameter residual is shown on the diagonal with the vertical dotted line
indicating the zero residual, and the joint marginal distributions
of parameter pairs are shown on the off-diagonals with the seven colour
contours corresponding to the 10, 30, 50, 68.3 (green solid line), 
80, 95 and 99\% confidence levels. The locations of the zero residuals are
indicated by the $\times$.

When the \sn\ is small, for both an exponential disc galaxy (see
Fig. \ref{corr_sn5_n1_eg}) and for an elliptical galaxy (see
Fig. \ref{corr_sn5_n4_eg}), the posteriors are not unimodal and are
spread out over a large range in parameter space. Although there is a
weak covariance between the magnitude, $r_e$, $n$, and the sky
background, uncertainties in the parameter inferences are largely
dominated by the Poisson random noise present in the image.  However,
the situation becomes very different at high \sn\ as shown in Figure
\ref{corr_sn100_n1_eg} for an exponential disc galaxy and in Figure
\ref{corr_sn100_n4_eg} for elliptical galaxy. Note that these figures have
the same scale as Fig. \ref{corr_sn5_n1_eg} and Fig. \ref{corr_sn5_n4_eg}. 
Now the posterior forms
a strong mode close to the true value and the morphology of the
posterior is determined by the parameter covariance present in the
\sersic\ model, which strongly depends on the \sersic\ index.  There
is a stronger parameter covariance among magnitude, $r_e$, $n$, and
the sky background for the high \sn\ elliptical galaxy than for the
low \sn\ exponential disc galaxy, which can lead to larger errors when
marginalising the posterior distribution.

The Bayesian-based \galphat\ procedure explicitly incorporates the 
parameter covariance present in the \sersic\ model.
Furthermore, it enables us to utilise the entire posterior distribution 
for a galaxy population to reliably test hypotheses based on that population.
This is practically feasible using \galphat, as we will demonstrate in
following sections.

\subsection{Model covariance and bias}

Using the posterior distribution from the converged Markov chain, we
illustrate the inherent covariance between parameters by showing joint
distributions of selected parameter combinations for the \sersic\
model.  To emulate a catalogue analysis, we generate an ensemble of
galaxies with astronomically representative model parameters for each
bin in observational conditions.  The dependence on observational
conditions are explored in the following several sections.

\subsubsection{Effects of galaxy \sn}
\label{snsection}

We characterise the posterior distributions of 500 images for \sersic\
models (100 in each \sn\ bin with a range of structural parameters,
see \S\ref{datasectionsn}).  Recall that these \sersic\ models have 8
free parameters: the centroid coordinates ($x,y$), the magnitude (MAG), the
half light radius $r_e$, the \sersic\ index $n$, the axis ratio $q$,
the position angle PA, and the sky background (sky).  We use the last 25000
states to characterise the posterior.  We constructed an ensemble
posterior distribution by pooling the sampled distribution for all the
images in each \sn\ bin.
 
\begin{figure}
\centering
\epsfig{figure=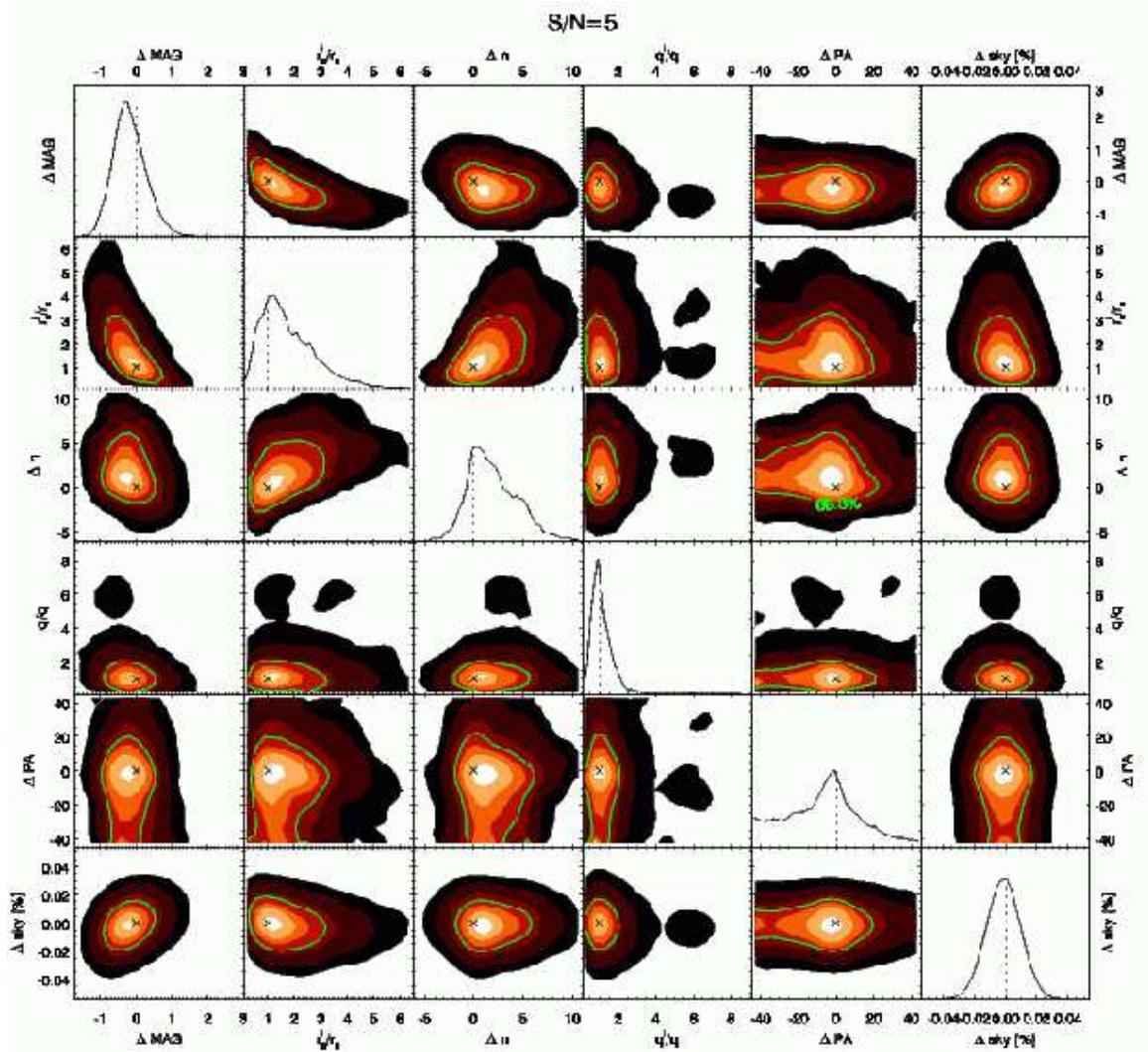,scale=0.7}
\vspace{5pt}
\caption{Posterior error distributions for the ensemble of galaxies
  with $\sn=5$.  The full marginal error distribution for each model
  parameter is shown on the diagonal. Joint marginal pairs of
  parameters are shown on the off-diagonals.  \galphat\ inferred
  parameters $r_e^i$ and $q^i$ are scaled by their input values and
  the other parameters are differences from their input values.  The
  fractional sky background errors are percentages.  The seven colour
  contours represent 10, 30, 50, 68.3, 80, 95 and 99\% confidence
  levels and the green solid line is the 68.3\% confidence level. The
  locations of the input values are indicated by vertical dotted lines 
  for the diagonal and $\times$ symbols for the off-diagonals.
  The values of magnitude, $r_e$, $n$ and sky background are strongly
  correlated.  Although the constraints are tighter with increasing
  galaxy \sn, the strength of the parameter covariance increases with
  increasing galaxy \sn\ (see
  Figs. \ref{corr_sn10}--\ref{corr_sn100}).}
\label{corr_sn5}
\end{figure}

\begin{figure}
\centering
\epsfig{figure=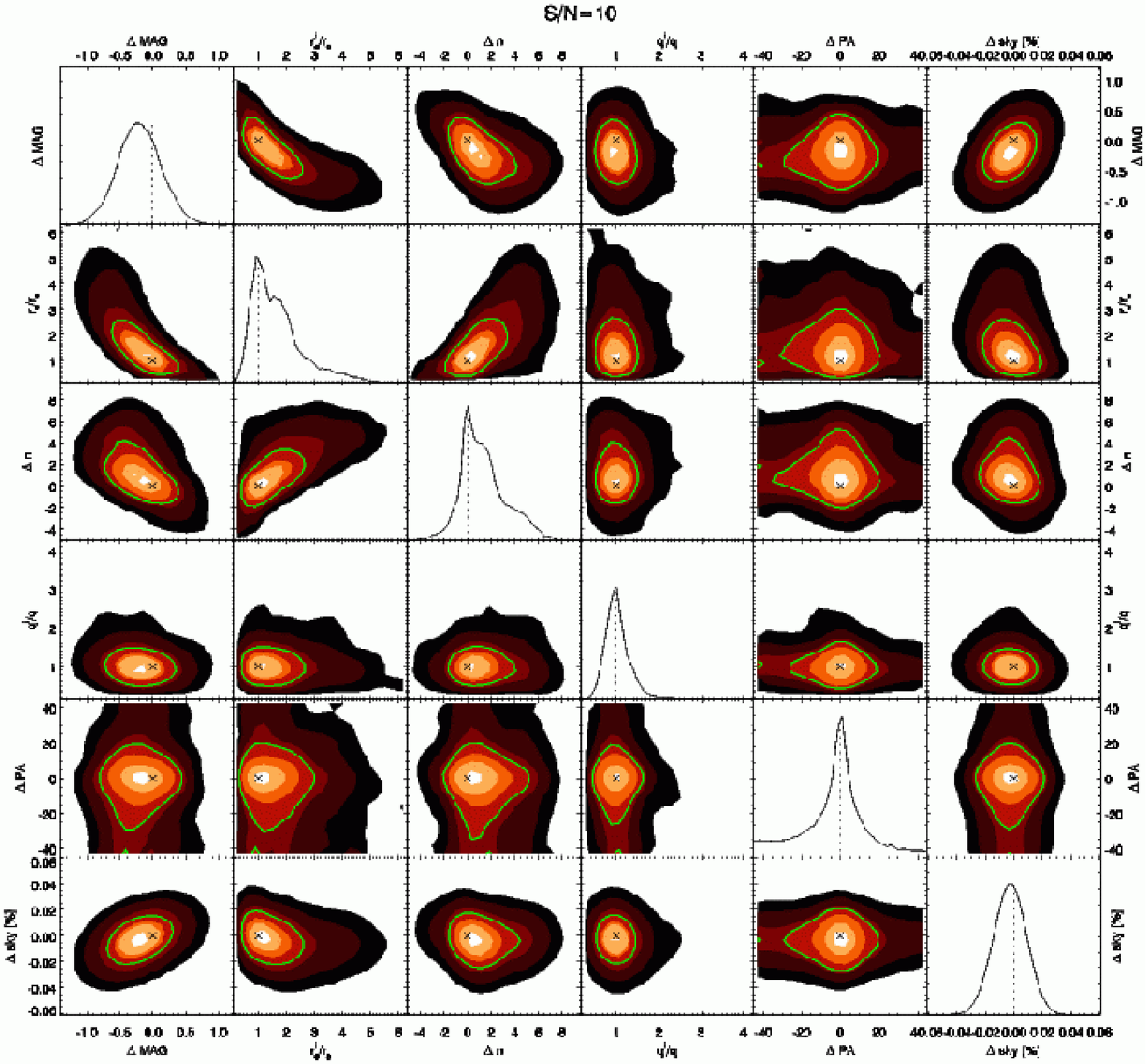,scale=0.6}
\caption{Ensemble parameter error posteriors for galaxies with \sn$=10$. See
  caption for Fig. \ref{corr_sn5}.}
\label{corr_sn10}
\end{figure}

\begin{figure}
\centering
\epsfig{figure=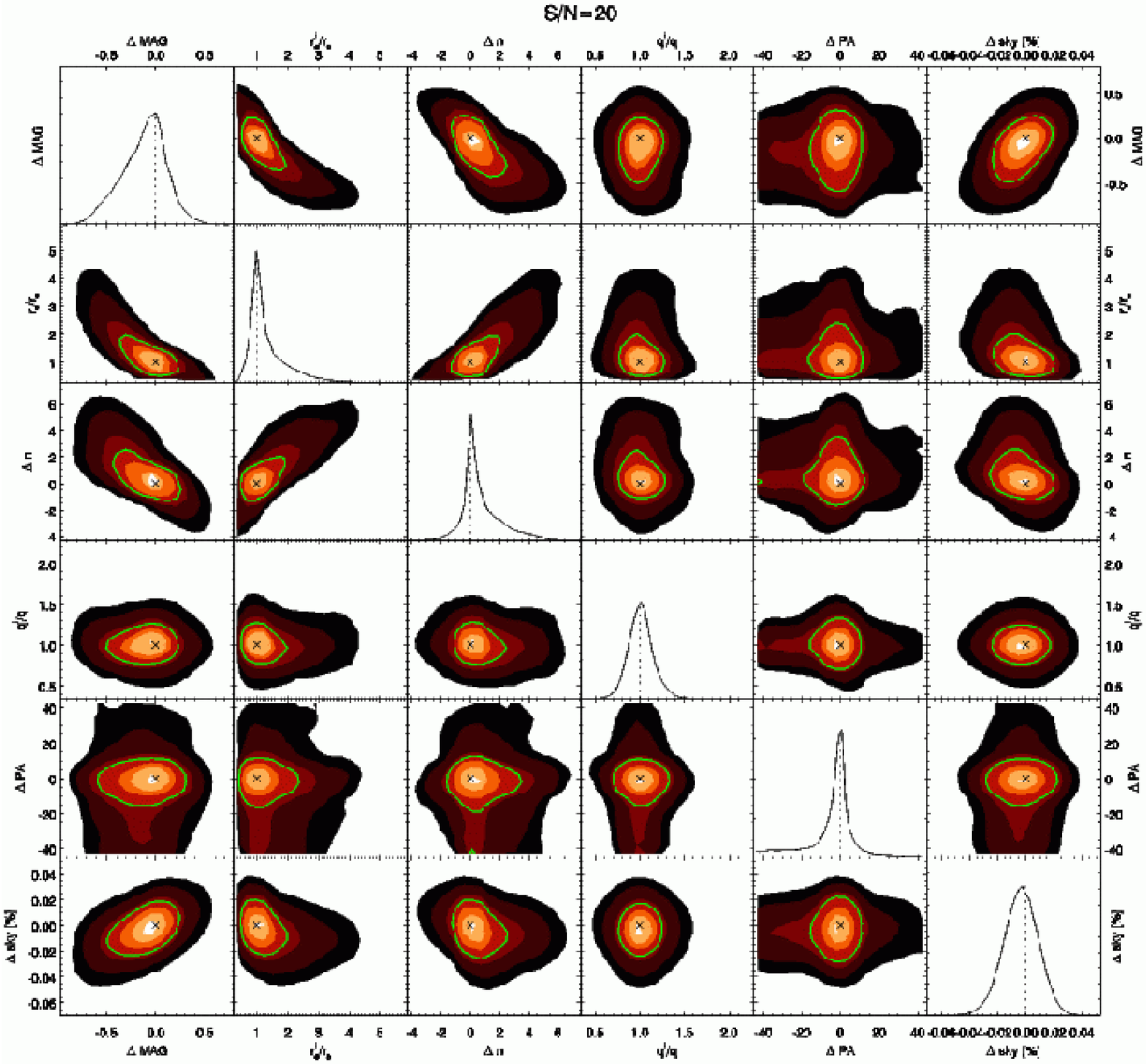,scale=0.6}
\caption{Ensemble error parameter posteriors for galaxies with \sn$=20$. See
  caption for Fig. \ref{corr_sn5}.}
\label{corr_sn20}
\end{figure}

\begin{figure}
\centering
\epsfig{figure=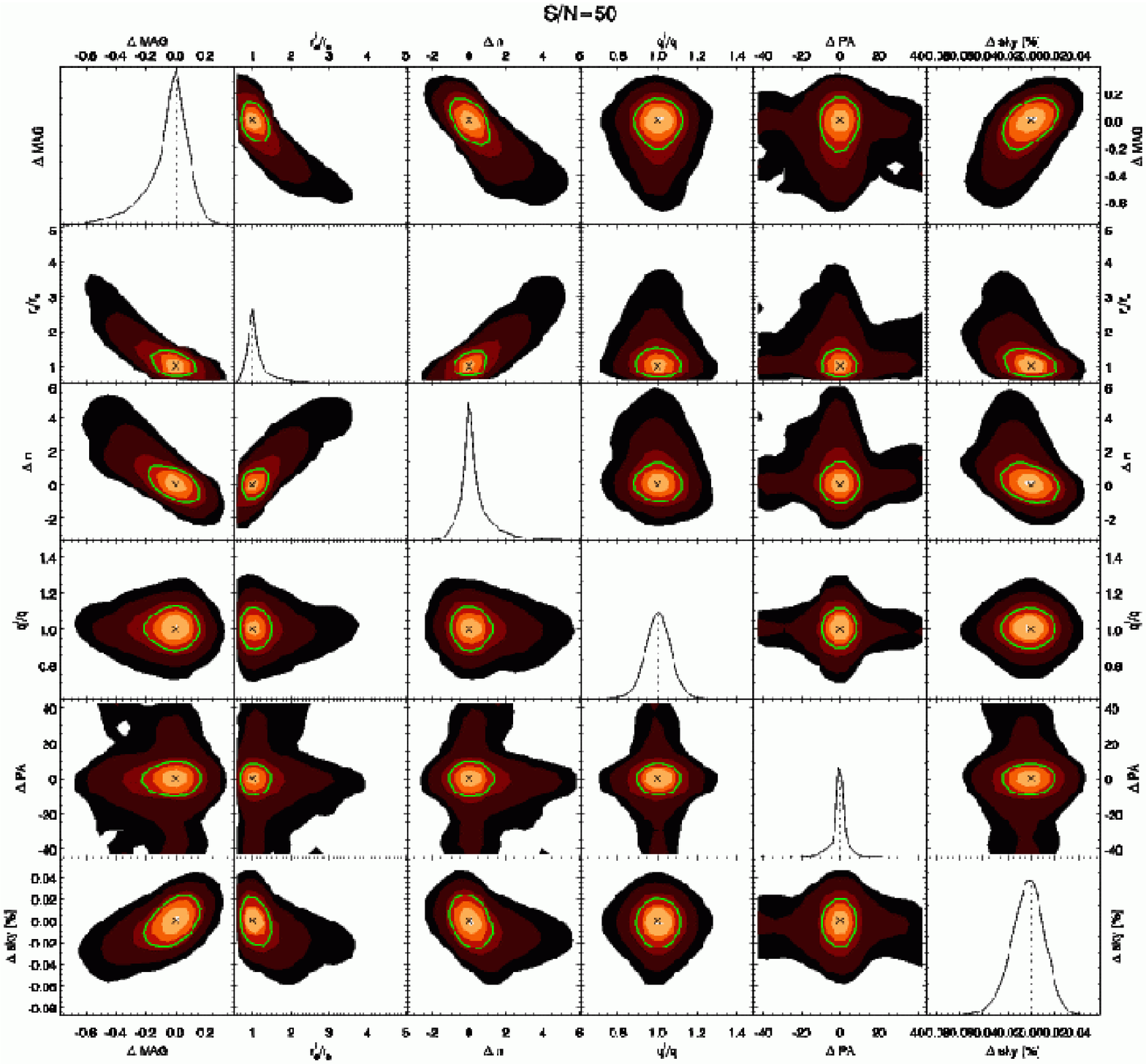,scale=0.6}
\caption{Ensemble error parameter posteriors for galaxies with \sn$=50$.
See caption for Fig. \ref{corr_sn5}.}
\label{corr_sn50}
\end{figure}

\begin{figure}
\centering
\epsfig{figure=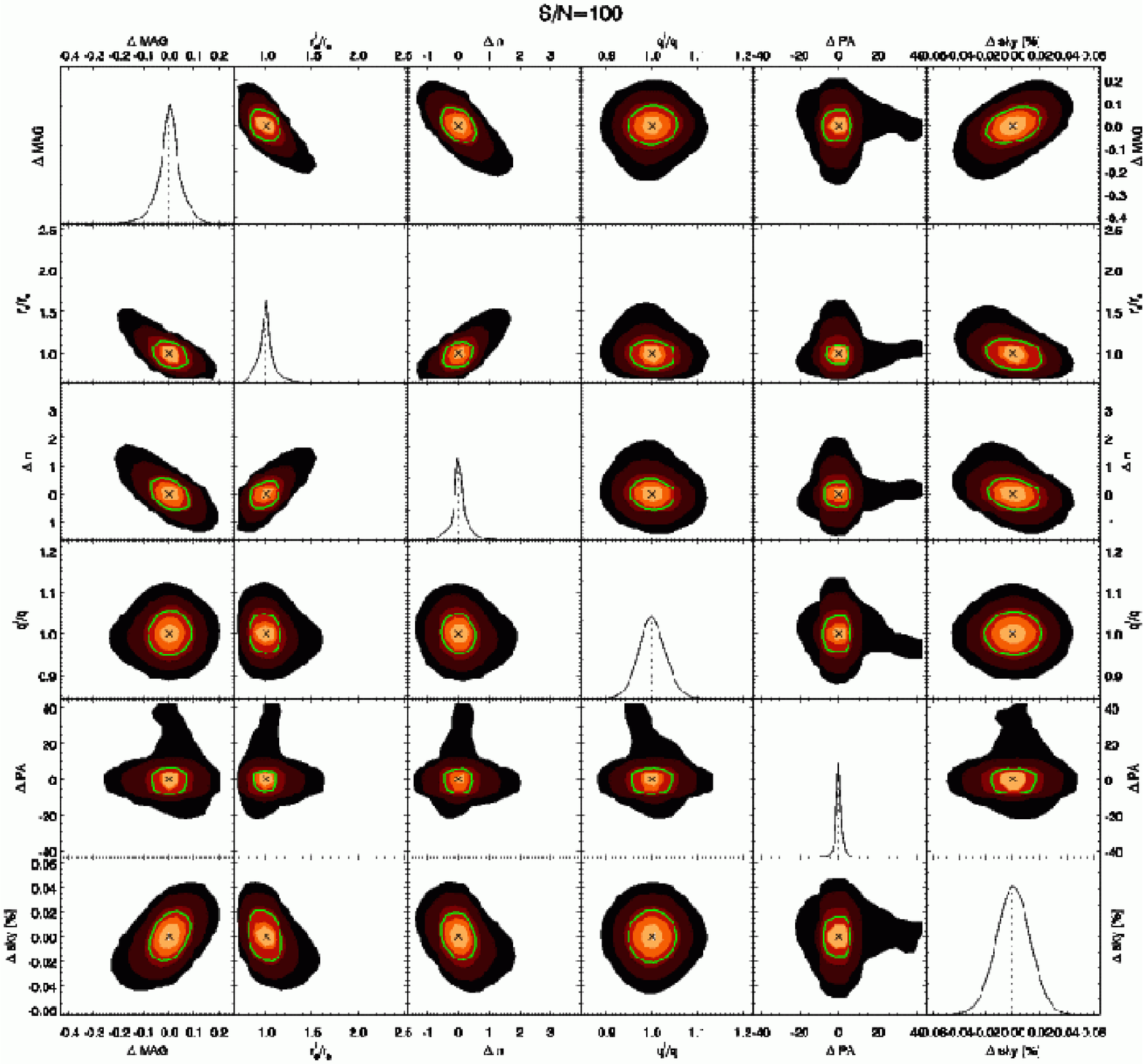,scale=0.6}
\caption{Ensemble error parameter posteriors for galaxies with \sn$=100$. See caption for Fig. \ref{corr_sn5}.}
\label{corr_sn100}
\end{figure}

We show all the marginalised distributions for errors in the
magnitude, $r_e$, $n$, $q$, PA, and the sky background for each \sn\
bin in Figures \ref{corr_sn5}, \ref{corr_sn10}, \ref{corr_sn20},
\ref{corr_sn50} and \ref{corr_sn100}.  Hereafter,
we use the superscript $i$ to denote the \galphat\ inferred value. The
parameters $r_e$ and $q$ are plotted as fractional changes scaled by
their input values and the values of the other parameters are the
differences from their input values.  We quote the sky background
error as the fractional percent error, i.e. the fractional change
scaled by the input sky background multiplied by 100. Each diagonal 
subplot is the full marginalised ensemble posterior error distribution 
of the corresponding parameter with a vertical dotted line indicating 
the location of the input value. Each off-diagonal subplot is the joint 
distribution of ensemble error posteriors for the corresponding parameter pair.
The seven contour levels are the 10, 30, 50, 68.3, 80, 95
and 99\% confidence levels, and the green solid line marks the 68.3\%
confidence level, corresponding to a ``one-sigma'' normal
confidence. The black crosses indicate the locations of the input
values.

Here, we use galaxies with $6 < r_e < 20$ pixels, larger than the PSF
that has a HWHM of 1.48 pixels to reduce the effects of resolution
(the posterior distributions of small galaxies are described in
\S\ref{resection}).  For this
sample, $q$ and PA are not covariate with the other parameters. However,
the values of the magnitude, $r_e$, $n$, and the sky background are obviously
covariate, and the covariance becomes stronger with increasing \sn.  The origin
of this covariance is straightforward to understand.  For a given
surface brightness profile, a magnitude underestimate (i.e. a luminosity
overestimate) results in an overestimate of $r_e$ to better
match the observed brightness distribution.  Since the \sersic\ model
parameters $r_e$ and $n$ have a positive correlation
\cite[e.g.][]{trujillo2001}, $n$ is also overestimated. Similarly, the sky
background is underestimated to help compensate for the underestimated
magnitude. This argument also holds exactly in the opposite direction
for an overestimated magnitude.  The shape of the joint posterior
distributions in Figures \ref{corr_sn5}--\ref{corr_sn100} show that the
strength of the parameter covariance depends on galaxy \sn\ (and
\sersic\ index $n$ as will show in \S\ref{sersicnsection}).

In concert with our intuition, the confidence regions for $q$ and PA
shrink with increasing galaxy \sn.  For example, the asymmetric
heavy-tailed residual posterior distribution in $\Delta PA$
clearly seen in Figure \ref{corr_sn5} becomes symmetric as \sn\ increases
(Figs. \ref{corr_sn10}--\ref{corr_sn100}). This tail has its origin in
the ambiguity of $PA$ for $q\approx1$.

The covariance of magnitude, $r_e$, $n$ and sky background also
changes with galaxy \sn.  For low \sn\ (\sn$=5$,
Figure \ref{corr_sn5}), pairs of these parameters exhibit a clear covariance;
the sky background is strongly covariant only with the magnitude.
Also notice that the marginalised distributions of the errors in the
magnitude and $n$, and of fractional errors in $r_e$ are not normal as is
conventionally assumed and that the 68.3\% confidence region is not elliptical.
This non-normal behaviour results from the lack of information at low
\sn\ to constrain one or more of the covariate parameters.

As the \sn\ increases, the covariance of the magnitude, $r_e$, $n$ and
the sky background increases while the confidence regions decrease
(see Figures \ref{corr_sn10}--\ref{corr_sn100}) and the asymmetry of
the marginalised posterior distributions vanishes.  The posterior
distribution is dominated by the likelihood function, which is sharply
peaked and approaching its asymptotic form.  In addition, the strength
of the correlation depends on $n$ owing to the strong correlation of
$n$ with the sky background.  This can be seen in the joint posteriors
of $n$ and the sky background in Figures
\ref{corr_sn20}--\ref{corr_sn100} where the confidence level contours
appear to be mixture of different covariance ellipses from groups of
galaxies with different $n$, as will be shown in the next section.

\subsubsection{Effects of \sersic\ index $n$}
\label{sersicnsection}

\begin{figure}
\center
\epsfig{figure=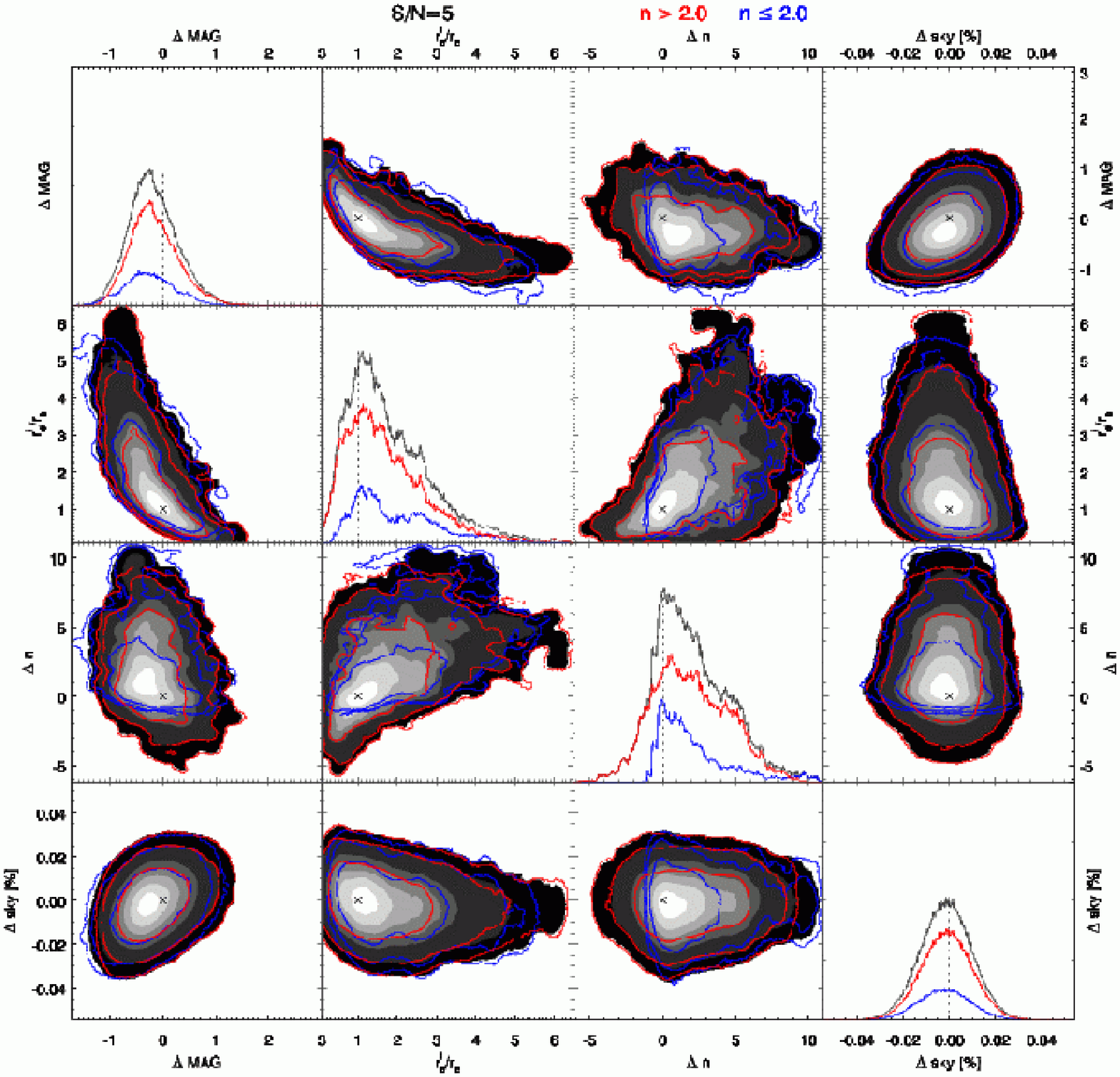,scale=0.6}
\vspace{5pt}
\caption{Posterior error distribution for magnitude, $r_e$, $n$ and sky
  background with $\sn=5$ from Figure \ref{corr_sn5} separated by
  $n>2.0$ (red) and $n\le2.0$ (blue).  Confidence levels are 68.3,
  95.4 and 99.7\% and the input value is marked by a $\times$
  and a vertical dotted line for joint and marginal distribution respectively.
  Grey contours and curves represent the total sample.  The parameter
  covariance is strong when $n$ is large.}
\label{corr_sersicn_sn5}
\end{figure}

\begin{figure}
\center
\epsfig{figure=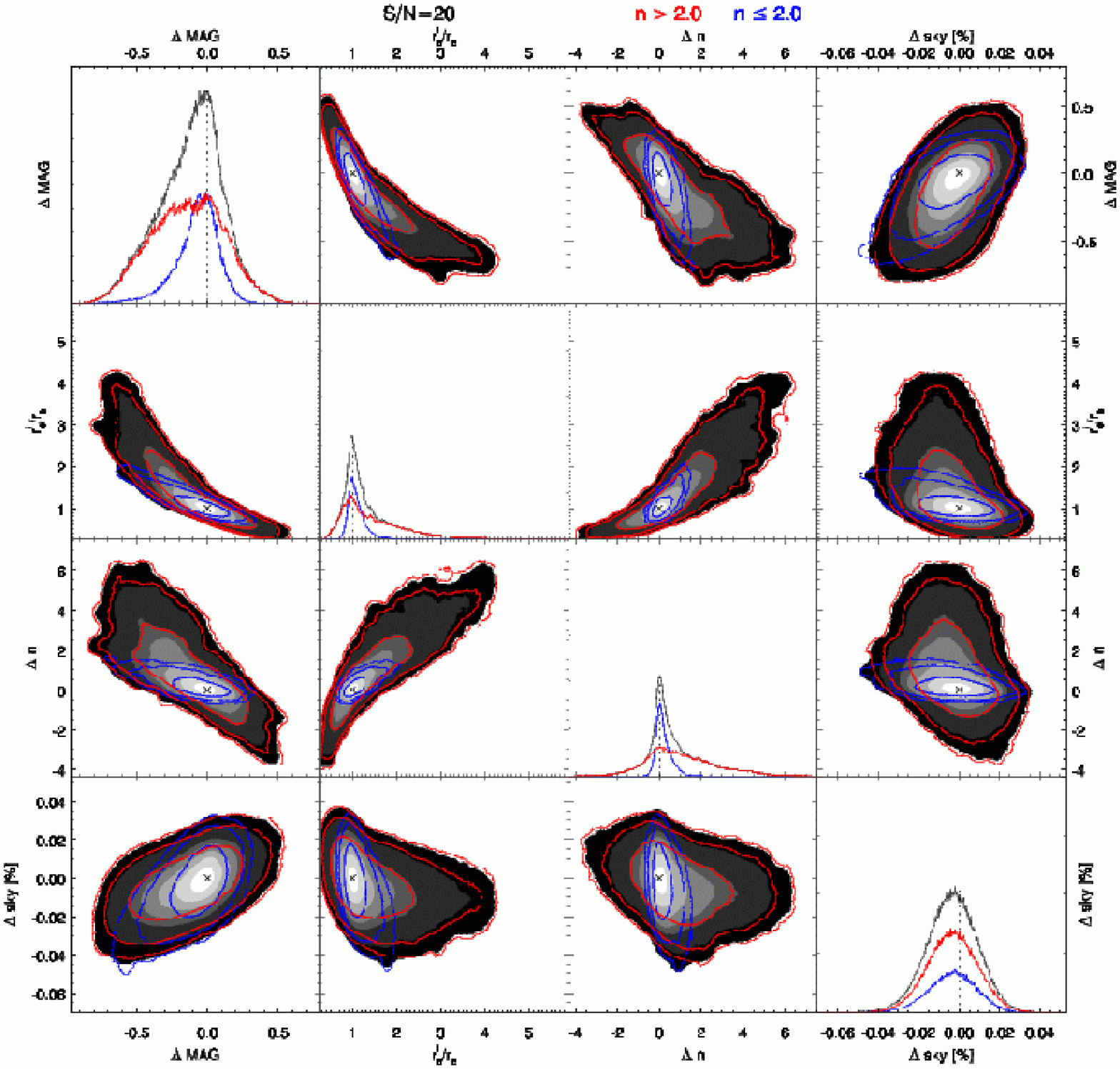,scale=0.6}
\vspace{5pt}
\caption{Posterior error distribution for selected parameters with \sn$=20$
  from Fig. \ref{corr_sn20}. See caption for
  Fig. \ref{corr_sersicn_sn5}.  }
\label{corr_sersicn_sn20}
\end{figure}

\begin{figure}
\center
\epsfig{figure=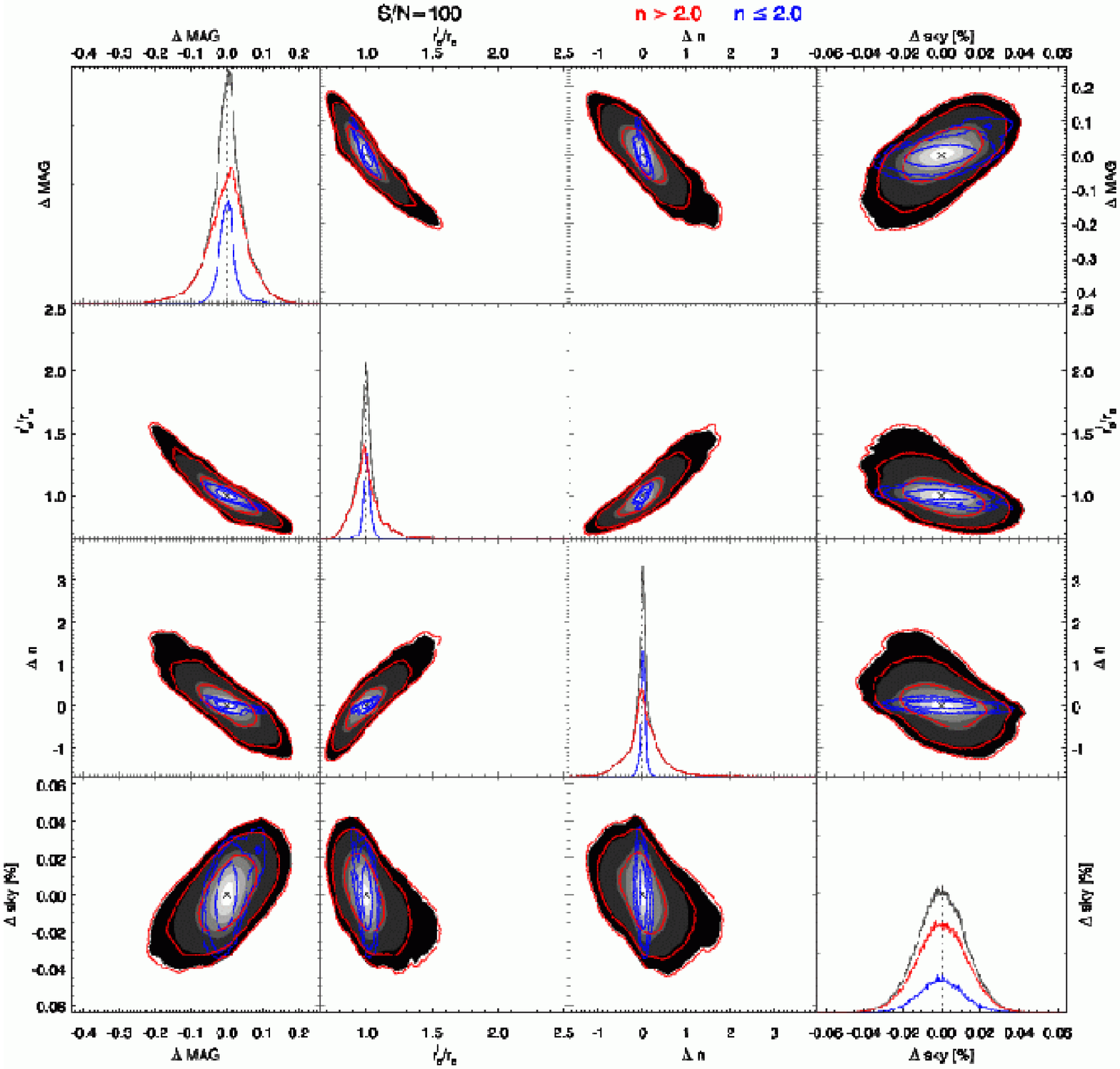,scale=0.6}
\vspace{5pt}
\caption{Posterior error distribution for selected parameters with \sn$=100$
  from Fig. \ref{corr_sn100}. See caption for
  Fig. \ref{corr_sersicn_sn5}.  }
\label{corr_sersicn_sn100}
\end{figure}

To investigate trends with the \sersic\ index $n$, we selected four model
parameters to study in depth, magnitude, $r_e$, $n$ and sky
background, for $\sn=5, 20, 100$.  We divide the samples in each \sn\
bin into two groups: $n>2.0$ and $n\le 2.0$.  We show the marginalised
posterior error distributions for each group in Figures
\ref{corr_sersicn_sn5}--\ref{corr_sersicn_sn100}.  The contours and
curves for the samples with $n\le 2.0$ and $n > 2.0$ are shown by the
blue and red colours, respectively.  The contour levels for each group
corresponds to the 68.3, 95.4, and 99.7\% confidence levels.  The
black crosses are the locations of the input values and the grey
contours and grey curves show the total sample as in Figures
\ref{corr_sn5}, \ref{corr_sn20}, \ref{corr_sn100}.

For \sn$=5$ (Fig. \ref{corr_sersicn_sn5}), the marginalised error posterior
for $n\le 2.0$ has a sharp truncation on the left hand side, owing to the
prior distribution boundaries of 0.5 and 11.99 on $n$.  Otherwise,
the posterior error distributions of these two groups are similar.  As
expected for low \sn, the errors are dominated by random
statistical errors and parameter covariance is not significant.  However, for
higher \sn\ (see Figs. \ref{corr_sersicn_sn20} and
\ref{corr_sersicn_sn100}), the differences between low and high values
of $n$ are significant.  When the \sn$\ge 10$, the confidence regions for
$n>2.0$ galaxies (red) are larger than for those with $n\le 2.0$ (blue).  In
part this is a consequence of the degeneracy between the sky
background and the extended profile for larger values of $n$, which makes the
morphological parameters for small $n$ galaxies better constrained.
One can see the larger covariance for $n>2.0$ in
the joint posterior error distributions shown in
Figures \ref{corr_sersicn_sn20} and \ref{corr_sersicn_sn100}.

Moreover, covariance with the sky background significantly affects the
inference of the magnitude and $r_e$.  For example, when $\sn=100$, a
$\pm0.04$\% variation in the sky background estimate may lead to an uncertainty
in the magnitude of up to $\pm0.2$ for elliptical galaxies, i.e. those
with $n>2.0$.  Conversely, ignoring the sky background uncertainty can
induce a significant bias in the other parameters that would be
difficult to assess only using the best-fit parameters from a simple
$\chi^2$ minimisation.  Our approach incorporates the parameter
covariance and random statistical uncertainty in any subsequent
inference.  Finally, we note that the parameter distributions of both
large and small $n$ galaxies is weakly biased at worst. This
demonstrates that the \galphat-inferred posterior maximum reliably
recovers the true input parameters.  We expect that careful attention
to prior distributions could reduce the bias for lower values of
\sn\ as well.

\subsubsection{Effects of galaxy $r_e$}
\label{resection}

The intrinsic shapes of small galaxies are obscured by the PSF.
Although any pixelisation method induces some bias for any sized
galaxy, if $r_e$ is comparable to or smaller than the PSF width, the
axis ratio $q$ will be be unrecoverable since the central pixels will
contain most of the flux.  Moreover, numerical techniques without
explicit error control when computing $I(x,y)$ (see eq. \ref{eq:Ijk})
may fail to produce an accurate theoretical prediction for the model
flux in the central pixels.  \galphat\ naturally addresses both
problems.  Firstly, the Bayesian inference produces an estimate
conditioned on the true value of the PSF and, therefore, will produce a
posterior distribution consistent with all the possible models that yield
the observed profile \emph{after} convolution with the PSF.  Secondly,
the \galphat\ model images are generated by interpolating a
numerically integrated table that is accurate over the entire area
of the galaxy with a numerical error tolerance set by the user.  In
particular, both the inner and outer profiles are well-resolved
because the tabulated grids are independent of scale.  In
other words, it is not possible for \galphat\ to produce a poor
estimate of the central pixel values for any value of $r_e$.

\begin{figure}
\centering
\epsfig{figure=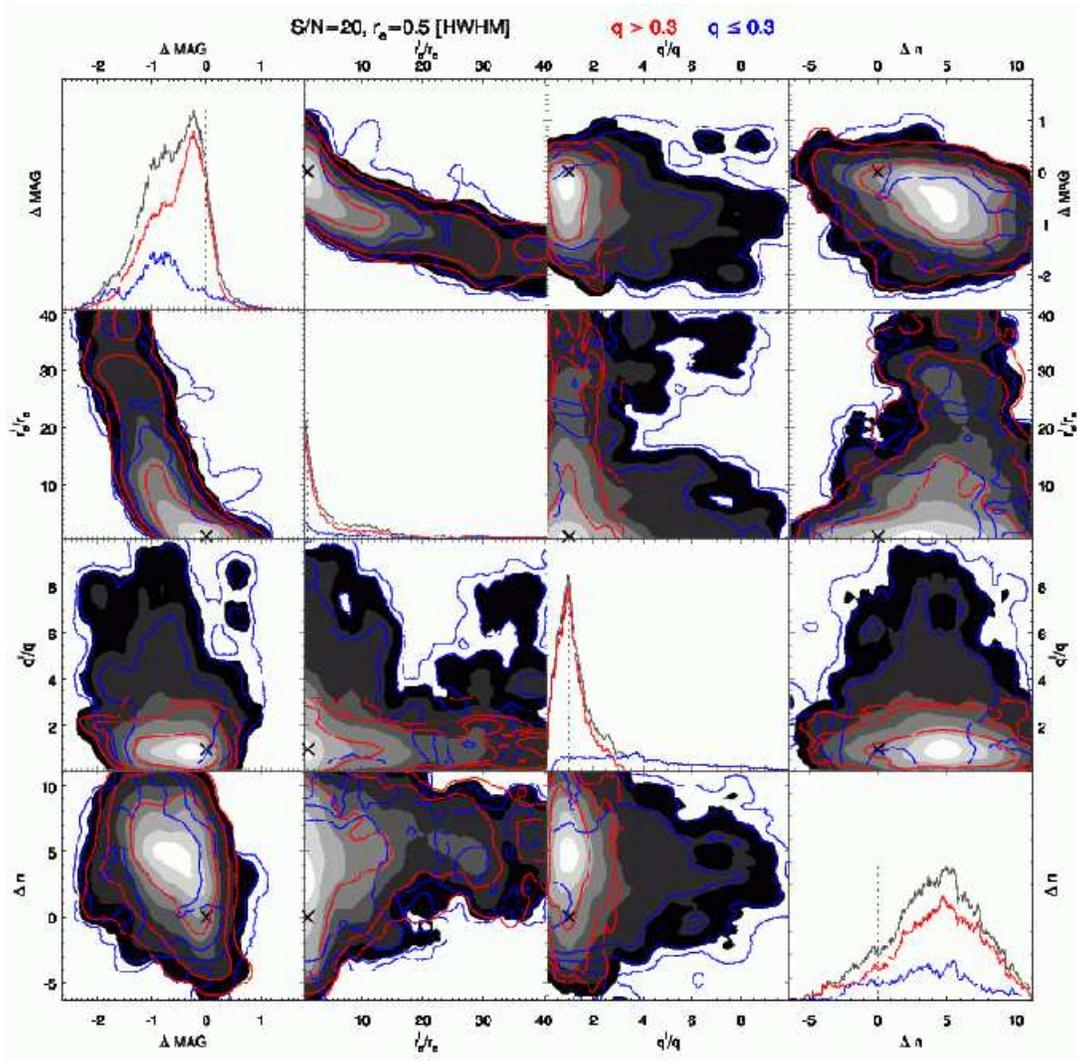,scale=0.6}
\vspace{5pt}
\caption{Posterior error distributions for magnitude, scaled $r_e$,
  scaled $q$ and residual $n$ with $\sn=20$ and 
  $r_e=0.5\ \mbox{PSF HWHM}$,
  separating samples with $q>0.3$ (red) and $q\le0.3$ (blue). Input
  values are shown by a $\times$.  Contours are 68.3, 95.4 and 99.7\%
  confidence levels.  The grey scale contours and curves represent
  the total sample.  When a galaxy is close to edge-one (i.e. $q<0.3$)
  and smaller than the PSF, the marginalised posterior of $q$ is
  almost flat and the posterior maximum of magnitude is
  negatively biased (i.e. the luminosity is overestimated). }
\label{corr_q_sn20_re1}
\end{figure}

\begin{figure}
\centering
\epsfig{figure=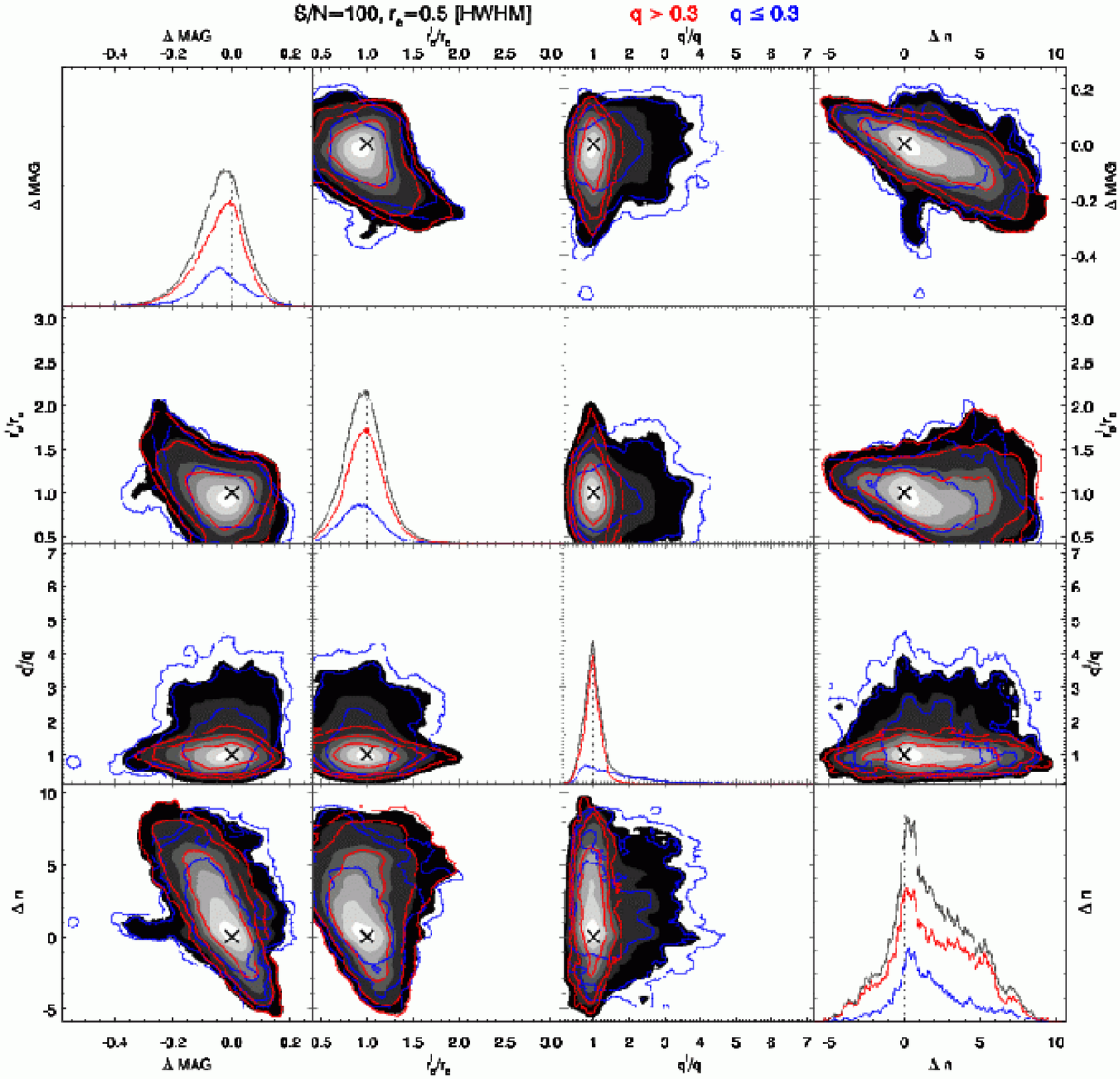,scale=0.6}
\vspace{5pt}
\caption{Posterior error distributions for selected parameters with
  $r_e=0.5\ \mbox{PSF HWHM}$ and \sn=$100$.  See caption for
  Fig. \ref{corr_q_sn20_re1}.  }
\label{corr_q_sn100_re1}
\end{figure}

\begin{figure}
\centering
\epsfig{figure=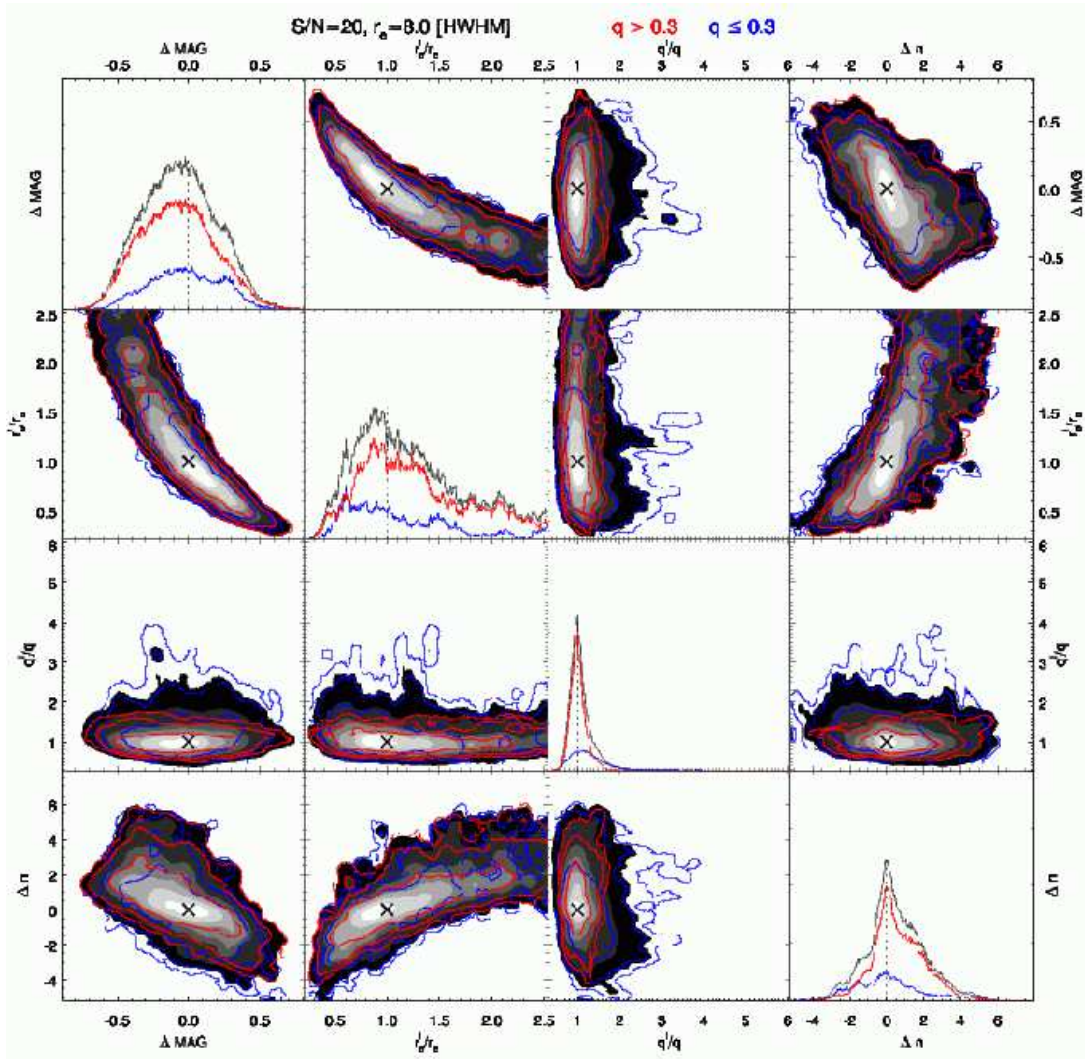,scale=0.6}
\vspace{5pt}
\caption{Posterior residual distributions for selected parameters with
  $r_e=8.0\ \mbox{PSF HWHM}$ and \sn=$20$.  See caption for
  Fig. \ref{corr_q_sn20_re1}.  }
\label{corr_q_sn20_re5}
\end{figure}

\begin{figure}
\centering
\epsfig{figure=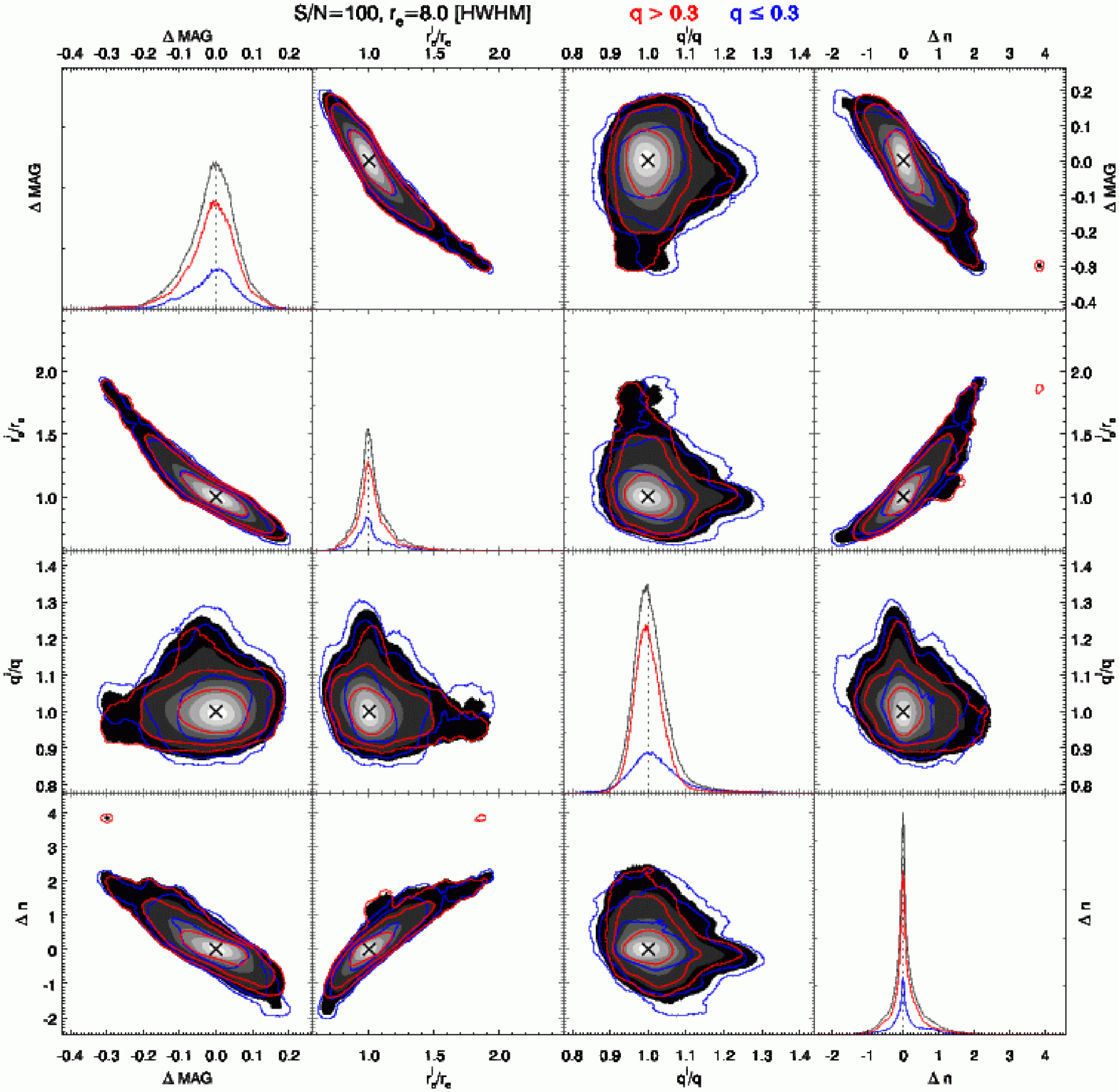,scale=0.55}
\vspace{5pt}
\caption{Posterior error distributions for selected parameters with
  $r_e=8.0\ \mbox{ PSF HWHM}$ and \sn=$100$.  See caption for
  Fig. \ref{corr_q_sn20_re1}.  }
\label{corr_q_sn100_re5}
\end{figure}

Even when accurately determined, the true value of $q$ affects the
other structural parameters because a small galaxy with a small $q$
takes up fewer image pixels.  Similar to \S\ref{sersicnsection}, we
investigate these effects in Figures
\ref{corr_q_sn20_re1}--\ref{corr_q_sn100_re5} by dividing the galaxy
sample into two groups: $q > 0.3$ (red) and $q\le 0.3$ (blue), and by
examining the posterior error distributions of magnitude, $n$, $r_e$,
and $q$.  We also select two bins in size: $r_e=0.5$ times the HWHM of
the PSF and $r_e= 8.0$ times the HWHM of the PSF, for two different
values of \sn: 20 and 100.

Figures \ref{corr_q_sn20_re1} and \ref{corr_q_sn100_re1} show the
posterior error distributions for galaxies with
$r_e=0.5$ times the PSF HWHM for a
\sn\ of 20 and 100, respectively.  The three contours on the
background grey contours, which indicate the total sample, denote the 68.3,
95.4 and 99.7\% confidence levels for galaxies with $q>0.3$ (red) and $q\le0.3$
(blue).  Since the PSF dilutes any evidence of
the intrinsic shape inside the PSF FWHM, the posterior error
distributions of $r_e$ and $q$ have significant probability density at
larger values than the input values.  Moreover, the magnitude is
covariant with $r_e$ and $q$.  As a consequence, the posterior
distribution of magnitude has significant probability at values
smaller than the input value.  Similarly, the bias in the \sersic\
index $n$ is exacerbated by its covariance with $r_e$.

In Figure \ref{corr_q_sn20_re1}, the posterior error distribution for
$r_e$ has a tail to large $r_e$ for both $q > 0.3$ and $q\le
0.3$.  As a result, the maximum of the posterior distribution
in $n$ is also similarly biased to large $n$ for both groups.
The bias in the two groups for $q$, however, is
dramatically different: for $q\le 0.3$ (blue) the distribution is
approximately uniform over a large range, while for $q > 0.3$ (red) the
distribution has a mode centred near 1.  This is an artifact of the
PSF convolution; the PSF naturally makes the galaxy appear rounder.
Therefore, the bias in $q$ for rounder galaxies is modest but
the bias for flat, intrinsically edge-on galaxies is large.

The residual magnitude distribution for the entire sample (grey) is
slightly negatively biased owing to the excess posterior probability
of $r_e$ at larger parameter values than the true value.  The bias
differences in $q$ for the two groups leads to a bias difference in
magnitude: the bias for $q\le 0.3$ (blue) is larger than the bias for
$q > 0.3$ (red).  As described above, the posterior distribution 
for $q$ are biased
upwards as $q$ decreases.  On the other hand, because $q$ is poorly
constrained, the luminosity can be adjusted in a variety of ways to
achieve the same surface brightness for a given $r_e$ by making both
$q$ and the luminosity either large or small (see
eq. \ref{enclosedL}). This results in both a large spread in magnitude
as well as a magnitude underestimate owing to a luminosity that is
overestimated to compensate for the upward bias in the posterior 
distribution of $q$.

Figure \ref{corr_q_sn100_re1} shows the posterior error distributions
for $\sn=100$.  As anticipated, the posterior distributions are better
confined in parameter space and the biases of the posterior maxima are
significantly reduced, e.g. the error posterior for $n$ has a mode at
0.  The posterior maxima for $q > 0.3$ galaxies (red) show no strong
biases.  The posterior error distribution for $q$ when $q \le 0.3$
(blue) reveals an extended tail and, in contrast to Figure
\ref{corr_q_sn20_re1} for low \sn, it has a mode below 1.

As $r_e$ increases beyond the PSF size, we expect these biases to
decrease.  Figures \ref{corr_q_sn20_re5} and \ref{corr_q_sn100_re5},
which show the results when $r_e=8.0\ \mbox{HWHM}$ of the PSF for a
\sn\ of 20 and 100, respectively, confirms this expectation.  The bias 
of the posterior maximum is modest and the differences between the
$q > 0.3$ and $q\le 0.3$ groups disappear.  Also, for fixed $r_e$, the
confidence regions decrease with increasing \sn, as expected.

\subsubsection{Effects of image size}
\label{skysection}

We have seen that \sersic\ model parameters, such as the magnitude, $n$, and
$r_e$ are covariant with the sky background.  Therefore, an inaccurate
sky background determination may bias the inferred galaxy structural
parameters.  To circumvent this, researchers have measured the sky
background independently from their model fits.  If the sky background
measurement is not more precise than the model fit itself, this
procedure has two disadvantages: 1) because of the covariance, the
subsequent parameter posterior distribution will be biased; 2) characterisation of
the covariance will no longer be part of the posterior distribution.

The Bayesian MCMC approach implemented by \galphat\ enables us to take
all the parameter covariances into account in our galaxy modelling.
In particular, we may characterise the influence of parameters such as
the sky background.  In this section, we explore the influence of the
blank-sky fraction in the image.  Assume that one must analyse a
particular source on a large image.  How much of the frame should one
keep for one's parameter inference?  Retaining a large fraction of
blank sky area is better to accurately determine the sky background.
However, a larger fraction of blank sky implies a larger image size
and more computation time.  Of course, truly blank sky is rare in
Nature.  Nonetheless, an understanding of the trade-off between an
accurate sky background determination and a fast model image
generation is necessary to design an efficient analysis.  We test the
dependence on image size by generating 10 simulated galaxies with
$\sn=50$, $r_e=10$ pixels, $n=4$, $q=1$, and a 500 [ADU] Poisson sky
background and model them using different size image regions specified
by 4, 8, 12, 16 and 20 times $r_e$. We add 30,000 converged MCMC
states from each galaxy to obtain the posterior error distribution.

\begin{figure}
\centering
\epsfig{figure=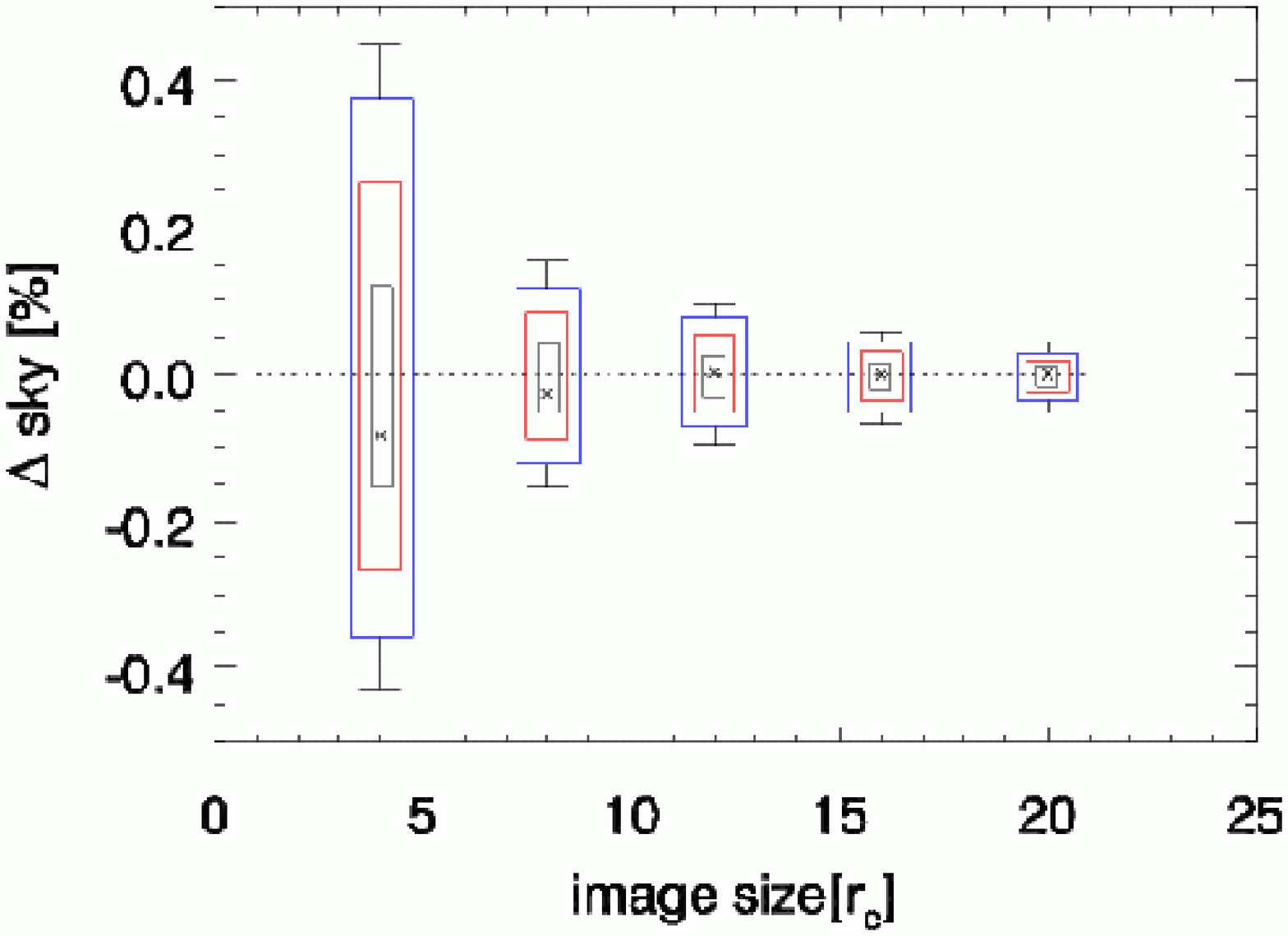,scale=0.6}
\vspace{5pt}
\caption{Posterior error distributions of the sky background for images 
  with different sizes. Ten galaxies with the same \sersic~model parameters but 
  different random noise, are modelled with different size image
  regions, from 4 to 20 times the input $r_e$. 
  The galaxies have \sn$=50$, $n=4$, $r_e=10$, $q=1$ and a Poisson distributed 
  sky background with 500 ADU.  Within each bin, an ensemble of parameter error
  posterior from the 10 galaxies is shown with posterior maximum (cross symbol),
  68.3\% (grey box), 95.4\% (red box) and 99.7\% (blue box) confidence levels. 
  Minimum and maximum data values are indicated by the error bars.  
  Although the sky background posterior is nearly symmetric regardless of the
  blank sky fraction in the image, the posterior maximum becomes slightly
  biased downwards with increasing confidence intervals as the blank sky
  fraction decreases. 
}
\label{galimsize}
\end{figure}

\begin{figure}
\centering
\epsfig{figure=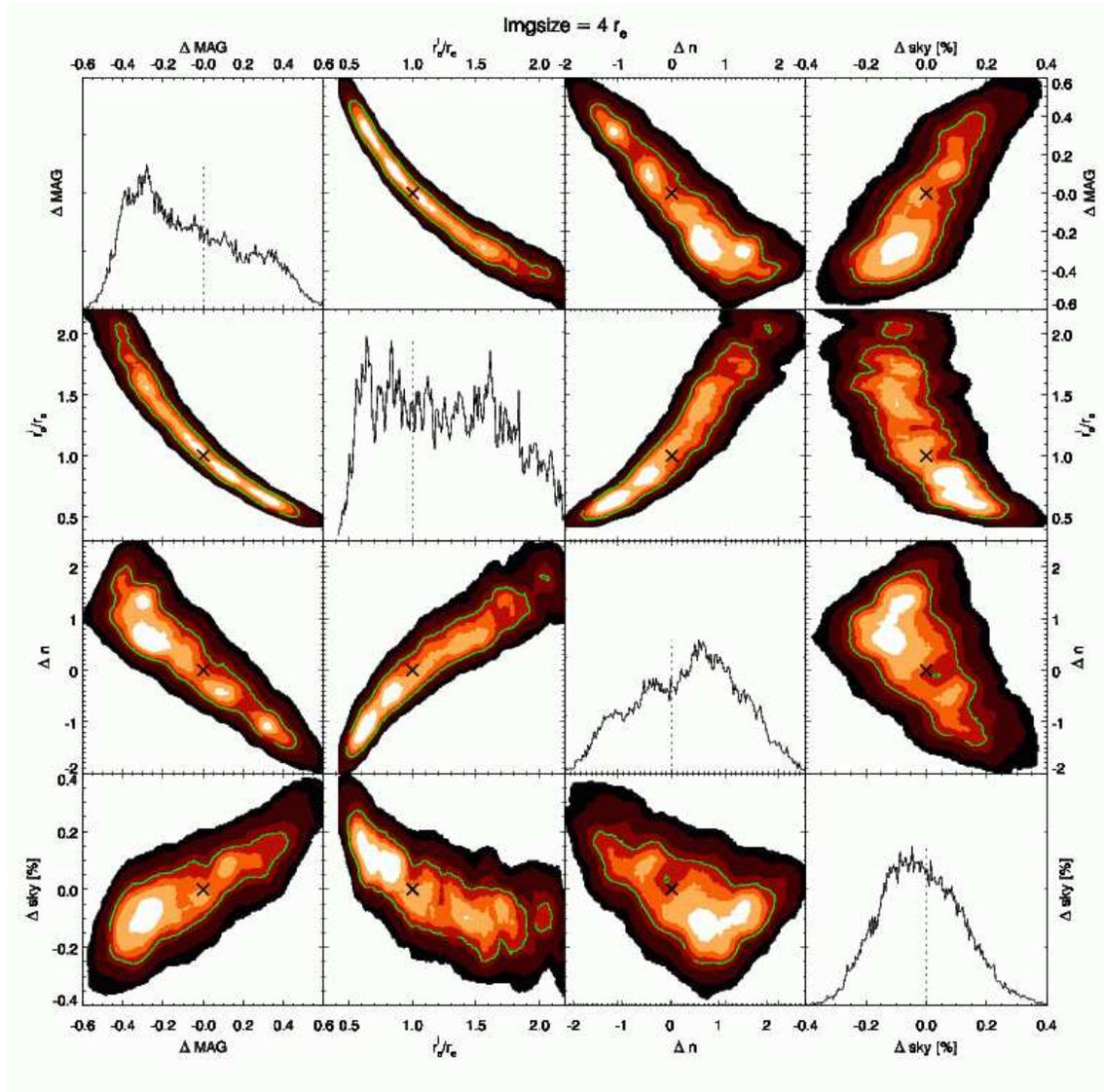,scale=0.55}
\vspace{5pt}
\caption{Error posteriors of magnitude, $r_e$, $n$ and sky 
  background when the image region is $4r_e$.
  The green line highlights the 68.3\% confidence level.
  Note that a $0.1\%$ bias in the sky background posterior maximum leads to
  biases in the other parameters, e.g. -0.3 in magnitude, and the
  inference for $r_e$ is very weak.
}
\label{galimsize_re1_corr}
\end{figure}

\begin{figure}
\centering
\epsfig{figure=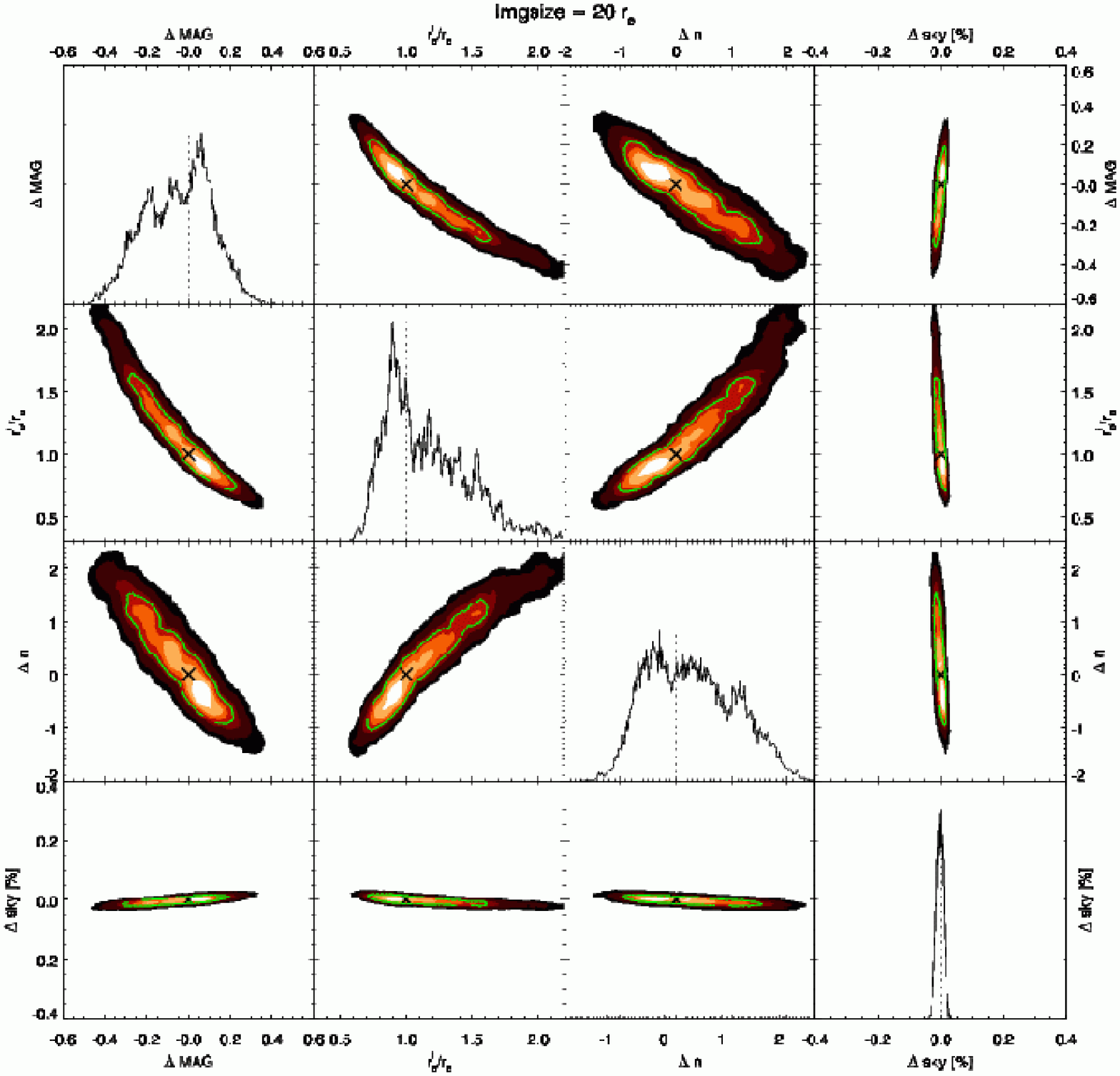,scale=0.6}
\vspace{5pt}
\caption{Error posteriors of magnitude, $r_e$, $n$ and sky    
  background when the image size is $20r_e$. 
  See caption for Fig. \ref{galimsize_re1_corr}.
  Note that the parameter inference becomes more reliable with 
  increasing image size.  
}
\label{galimsize_re5_corr}
\end{figure}

Figure \ref{galimsize} shows the marginalised error posterior
distributions for the sky background of the \sersic\ model as a
function of image size. The cross indicates the posterior maximum and
the 68.3, 95.4 and 99.7\% confidence levels are indicated by the grey,
red and, blue boxes, respectively.  The minimum and maximum data
values are indicated by the error bars.  Although the \galphat\
inference of the sky background has a symmetric posterior error
distribution around zero, regardless of the blank sky fraction in the
image, the posterior maximum is below the true sky value as the blank
sky fraction decreases.  Since the galaxy surface brightness model,
i.e. a \sersic\ profile, is degenerate with the sky background in
the outer regions, a small image containing a relatively large galaxy
may introduce a bias in the posterior distribution of the sky background 
and hence also affects the inference of the other parameters.  
In other words, owing to the \sersic\ model surface brightness and 
sky background degeneracy, either increasing the galaxy luminosity and 
decreasing the sky background or decreasing the galaxy luminosity and 
increasing the sky background can match the background level of the image. 
However, this flexibility in a \sersic\ profile to produce the sky background 
allows a galaxy to include part of the sky background flux and thus the sky
background posterior tends to be biased downwards as we show in 
Figure \ref{galimsize_re1_corr}.
This sky background bias that stems from not
enough information about the blank sky in the image becomes more
significant with increasing $n$.  The \galphat\ inference of the sky
background for this particular case of $n=4$ is not biased unless the
image size is smaller than $12r_e$ and the bias in the sky background 
posterior maximum is very small, only
$0.1\%$, even when the image only extends to $2r_e$.

However, it is remarkable that this tiny sky background bias leads to
significant biases in the inferences of the other parameters, i.e. the
magnitude, $r_e$, and $n$, as shown in Figure \ref{galimsize_re1_corr}
where we plot the error posteriors of the magnitude, $r_e$, $n$, and
the sky background for a galaxy with $n=4$ and an image size of
$4r_e$.  In addition to the strong parameter covariances already
illustrated in \S\ref{snsection} and \S\ref{sersicnsection}, notice
that the posterior maximum of the magnitude is biased low by 0.3 owing
to the very small ($0.1\%$) bias in the sky background posterior maximum 
and that the posterior distribution of $r_e$ is very broad.  If the image size
increases, these biases and weak inferences become less significant as
shown in Figure \ref{galimsize_re5_corr}, which shows the case when
the image size is $20r_e$.

Even this simple test using a small number of galaxies reconfirms the
importance of an accurate sky background.  In addition, it illustrates
that an accurate characterisation of measurement uncertainties is
essential to limiting bias.  Therefore, rather than fixing or
subtracting a sky background determined by a independent measurement,
one must model the galaxy image including the sky background and
characterise the full posterior distribution of the model parameters.
Although \galphat\ can handle the interactions of random uncertainties
and parameter covariance over a wide range of images sizes, the bias
is reduced with an image size of at least $10r_e$ for a galaxy with
$n=4$.

\subsubsection{Effects of assumed PSF errors}
Errors in the assumed PSF will lead to errors in the
\galphat\ inferred parameters.  We characterise PSF errors
by the difference in PSF FWHMs:
\[
\Delta \mbox{FWHM} =
  \frac{\mbox{FWHM}_{assumed} -
    \mbox{FWHM}_{true}}{\mbox{FWHM}_{true}}.
\]
We investigate \galphat-sampled posterior error distributions for
\sersic\ models using a PSF with $\Delta\mbox{FWHM} = -15\%, -5\%,
5\%, 15\%$.  From the sample of simulated galaxy images in
\S\ref{datasectionre}, which were convolved with a 2MASS-motivated
2.96 FWHM pixel PSF, we select galaxies with a $\sn=100$ and in four
$r_e$ bins of size 1, 2, 4 and 8 times the PSF HWHM of 1.48 pixels.

\begin{figure}
\centering
\epsfig{figure=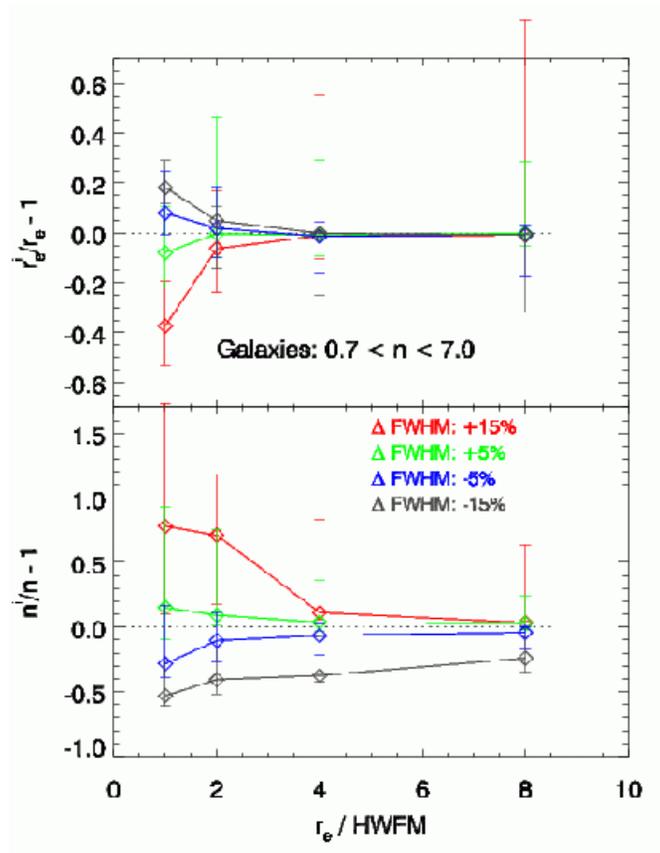,scale=0.6}
\vspace{5pt}
\caption{The posterior maximum errors of $r_e$ and $n$ for
  errors in the assumed PSF widths
  with $\Delta\mbox{FWHM} = 15\%, 5\%, -5\%, -15\%$
  as labelled.  The error bars show the $p=0.5$ confidence interval.
  The biases in the posterior maximum become smaller with increasing galaxy
  size.}
\label{psfvariation}
\end{figure}

We show the results in Figure \ref{psfvariation} where we plot the
posterior maximum of relative error in $r_e$ and $n$ with the error
bar indicating the $50\%$ confidence interval for the four different
$\Delta\mbox{FWHM}$s. If the PSF width were overestimated, we would
expect that the inferred galaxy size would be smaller than its true
size and that the inferred profile would be more concentrated, i.e. a
larger $n$ than the intrinsic value, and vice versa.  Figure
\ref{psfvariation} confirms these expectations. Also as expected, the
bias of the posterior maximum is less if the galaxy size becomes larger than
the PSF FWHM. The bias of the $r_e$ posterior maximum is larger when the PSF
is overestimated than when it is
underestimated as previously observed by \citet{macarthur2003}.
Moreover, notice that there is a systematic offset of the ensemble
posterior maximum of $n$ even if the galaxy's $r_e$ is large,
indicating that the bias of the $n$ posterior maximum owing to errors in the 
assumed PSF FWHM also depends on $n$ itself.

\begin{figure}
\centering
\epsfig{figure=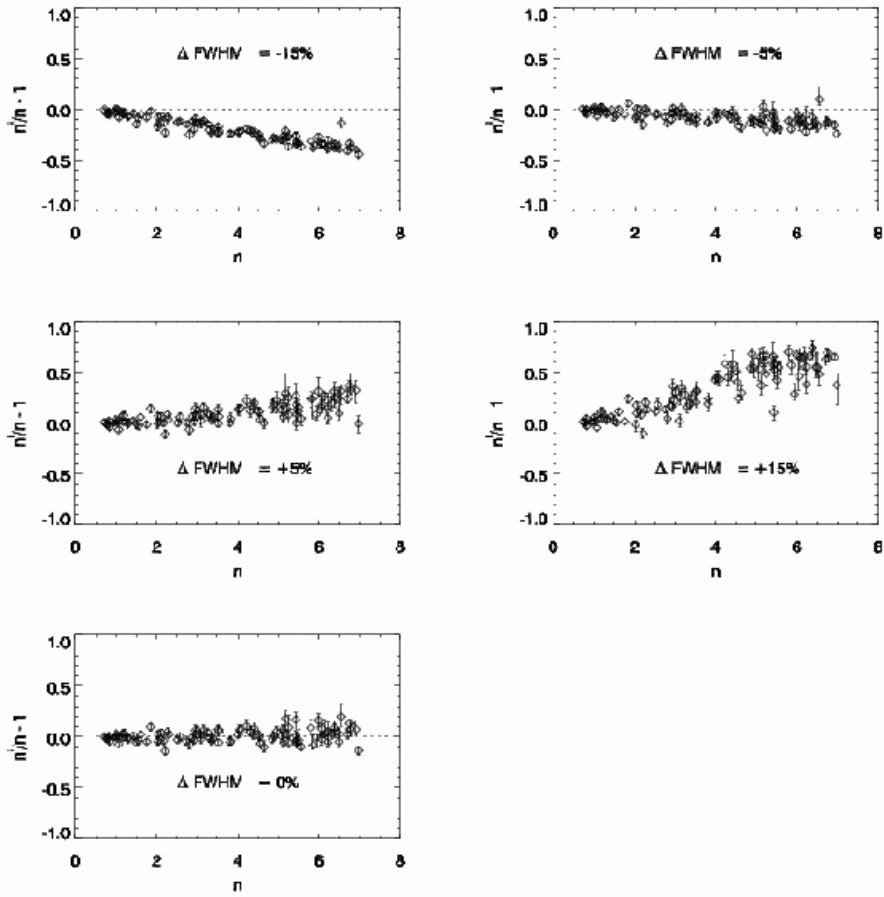,scale=0.6}
\vspace{5pt}
\caption{Systematic bias in the inferred galaxy \sersic\ index $n$ owing to 
  assumed PSF
  width errors based on the posterior error distributions for 100
  galaxies with $\sn=100$ and $r_e=8\ \mbox{HWHM}$ of the PSF using 4 different
  assumed PSF FWHM errors with $\delta\mbox{HWHM} = -15\%, -5\%, +5\%, +15\%$.
  Each diamond with error bar is the median and 50\% confidence range for
  the relative difference between the inferred value and the input
  value. The bottom left panel shows results for the correct PSF.  }
\label{galpsf}
\end{figure}

We characterise the $n$ dependent bias by investigating the posterior
error distribution of $n$ for the galaxies with $r_e=8\ \mbox{HWHM}$ 
of the true PSF.  Figure \ref{galpsf} plots the relative errors 
in the median values of the \galphat\ posterior samples ($n^i$) of $n$ 
as $\Delta n = \frac{n^i-n}{n}$ for each galaxy in each of the PSF error samples:
$\Delta\mbox{FWHM} = -15\%, -5\%, 0\% +5\%, 15\%$.  The median values
of $n^i$ are biased and this bias becomes large as $n$ increases.  When the
assumed PSF FWHM is overestimated, $n$ is overestimated with increasing $n$
since the larger PSF artificially extends the profile and conversely
when the assumed PSF FWHM is underestimated, $n$ is underestimated
with increasing $n$ since the smaller PSF artificially contracts the
profile. The biases and uncertainties in $n^i$ for PSF FWHM
overestimation are larger than for PSF FWHM underestimation.  For
example, the inferred value $n^i$ for an elliptical galaxy with $n=4$
will be smaller than the true value by 20\% if the
assumed PSF FWHM is smaller than the correct PSF by 15\%, but it will
be larger by 30\% with considerable scatter if the assumed PSF FWHM is
larger than the correct PSF by 15\%.  If the correct PSF is used then
the inference of $n$ is unbiased but the scatter still increases with
$n$.

\subsection{Run time}
\label{runtimesection}

\begin{table}
\centering
\caption{\galphat\  wall clock time}
\begin{tabular*}{0.99\textwidth}{@{}llllll@{}} \hline\hline
Image/model & samples & CPU & Processors & wall clock time & MCMC algorithm \\ \hline
$226\times226$     & 40,000  & Quad-core AMD Opteron & 8  & $1.5$ hr & hierarchical tempering parallel chain \\
\sersic &         & 2613 MHz              &    &               & \\ \hline
$100\times100$     & 40,000  & Quad-core AMD Opteron & 8 & $0.4$ hr & hierarchical tempering parallel chain \\
\sersic     &         & 2613 MHz              &    &               & \\ \hline
\end{tabular*}
\label{runtime}
\end{table}

The \galphat\ run time depends on the MCMC algorithm, the complexity
of model, the size of model image, the desired number of converged
samples, and the number of model components.  Although it is difficult
to characterise the \galphat\ run time for each dependency, we find
that the dependence on the number of converged MCMC states and on the
size of the image, i.e. number of pixels, are approximately
linear. However, the dependence on the MCMC algorithm (i.e. the
temperature ladder and the number of chains, which depends on the
parameter dimensionality) and the model complexity (i.e. the number of
dimensions, the number of model components, and the parameter
covariance in the chosen model family) are nonlinear.  Therefore, it
requires some experimentation to find the best strategy for each
application. By cross-checking with different MCMC algorithms
e.g. simulated tempering \citep{neal1996} or differential evolution
\citep{braak2006} with tempering, for a subset of galaxy samples, we
were able to tune the parameters for the PHS algorithm used in our
study.

Table \ref{runtime} shows an example of \galphat\ run time for two
galaxy images: one is typical of a 2MASS galaxy image and the other is a
relatively large image, which we choose intentionally to demonstrate
\galphat 's feasibility in a marginal case as could occur for a 2MASS
galaxy image, since we plan to analyse 2MASS galaxies in the future.
Table \ref{runtime} lists the total wall clock times of two example
simulations for 40,000 converged states in $226\times226$ and
$100\times100$ pixel images for the \sersic\ model. The number of
chains used for \sersic\ modelling is 8.  The BIE assigns one
processor per chain.  The run time is the total time for
obtaining the 40,000 converged samples using the PHS MCMC algorithm.
The necessary number of converged samples depends on the application.
Tests are carried out using the University of Massachusetts
Astronomy department HPC cluster.

For a \sersic\ model, 40,000 converged samples for one galaxy with a
typical image size of $100\times 100$ can be obtained within 25
minutes of wall clock time using 8 processors. This of course means it
would take 3.2 hours on a single quad core processor or 0.5 days on a
single core processor.  This may seem computationally impractical but
multi-core processors and multi-processor machines are becoming the
norm and of course processor speeds and machine sizes are increasing
all the time.  For example, even now, each of our cluster nodes had 16
processors.  Employing 32 nodes of our cluster for one day would yield
the posterior distributions of about 3000 galaxies.  This means
that 10,000 2MASS galaxies can be analysed using a single
\sersic\ model in
less than 4 days using 32 of our 16 core nodes, demonstrating the
current feasibility of this approach.

\section{Summary}
\label{conclsection}

We introduce \galphat, a Bayesian galaxy morphology analysis package,
designed to efficiently and reliably generate the probability
distribution of galaxy surface brightness model parameters from an
image.  We emphasise that the morphological analysis is a stepping
stone in a larger chain of inference on theories of galaxy formation
and evolution.  Therefore, we believe that it is productive to
consider the determination of galaxy morphology in the context of
hypothesis testing and model comparison, and this demands the full
posterior probability distributions for each galaxy in the ensemble.

In this section, we recap the history of morphological analyses,
summarise the technical improvements offered by \galphat, list major
findings from our performance tests, and briefly describe our future
work.

\subsection{Recap}

Our approach offers a number of significant advantages for estimating
surface brightness profile parameters.  First, the topology of the
likelihood function is almost certain to be multimodal in the
high-dimensional space.  Downhill optimisation techniques demand
precise prior information that is not generally available.  In
addition, one should have a way of assessing the significance of this
\emph{global} extremum with respect to nearby local and possibly
unanticipated modes.  Using the various tempering algorithms available
in the BIE, our tests have demonstrated that we can achieve a
steady-state distribution and the simulated posterior will include any
possible multiple modes supported by the prior distribution.  Given
the posterior distribution, we may then precisely estimate the
confidence levels.  Secondly, the model will often have correlated
parameters.  As illustrated in \S\ref{casestudy}, any hypothesis
testing that uses the ensemble of best-fit parameters will be
affected by these correlations. The full posterior distributions from
\galphat\ identify these correlations and incorporate them in
subsequent inferences.

We can use posterior simulations over ensembles of images to test the
significance of cluster and field environments on galaxies as
evidenced in their photometric parameters.  A more elaborate model
might include angular harmonics of the light distribution
(e.g. galactic bars and spiral arms); we could use Bayes factors to
assess the support in the data for various harmonics.  We could do
this for an entire set of galaxy images in multiple bands
simultaneously.  This is much more natural and likely to be more
powerful than the standard practice of cataloguing parameters and
attempting to look for correlations in scatter plots.

Furthermore, the adoption of specific functional families that
resemble galaxy profiles, e.g. \sersic, {\em may} not provide the best
discrimination in attempting to interpret the correlation between
derived parameters and hypotheses.  Our extensive study of \sersic\
profiles undertaken for this paper convinces us that the strong
inter-parameter covariances weaken the inference.  This suggests that
families of orthogonal functions conditioned to match the outer galaxy
profile might be a more productive choice \cite[e.g.][]{spergel2010}.
A complete set of orthogonal functions might be straightforwardly
transformed to match a fiducial profile, opening up a wide range of
possible applications for characterising galaxy properties.  With
carefully chosen prior distributions, we can use Bayes factors to test
the significance of multiple component models and models with
different functional forms.  This analysis can straightforwardly
answer questions such as: is the standard model (\sersic\ and
exponential disc profiles) preferred to all competing models (models
with cores, dark-matter profiles) and vice versa? Is there a
particular alternative model that is supported more significantly by
the data?  We will explore some of these possibilities in our
follow-up paper \citep{yoon2010b}.

\subsection{\galphat\ features}

\galphat\ is built on the Bayesian Inference Engine, \bie\
\citep{weinberg2010}, an object-oriented optimised parallel platform
for Bayesian computation.  The \bie\ implements a variety of
algorithms and tools that can be chosen at run time to match the
problem at hand.  For example, in the tests presented here, we have
found that the PHS algorithm \citep{rigat2008} provides the fastest
convergence while efficiently exploring the parameter space.  Other
algorithms may excel for different model families or for those with
additional components.  In addition, the \bie\ enables collaboration:
new algorithms developed for the \bie\ become available for the
community.

These methods require more frequent likelihood evaluations than
competing approaches, so optimisation is essential.  \galphat\
incorporates the following key features: 1) pre-computed images using a
scale-free relocatable interpolation scheme with strong error control;
subsequent image generation accurately represents the surface
brightness integrated over every pixel and is very fast; and 2) a Fourier
shift-theorem-based rotation algorithm that is both accurate and much
faster than the often-used interpolation methods.  Although we explore
\sersic\ models in this paper, \galphat\ can be applied to a wide
variety of parametric families, limited ultimately by the physical memory
available to store the lookup tables.  The likelihood calculation time
for a \sersic\ model is less than 0.1 sec for an image size of
$190\times190$ pixels.

\subsection{GALPHAT performance on \sersic\ profiles: parameter
  covariance and bias}

A summary of our major findings are as follows:
\begin{enumerate}
\item \emph{Galaxy \sn.}\ Computation of the posterior distribution
  enables assessment of parameter covariance that includes the full
  error model.  For the \sersic\ model, the magnitude, $r_e$, $n$ and the sky
  background are strongly correlated, and the strength of the
  correlation increases with \sn.  The covariance of the sky
  background with the other model parameters, i.e. the magnitude,
  $r_e$, and $n$, shows that a reliable inference requires an accurate
  characterisation of the background noise.  We fully expect that
  analogous trends will obtain for most parametric models.

\item \emph{\sersic\ index $n$.}\ The parameter correlation increases with
  increasing \sersic\ index $n$.  As a consequence, the confidence intervals for
  magnitude, $r_e$, and $n$ for galaxies with $n>2.0$ are roughly
  three times larger than those for $n\le2.0$.  However, the marginalised
  error posteriors of the sky background for these two groups do not have
  significantly different widths. Again, this underscores the need for
  the entire posterior distribution for subsequent inferences based on
  morphological quantities.

\item \emph{Galaxy $r_e$.}\ If $r_e$ is smaller than the PSF HWHM,
  most parameters are poorly constrained and the posterior maximum is
  biased.  For example, the posterior distribution of $r_e$ has a
  significant probability at larger $r_e$ than the true value for a
  non-informative prior.  Similarly, the inference of axis ratio $q$
  will be positively biased for intrinsically small $q$.  Both biases
  lead to an overestimate of the total flux.  As a galaxy's $r_e$ becomes
  larger compared with the PSF HWHM, this bias decreases.  Also, the
  bias decreases with increasing \sn\ even if $r_e$ is smaller than
  the PSF HWHM.

\item \emph{Image size.}\ We find that the inferred value for the sky
  background is nearly symmetric about the true value even for an
  $n=4$ \sersic\ profile with an image size of $8 r_e$. The background
  bias disappears and the uncertainty drops dramatically for image
  sizes larger than approximately $10 r_e$. The covariance of sky
  background with galaxy surface brightness parameters motivates
  including the sky background in the overall model to ensure that the
  correlations are represented in the posterior probability
  distribution.

\item \emph{PSF variation.}\ Errors in the assumed PSF width introduce
  a bias in the posterior distributions of galaxy size $r_e$ and 
  \sersic\ index $n$. If the PSF FWHM is overestimated,
  the galaxy's $r_e$ is underestimated and its $n$ is overestimated, and vice
  versa. The bias in the inferred value of $n$ increases with
  $n$.  The bias increase is larger
  if the PSF width is overestimated rather than underestimated.

\item \emph{Run time.}\ \galphat 's\ run time depends on the Monte Carlo
  algorithm, the desired size of the converged sample, the image size, and
  the model complexity (i.e. the number of model parameters and the 
  parameter covariance) and the computational complexity.  
  Based on extensive experiments, we found
  that the PHS MCMC algorithm efficiently sampled the posterior for most of
  our \sersic\ model tests.  The typical wall clock times for generating 
  40,000 converged posterior MCMC samples of galaxies with 
  image sizes of $226\times226$ and $100\times100$ pixels, using \sersic\ model 
  are about 1.5 (with 8 CPUs) and 0.4 (with 8 CPUs) hours, respectively, 
  using 2GHz AMD quad-core Opteron processors.

  Although the optimised algorithms used in \galphat\ significantly
  improves the likelihood computation time, it is still much slower
  than other conventional galaxy image decomposition algorithms.
  However, the existence of posterior probability distributions for an
  ensemble of galaxies enables reliable inferences for models of
  galaxy formation and evolution, and we feel that this more than
  compensates for the increased computational overhead.  Moreover, the
  overhead will continue to decrease with the increasing availability and
  performance of HPC-class facilities.

\end{enumerate}

\subsection{Future work}

We have investigated a two-component bulge and disc model composed of
\sersic\ bulge with varying $n$ and exponential disc component with
$n=1$ with mutual independent prior distributions.  This naive model
exhibits strong correlations between multiple dimensions in parameter
space, and these occasionally lead to poorly mixing Markov chains.
From this preliminary study, we are convinced that a thoughtful prior
distribution that removes astronomically unnatural regions of
parameter space is required for production work.  \citet{yoon2010b}
investigate \galphat\ two-component bulge and disc model fits and introduce
an informative prior based on typical 2MASS survey data. Using
both bulge and disc two-component and single \sersic\ models, we study
the sensitivity of prior-distribution choice on the inference of
galaxy parameters, demonstrate the practicality of model selection,
the preference of a two-component bulge and disc model versus a single
\sersic model, using Bayes factor analyses,
and illustrate the application of posterior distributions from an ensemble
of galaxy images to large-scale inference problems.

In addition, we are currently using \galphat\ to study the morphology
of a \Ks-band magnitude limited sample of 2000 2MASS galaxies in the
SDSS footprint, and will accurately characterise the luminosity
functions of each component, i.e. the bulge and the
disc, separately, and investigate any intrinsic correlations between
the model parameters and any correlations with external galaxy
properties such as star formation rate and environment in a
forthcoming paper.  We hope that \galphat\ will become a well used
tool to aid in our understanding of galaxy properties in the near future.

\section*{Acknowledgements}

This work was supported in part by NSF IIS Program through award
0611948 and by NASA AISR Program through award NNG06GF25G.  We thank
Craig West for his support of the UMASS Astronomy HPC Linux cluster. We also
thank Daniel McIntosh and Yicheng Guo for providing sample images and
useful discussions in the early stages of this work.

\bibliography{galphat1}
\bibliographystyle{mn2e}

\label{lastpage}
\null
\end{document}